\documentclass[usenatbib]{mnras}

\usepackage[T1]{fontenc}
\usepackage{ae,aecompl}
\usepackage{graphicx}	
\usepackage{amsmath}	
\usepackage{amssymb}	
\usepackage{subfig}
\usepackage{float}
\usepackage{textcomp}
\usepackage[usenames,dvipsnames]{color}
\title[Absorption systems at $z\sim2$ as a probe of CGM]{Absorption systems at $z\sim 2$ as a probe of the circumgalactic medium: a probabilistic approach}

\author[Mongardi et al.]
{C. Mongardi$^{1}${\thanks{email: astro386@hotmail.it}}, M. Viel$^{2,3,4}${\thanks{email: viel@sissa.it}}, V. D'Odorico$^{3}${\thanks{email: valentina.dodorico@inaf.it}}, T.-S. Kim$^{3,5}$, P. Barai$^{3,6}$, G. Murante $^{2}$,\and P. Monaco$^{1,3,4}$\\
 $^{1}$  Astronomy Unit, Department of Physics, University of Trieste, via Tiepolo 11, I-34131 Trieste, Italy\\ 
 $^{2}$  SISSA-International School for Advanced Studies, Via Bonomea 265, I-34136 Trieste, Italy\\ 
 $^3$ INAF\_Osservatorio Astronomico di Trieste, Via G.B. Tiepolo 11, I-34131 Trieste, Italy \\ 
 $^4$ INFN/National Institute for Nuclear Physics, Via Valerio 2, I-34127 Trieste, Italy\\
 $^5$ Department of Astronomy, University of Wisconsin, 475 N. Charter St., Madison, WI 53706, USA\\
 $^6$  Instituto de Astronomia, Universidade de Sao Paulo (IAG-USP), Rua do Mat\~{a}o 1226, BR-05508-090 Sao Paulo, Brazil}

\date{Accepted: 2018 May 14. Received: 2017 June 26. In original form: 2018 May 11}

\pubyear{2018}

\begin{document}
\label{firstpage}
\pagerange{\pageref{firstpage}--\pageref{lastpage}}
\maketitle

\begin{abstract}
  We characterize the properties of the intergalactic medium (IGM)
  around a sample of galaxies extracted from state-of-the-art
  hydrodynamical simulations of structure formation in a cosmological volume of 25 Mpc comoving at
  $z\sim 2$. The simulations are based on two different subresolution
  schemes for star formation and supernova feedback: the MUlti-Phase
  Particle Integrator (MUPPI) scheme and the Effective Model. 
  We develop a  quantitative and probabilistic analysis based on the apparent optical depth method of the properties of the absorbers as a function
  of impact parameter from their nearby galaxies: in such a way we probe different
  environments from circumgalactic medium to low-density filaments.
  Absorbers' properties are then compared with a spectroscopic  observational data set obtained from high-resolution quasar spectra. 
  Our main focus is on the N$_{\rm {C\, \textsc{iv}}}$-N$_{\rm {H\, \textsc{i}}}$ relation around simulated galaxies: the results obtained with MUPPI and the Effective  model are remarkably similar,
  with small differences only confined to regions at impact parameters $b = [1-3] \times r_{\rm {vir}}$.
  Using $\mbox{C\,{\sc iv}}$ as a tracer of the metallicity, we obtain
  evidence that the observed metal absorption systems have the highest
  probability to be confined in a region of 150-400 kpc around
  galaxies. Near-filament environments have instead metallicities too low to be
  probed by present-day telescopes, but could be probed by future
  spectroscopical studies.
  Finally we compute \mbox{C \,{\sc iv}} covering fractions which
  are in agreement with observational data.
    


\end{abstract}




\section{Introduction}
\label{s:intro}

The matter that fills the space between galaxies, the so-called intergalactic medium (IGM), is distributed
as a cosmic web of filaments and sheets: a component which has too low densities to be probed in emission, but can be observed through absorption lines in the spectra
of luminous background sources. There are still significant uncertainties
related to galaxy formation and evolution and to the modeling of the baryon's physical processes. 
In this context, the study of the IGM and of its evolution is of great relevance.
First of all, the IGM is the reservoir of baryons from which galaxies form and it is the fuel necessary for sustaining
star formation (SF), while at the same time it is being replenished with both newly accreted intergalactic gas and chemically
enriched materials from galaxies, carrying the imprints of galactic feedback.
Therefore, intergalactic space provides a critical laboratory for studying the baryon cycle that regulates SF
and galaxy growth.

In particular, IGM metal enrichment has gained a lot of interest after the second half of the 1990s
\citep[e.g.][]{Nath1997,Murakami1997,Dave1998,Gnedin1998}. In fact, 
it was originally thought that the gas infalling onto galaxies had a primordial composition, as heavy elements can be produced
only inside stars and SF is not present in the low-density and high-temperature IGM.
Before the era of high-resolution spectroscopy combined with large telescopes, in fact, quasars (QSO) absorption
lines were classified into two categories: 1) metal-enriched high \mbox{H \, \sc{i}} column density absorbers; and 2) metal-free low  \mbox{H \, \sc{i}} column
density absorbers (N$_{\rm {H\, \textsc{i}}} < 10^{17} $ cm$^{-2}$). Thus, confirming the above-mentioned idea.

With the advent of powerful spectrographs, like HIRES (HIgh Resolution Spectrometer) at the Keck telescope and UVES (Ultraviolet and Visual Echelle Spectrograph) 
at the ESO-VLT (European Southern Observatory-Very Large Telescope),
absorption lines of the triply ionized carbon doublet \mbox{C\,{\sc iv}} ($\lambda\lambda$ 1548.204, 1550.778), 
the most common metal transition found in QSO spectra,  have been observed at $z\sim 3$  associated with {\mbox{H\,{\sc i}}} lines of the so-called Ly$\alpha$ forest with 
N$_{\rm {H\, \textsc{i}}} \geq 10^{14.5} $ cm$^{-2}$ \citep{Cowie1995,Tytler1995,Songaila1998}.


Different scenarios have been proposed for the IGM metal-enrichment, such as an early-enrichment by Population III stars at 
very high redshift (10$<z<$20, e.g. \citet[][]{OstrikerGnedin1996,HaimanLoeb1997}), dynamical removal by galaxy mergers
\citep[e.g.][]{GnedinOstriker1997,Aguirre2001}, or galactic outflows/winds at lower redshift ($z\sim$2-3, e.g.
\citet[][]{Dave1998,Aguirre2001b,Schaye2003,SpringelHernquist2003,Murray2005,OppenheimerDave2006}). 

In the first scenario, outflows from early protogalaxies drive the enrichment: metals can easily escape from the shallow potential 
wells, polluting large comoving pristine regions of the IGM. The enriched regions evolve following the Hubble flow likely
resulting in a uniform distribution of metals far from galaxies at the studied redshift ($z\sim$2-3).

The last scenario predicts a metallicity-overdensity relation \citep{Aguirre2001,Dave1998,OppenheimerDave2006}, resulting
in a non-uniform metal pollution of the IGM.  The dynamical removal of metal-enriched gas by galaxies mergers has attracted less interest, as it cannot
account for observed present-day mass metallicity relation \citep{Aguirre2001}.

Most probably, IGM metal pollution is not due to a single mechanism, although the late enrichment scenario could have
a dominant effect, as it seems to have the greatest observational support. In fact, recent observations of local starbursts
\citep[e.g.][]{Martin1999,Martin2005,Rupke2005,Combes2006,Georgakakis2007,Lemaux2014,Hinojosa2016} and also the detection of strong outflows around
almost all galaxies at high redshift \citep[e.g.][]{Heckman2000,Pettini2001,Frye2002,Shapley2003,Martin2005} have provided
new insights into the nature and impact of supernova (SN)-driven winds, which can result in very fast and energetic outflows.
From an observational point of view, several studies have been carried out, in order to test which enrichment scenario has
a dominant effect and to study IGM/galaxies interactions. Efforts have been concentrated on two main approaches: the investigation of the level of pollution in the
low-density gas \citep[e.g.][]{Ellison2000,Schaye2003,Aracil2004,Dodorico2016} and the study of galaxy/absorbers relations.
The latter is based on the detection of metal absorption lines in 
QSO spectra correlated with the position and redshift of galaxies in the same field of view, both at high \citep{Adelberger2005,
Steidel2010,Crighton2011,Turner2014,Turner2015} and low redshifts \citep{Prochaska2011,Tumlinson2011,Werk2013,Bordoloi2014,LiangChen2014,Borthakur2015,Borthakur2016}.

These studies, in general, allow only a statistical insight into the problem. The few most ambitious surveys for weak
associated metal absorption features are not yet able to reach the shallow densities characteristic of the IGM, while the bulk 
of observations probes only the denser portions of the Universe, by detecting systems with $\log(\rm N_{\rm {C\, \textsc{iv}}}/\mathrm{cm}^{-2}) \gtrsim 12$ 
associated with $\log(N_{\rm {H\, \textsc{i}}}/\mathrm{cm}^{-2}) \gtrsim 14.5$, not providing sufficient evidence to discriminate between
the enrichment scenarios. On the other hand, galaxy/absorbers relations have mainly investigated regions around galaxies limited 
to a few hundreds of proper kpc.
Moreover, galaxy surveys are flux limited, introducing an unavoidable bias in galaxy/absorbers relations,
as the faintest galaxies are missed, resulting in a sparse picture of the circumgalactic medium (CGM)/IGM and of the true origin of the absorbers.
Numerical  hydrodynamical simulations that incorporate the relevant physical processes have emerged as a powerful tool to investigate these issues \citep{theuns02,cen05,tescari09,tornatore10,tescari11,barnes11,stinson12,dellavecchia12,barai13,oppe13,crain13,hirschmann13,voge14,ford14,schaye15,dave17}.

In this work, we aim at characterizing the properties of the CGM/IGM around galaxies at high redshift, using hydrodynamical
simulations compared with observational data taken from the literature. We focused on an analysis of a sample of simulated
galaxies and the environment around them using two different simulations. The goal was to test different subgrid schemes, in
order to better understand the chemical properties of the environment around galaxies and in particular to quantify the
physical scales up to which observed state-of-the-art absorption systems can be found.
This paper presents a probabilistic approach to the galaxy/IGM interplay that has the goal to provide the first step towards 
a characterization of the intimate relation between the two.

In this work, we define the CGM as the region surrounding the galactic halo up to a physical distance from galaxy centre of $\sim$300-400 kpc,
(which at $z\sim 2$ corresponds to few times the virial radius for a galaxy with a halo mass of $\sim 10^{10}-10^{11} \, \rm M_\odot$), while 
the IGM refers to larger distances and smaller overdensities. 

This paper is organized as follows. Section~\ref{s:sim} describes the set of used simulations. Section~\ref{s:data} reports
a description of the observational samples used to compare with.
In Section~\ref{s:simdatasample}, we describe the selection of the sample of simulated galaxies (Appendix A contains the details for the extraction of mock spectra). In Section~\ref{s:results}, the main results
of this work are presented, while in Section~\ref{s:conc}, the summary of our main conclusion is reported. In 
Appendix~\ref{s:Ncomparison},
we described the method used to compute column densities for simulated absorption systems and we discuss the implications of 
using different methods in the simulation, when comparing with observational data, whose properties are obtained using
Voigt Profile Fitting codes. In Appendix~\ref{ss:CDDF}, we report the column density distribution functions (CDDFs) of \mbox{H\sc{i}} and \mbox{C\sc{iv}} absorption 
systems used in Appendix~\ref{s:Ncomparison}. In Appendix~\ref{s:figures}, we report the numerous figures, whose discussion is reported in 
Section~\ref{ss:CompMUPPIEFF}.

%
\section{Simulations}
\label{s:sim}
Our simulations were performed using the \textsc{gadget-3} code \citep{Springel2005} which uses a \textsc{treepm} (particle mesh) gravity solver algorithm and the gas dynamics is computed with smooth particle hydrodynamics (SPH). 

Two cosmological volumes of size  $L_{\rm box} = 25$  Mpc comoving are evolved, starting from the same initial conditions, from $z=199$ up to $z=0$ with periodic boundary
conditions. 
Each volume uses $N_{\rm part}$ = 2 $\times$ 256$^3$ particles in the initial condition, half of which are dark matter (DM) particles and half are gas 
particles. 
DM and gas particle masses have the following values: $m_{\rm DM}$ = 2.68 $\times$ 10$^7$ M$_\odot$ and $m_{\rm gas}$ = 5.36 $\times$ 10$^6$ M$_\odot$,
respectively. 

In both models, we
consider energy feedback driven by SNe [while active galactic nucleus (AGN) feedback is not present];  adopt SF models with
metal-dependent cooling using the prescriptions of \citet{Wiersma2009};  assume the gas is dust-free, optically thin
and in photoionization equilibrium, heated by a uniform photoionizing background [cosmic microwave background plus the \citet{HaardtMadau2001}
model for the ultraviolet (UV)/X-ray];  use the chemical enrichment and stellar evolution model by \citet{Tornatore2007}, where
each star particle is treated as a simple stellar population and the production of 11 chemical elements 
(H, He, C, Ca, O, N, Ne, Mg, S, Si, and Fe) is followed. Different stellar yields are considered: from Type Ia SN \citep{Thielemann2003}, Type II SN 
\citep{WoosleyWeaver1995} and asymptotic giant branch 
stars \citep{van_den_HoekGroenewegen1997}, while stellar lifetimes are taken from \citet{PadovaniMatteucci1993}. 
A stellar initial mass function by \citet{Kroupa1993} in the range [0.1-100] M$_{\odot}$ is used.

The two boxes differ in the baryonic interaction prescriptions governing SF and stellar/SN 
feedback, which are subresolution phenomena and, thus, they have to be modelled ad hoc. 
One simulation is described in \citet{Barai2015}, from where we used the run E25cw. The other one is very similar to the run M25std from \citet{Barai2015}, but
with slightly different stellar yields and feedback parameters: $f_{\rm b,out}$ = 0.2, $f_{\rm b,kin}$=0.5, and $P_{\rm kin}$= 0.02. 
(The parameters $f_{\rm b,out}$, $f_{\rm b,kin}$ and $P_{\rm kin}$ will be described in the next subsection.)
Two different sub-resolution models were adopted: the MUPPI model \citep[MUlti-Phase Particle Integrator;][]{Murante2010,Murante2015} in run M25std, 
and the Effective Model \citep{SpringelHernquist2003} in run E25cw, whose main features are described in the following subsections. 
We refer the reader to the above mentioned papers for further details on the implementation of the two models. 

A flat $\Lambda$ cold dark matter model is used with the following parameters: $\Omega_{\rm m}$ = 0.24, $\Omega_{\Lambda}$ = 0.76, 
$\Omega_{\rm b}$ = 0.04, H$_0$ = 72 km s$^{-1}$ Mpc$^{-1}$ and $\sigma_8=0.83$.
%
%
%

\subsection{MUPPI Model}
\label{ss:MUPPI}
In the MUPPI model, 
whenever a gas particle's density is higher than a threshold value $n_{\rm {thr}}=0.01$ cm$^{-3}$ and its temperature is below the threshold 
$T_{\rm {thr}}$=5 $\times$ $10^4$ K, it enters a multiphase regime. 
The multiphase gas particle is composed of a hot and a cold phase in thermal pressure equilibrium ($n_{\rm c} T_{\rm c} = n_{\rm h} T_{\rm h}$), 
plus a virtual stellar component. 
The temperature of the cold phase is kept fixed at T$_{\rm c}$ = 300 K, while the hot phase T$_h$ is computed from the particle's entropy.

A fraction of the cold gas, $f_{\rm mol}$, is considered to be in the molecular phase. 
It is the reservoir of material available for SF. 
It is computed following the observed relation by \citet{BlitzRosolowsky2006} between the ratio of molecular to atomic gas surface 
densities and the external pressure exerted on molecular clouds. 
The external pressure is the mid-plane pressure of a thin disc composed of gas and stars in vertical hydrostatic equilibrium. 
In the MUPPI model, the hydrodynamical pressure of gas particles is used in place of the external pressure. 
This enables us to derive the following simple relation for computing $f_{\rm mol}$:
\begin{equation}
 f_{\mathrm {mol}} = \frac{1}{1+P_0/P} . 
\end{equation} 
Here P is the pressure of the gas particle and $P_0$ is the pressure at which half of the cold gas is molecular and is set to the 
value $P_0/k_B$ = 35000 K cm$^{-3}$ \citep{BlitzRosolowsky2006}. 

The SF rate is directly proportional to $f_{\rm mol}$ and is given by this equation:
\begin{equation}
 \dot{M}_{\rm {SF}}=f_\star \frac{f_{\mathrm {mol}} M_{\mathrm c}}{t_{\mathrm {dyn}}} .
\end{equation} 
Here, $f_\star$ is the SF efficiency, M$_c$ is the gas mass in the cold phase and 
$t_{\rm {dyn}}$ is the dynamical time, $t_{\mathrm {dyn}}= \sqrt{3 \upi / 32 G {\rho_c}}$, with $\rho_c$ being the cold phase density. 

It is important to highlight that in this model, SF is dependent on disc pressure. 
Thus no Schmidt-Kennicutt relation is imposed in the MUPPI model. 
Rather, as demonstrated in \citet{Monaco2012}, the Schmidt-Kennicutt relation is naturally recovered from the model. 

Mass and energy exchanges between the three gas phases (hot, cold, and stellar) are described by a set of equations. 
When a gas particle enters the multiphase, all the particle's mass is assigned to the hot phase. 
Then, matter flows among the three phases in the following way: 
cooling deposits hot gas into the cold component; 
evaporation, happening under the action of hot SN bubbles, brings cold gas back to the hot phase; 
SF deposits mass from the cold phase into the stellar component; and
mass restoration from dying stars moves mass from the stellar component back to the hot phase. 

A star particle is produced following the stochastic star formation algorithm of \citet{SpringelHernquist2003}. 
A multiphase gas particle undergoing SF is turned into a collisionless star particle whenever a random number drawn uniformly from the 
interval [0,1] falls below the probability
\begin{equation}
 P=\frac{M_p}{m_\star} \left[ 1-\exp\left( - \frac{\Delta M_\star}{M_p}\right)\right] . 
\end{equation} 
Here $M_{\rm p}$ is the gas particle mass (hot mass + cold mass + stellar mass), and $\Delta M_\star$ is the cold gas mass transformed to 
stellar mass in a single SPH time-step. 
The mass of the star particle, $m_\star$, is defined as $m_\star = M_{\rm p} / N_{\rm g}$ with $N_{\rm g}$ being the number of generations of stars per 
gas particle (with $N_{\rm g}$ set equal to 4, $m_\star$=1.34 $\times$ 10$^6$ M$_\odot$). 

Energy feedback from SN is transferred to gas particles both in the form of thermal and kinetic energy. 
Thermal energy is given to the hot phase, which is lost by cooling and acquired from SN explosions. 
The thermal energy available for feedback from a single gas particle and returned to the hot phase is a fraction $f_{\rm {fb,out}}$ of the 
energy of a single SN. 
The model assumes to have one SN event per M$_\star$ = 120 M$_\odot$. 
A fraction of this energy is given to the local hot phase to sustain the high temperature of the particle itself. 
The remaining energy is redistributed to the hot phase of neighbouring gas particles within its SPH smoothing length and 
lying in a bicone of aperture $\theta$ = 60$^\circ$. 
The bicone axis is aligned along the least resistance direction of the gas density. 
This is to simulate the explosion of SN bubbles. 

Kinetic feedback is implemented in the following manner. 
When a particle exits from the multiphase regime, it is assigned a probability P$_{\rm {kin}}$ to become a wind particle. 
Then, it can receive kicks from neighbouring gas particles for a given time $t_{\rm {wind}}$. 
The kinetic energy available for feedback from a single gas particle is a fraction $f_{\rm {b,kin}}$ of the energy of a single SN. 
It is distributed to neighbouring wind particles in a cone within the gas particle's SPH smoothing length in a similar manner as 
for the thermal feedback. 
For each wind particle, the total kinetic energy available from all neighbouring kicking gas particles is first computed, as described 
above. 
Then the wind particle's kinetic energy is increased by this total amount. 
The velocity vector of the wind kick is oriented by energy weighting among all the kicking particles. 

Unlike other feedback prescriptions in the literature \citep{Oppenheimer&Dave2008,Schaye2010,SpringelHernquist2003}, the MUPPI model 
depends only on the local properties of the gas. 
The mass-loading factor or the velocity of the wind is not input parameters of the feedback model. 
However these quantities can be estimated empirically as described in \citet[estimated mass load factor of $\sim$1.5 and estimated average wind 
velocity of $\sim$600 km s$^{-1}$]{Murante2015}. 

A gas particle stays in the multiphase regime until its density reaches a value equal to 
or lower than 1/5 of the density threshold $n_{\rm {thr}}$. 
If the energy feedback is incapable to sustain the hot phase resulting in hot phase temperatures that stay lower than 10$^5$
K, the particle is forced to escape from the multiphase regime. 

The MUPPI model is able to produce realistic disc galaxies even at
relatively low resolution such as those used in this paper, as shown
in Murante et al (2015). In \citet[figs. 1 and 2]{Goz2017} a summary of the main
galaxy properties  found in the MUPPI  box used here is given. In particular, their fig. 1 shows that the box
contains a fair number of galaxies with relatively  small dynamical
values of  the bulge-over-total stellar mass ratio. In \citet{Barai2015}, the SF rate density of a number of different SF
and feedback model is shown (models used here are M25std and
E25cw, as said before).  The effective model has a larger SF at high
redshifts and a lower one at low ones. In the same paper, an analysis
of the outflows shows that MUPPI is more efficient than the effective
model in expelling the gas at high redshift, leaving it available for
low-redshift disc formation when it falls back, while the effective
model converts it in stars and produces less discs.
\subsection{Effective Model}
\label{ss:EFFModel}
In the Effective model, 
a multiphase gas particle is composed of a hot and a cold phase in thermal pressure equilibrium. 
Gas particles enter a multiphase regime whenever their density is higher than a threshold value $\rho_{\mathrm{SF}}=0.13$ cm$^{-3}$. 
This threshold is a SF density threshold, as SF prescription is not based on the \citet{BlitzRosolowsky2006} relation. 
It depends only on the cold gas mass, which is directly converted into stars on a characteristic time-scale, given by the formula:
\begin{equation}
  \dot{M}_{\mathrm {SF}}= \frac{ M_{c}}{t_\star}
\end{equation}
where $t_\star=t_{\star,0}\, \sqrt{\rho /\rho_{\mathrm{SF}}}$. 
A value of t$_{\star,0}$ = 2.1 Gyr was chosen by \citet{SpringelHernquist2003} in order to fit the Kennicutt relation. 

Mass and energy exchanges are regulated by the same physical processes described in the MUPPI section. 
Star particles are generated from gas particles using the stochastic scheme introduced by \citet{KatzWeinbergHernquist1996}. 

Energy feedback is given both in the form of thermal and kinetic energy. 
Differently from the MUPPI model, the two forms of feedback do not consider any distribution of the energy to neighbouring particles 
in a cone within the SPH smoothing length simulating the blowout of an SN bubble. 
Thermal feedback in the form of thermal heating and cloud evaporation is implemented.

Kinetic feedback uses the prescriptions of the {\it energy-driven wind} scenario with a constant wind velocity. 
The wind mass-loss rate is given by this formula:
\begin{equation}
 \dot{M}_w = \eta \dot{M}_{\rm{SF}}
\end{equation}
where $\eta$ is the mass-loading factor, which is set to $\eta$ = 2. 
A fixed fraction $\chi$ of the SN energy is converted into wind kinetic energy:
\begin{equation}
 \frac{1}{2}\dot{M}_w v_w^2=  \chi \epsilon_{SN} \dot{M}_{\rm{SF}} . 
 \end{equation} 
 Here $v_w$ is the wind velocity and $\epsilon_{SN}$ is the average energy released by SN for each $M_\odot$. 
 So we have: 
\begin{equation}
 v_w=\sqrt{\frac{2 \chi \epsilon_{SN}}{\eta}}
\end{equation}
In our run we use a fixed value $v_w$ = 350 km s$^{-1}$. 
Gas particles are given wind kick using a probabilistic approach (see Equation 10 in Barai et al. 2013 for details). 
Their velocity $v$ is incremented according to: 
\begin{equation}
 v'=v+v_w \hat{n}
\end{equation}
where $\hat{n}$ is the direction of the wind, preferentially chosen along the rotation axis of spinning objects.

\section{Observational data sample}
\label{s:data}
Here, we report a brief description of the observational data used.
The comparison with simulated \mbox{C\,{\sc iv}} and \mbox{Si\,{\sc iv}} absorption systems has been made using the observational
sample by \citet[][hereafter K16]{Kim2016}, which are described in Section~\ref{ss:dataKim}, while \mbox{O\,{\sc vi}} is compared with the observational
sample by \citet{Muzahid2012}, described in Section~\ref{ss:dataOx}. Constructed \mbox{C\,{\sc iv}} covering fractions are compared with those derived by
\citet{Landoni2016,Prochaska2014} and \citet{Rubin2015}, which are described in Section~\ref{ss:dataCov}.
\subsection{\mbox{C\,{\sc iv}} and \mbox{Si\,{\sc iv}} data sample}
\label{ss:dataKim}
Data for the comparison with the simulated \mbox{C\,{\sc iv}} (and also \mbox{Si\,{\sc iv}}) absorption are taken from K16.
This data set is chosen because it represents a homogeneous and
statistically significant sample of high resolution absorption lines.
The sample consists of 23 QSOs in the redshift range $2\lesssim z\lesssim$3.5, which were observed with
the VLT/UVES (21 spectra) and Keck/HIRES (2 spectra)
instruments.
Spectra were first presented in the following papers: \citet{Kim2004,Kim2007,Kim2013}, and \citet{BoksenbergSargent2015}.

Spectral resolution is $R\sim45000$ ($\sim$ 6.7 km s$^{-1}$), while typical signal-to-noise ratios (S/N) per pixel are of the order of 30$\sim$50 
for the Ly$\alpha$ forest and $\sim$100 for the \mbox{C\,{\sc iv}} region.
 
\mbox{H\,{\sc i}} and \mbox{C\,{\sc iv}} absorption lines detected in all the QSO spectra are in the redshift range 
$1.5\lesssim z\lesssim3.3$. 

Data points, used for comparison with simulations in Section \ref{s:results},  are integrated 
column densities representing \mbox{H\,{\sc i}} and \mbox{C\,{\sc iv}} $systems$, as defined in K16.
First, the entire normalized spectra were inspected to identify the \mbox{C\,{\sc iv}} doublets.
Then the absorption lines were fitted with Voigt profiles using the \textsc{vpfit} 
code\footnote{http://www.ast.cam.ac.uk/~rfc/vpfit.html}
\citep{Carswell2014}, decomposing the complex
absorption features into individual, single component by minimizing the normalized X$^2$ of $\sim$1.3.
The output of the fit consists of the physical parameters, such as the column density {in cm$^{-2}$}, the redshift and the Doppler
parameter {in km s$^{-1}$}, of the single fitted components.

{Then, the \mbox{C\,{\sc iv}} systems are defined as all the individually fitted \mbox{C \,{\sc iv}} 
components 
within the fiducial $\pm$150 km s$^{-1}$ interval centred at the \mbox{C \,{\sc iv}} flux minimum of a single or multiple 
\mbox{C \,{\sc iv}}
feature. The integrated column density is then 
calculated by adding up all the column densities of the \mbox{C \,{\sc iv}} components within $\pm$150 km s$^{-2}$.}

If the \mbox{C \,{\sc iv}} absorption extends over $\pm$150 km s$^{-1}$ or when two different absorption 
profiles are found to
lie within $\pm$150 km s$^{-1}$ but they are spread beyond the $\pm150$ km s$^{-1}$, then the velocity range
is extended by a step of 100 km s$^{-1}$ in order to include all the \mbox{C\,{\sc iv}} absorption.

The \textit{associated} \mbox{H \,{\sc i}} system is obtained by applying the same velocity range as the 
\mbox{C \,{\sc iv}} system. 
While the fiducial integration velocity range of $\pm$150 km s$^{-1}$ could be considered arbitrary, the choice of 
the velocity range does not affect the observed integrated N$_{\rm{ H\, \textsc{i}}}$ versus N$_{\rm{ C\, \textsc{iv}}}$ relation (see beginning of Section~\ref{ss:Nrelation1}
for a brief explanation of this relation) as long as it is larger 
than 100 km s$^{-1}$ (K16).
With this working definition, the \mbox{ C\,{\sc iv}}
centroid does not necessarily coincide with the \mbox{H \,{\sc i}} centroid and there is no 
evidence that \mbox{C \,{\sc iv}} and \mbox{H \,{\sc i}} systems are physically connected. This definition is thus more related
to volume-averaged quantities, commonly used in numerical simulations.

Finally, it is demonstrated in K16, that the N$_{\rm{ C\, \textsc{iv}}}$ versus 
N$_{\rm{ H\, \textsc{i}}}$ relation converges when considering velocity integration ranges greater than 70 km s$^{-1}$.

In order to be consistent with data, we used the same velocity integration range in the analysis of the simulated absorption
systems.

We compare the data only with one snapshot of the simulations at $z\sim1.94$, since we verified that there is no significant evolution in the N$_{\rm{ C\, \textsc{iv}}}$ vs 
N$_{\rm{ H\, \textsc{i}}}$ relation in the redshift range covered by the K16 sample (see also the discussion in K16).
\subsection{\mbox{O\,{\sc vi}} data sample}
\label{ss:dataOx}
Comparison with simulated \mbox{O\,{\sc vi}} absorption systems is performed with the observational sample by \citet{Muzahid2012}. The sample is formed by 18 QSOs
in the redshift range $2.1\lesssim z\lesssim$3.3, observed with the VLT/UVES. Typical S/N$\sim$30-40 and 60-70 are achieved in the wavelength range of 3300 and 5500
\AA{}. Spectral resolution is $R\sim$45000 (FWHM $\sim$6.6 km s$^{-1}$) over the entire wavelength range.

\mbox{H\,{\sc i}} and \mbox{O\,{\sc vi}} absorption lines detected in all the QSO spectra are in the redshift range 
$1.9\lesssim z\lesssim2.9$.

They define an absorption system by grouping together all the lines whose separation from the nearest neighbour is less than a linking length $v_{\rm {link}}$=100
km s$^{-1}$, following the prescriptions by \citet{Scannapieco2006}.
\subsection{\mbox{C\,{\sc iv}} covering fraction data sample}
\label{ss:dataCov}
For the comparison with the constructed covering fractions (see Section~\ref{ss:coveringfraction}), we used the observational 
samples of \citet{Rubin2015}, \citet{Landoni2016} and \citet{Prochaska2014}.
\citet{Landoni2016} and \citet{Prochaska2014} use QSO projected pairs, that is they use the spectrum of a background QSO to 
study the gaseous envelope of a foreground QSO host galaxy. 

The sample of \citet{Prochaska2014} consists of  427 QSO pairs, whose spectra were taken mainly from the BOSS (Baryon Oscillation Spectroscopic Survey) or obtained with
the Keck/LRIS (Low Resolution Imaging Spectrometer) instrument, with impact parameters in the range 39 kpc $<b<1$ Mpc and redshifts $1.8<z<3.5$.
The 18 QSO pairs by \citet{Landoni2016} were taken from the SDSS (Sloan Digital Sky Survey) and have impact parameters smaller than $\rm b<$200 kpc and redshifts $2.0<z<2.8$.
The main difference between these two samples is that \citet{Prochaska2014} perform a boxcar integration in a 3000 km s$^{-1}$ velocity window centred on the 
redshift of the foreground QSO, while \citet{Landoni2016} integrate on a window of 1200 km s$^{-1}$.

\citet{Rubin2015} use, instead, QSO pairs to study the diffuse gas in the CGM of 40 damped Lyman-$\alpha$ \citep[DLAs,][]{Wolfe1986} systems, that is the primary sightline probes
an intervening DLA system in the redshift range 1.6$<z<$3.6, while the secondary sightline is used to look for a 
\mbox{C\,{\sc iv}} absorption at the same redshift of the DLA, up to a distance of the order $b<$300 kpc.

\section{Simulated data sample}
\label{s:simdatasample}

\begin{table}
	\centering
	\caption{Properties of the sample galaxies of the MUPPI and the Effective Model Boxes. In the first column of the two
	tables, there is the object ID; in the second column, the halo mass of the galaxy; in the third column, the 
	stellar mass of the galaxy; and in the fourth column, the virial radius of the galaxy in physical units.}
	\label{tab:Galaxies}
		\begin{tabular}{lccr} 
		\hline
		\multicolumn{4}{c}{MUPPI model}\\
		\hline
		Obj. & M$_{h}$ & M$_\star$ & R$_{\rm vir}$\\
		     & (M$_\odot$) & (M$_\odot$)& (phys. kpc)\\
		\hline
		1 & 6.87$\times$10$^{11}$ & 2.37$\times$10$^{10}$  & 94\\
		2 &  4.03$\times$10$^{11}$ &1.13$\times$10$^{10}$  & 78\\
		3 & 4.02$\times$10$^{11}$  & 8.31$\times$10$^{9}$  & 78\\
		4 & 3.63$\times$10$^{11}$  & 1.11$\times$10$^{10}$  & 76\\
		5 &  2.74$\times$10$^{11}$& 1.24$\times$10$^{9}$ & 69\\
		6 & 2.60$\times$10$^{11}$ & 6.87$\times$10$^{9}$ & 68\\
		7 &  2.08$\times$10$^{11}$& 6.83$\times$10$^{9}$ & 63\\
		8 & 1.66$\times$10$^{11}$ & 5.92$\times$10$^{9}$ & 58\\
		9 &  1.43$\times$10$^{11}$& 5.04$\times$10$^{9}$ & 55\\
		10 & 1.06$\times$10$^{11}$ & 3.38$\times$10$^{9}$ & 50\\
		11 & 9.49$\times$10$^{10}$ &2.54$\times$10$^{9}$  & 48\\
		12 & 9.22$\times$10$^{10}$ & 1.95$\times$10$^{9}$ & 48\\
		13 &  9.12$\times$10$^{10}$&2.84$\times$10$^{9}$ & 48\\
		14 & 8.91$\times$10$^{10}$ &  2.53$\times$10$^{9}$& 47\\
		15 & 8.21$\times$10$^{10}$ &2.06$\times$10$^{9}$  & 46\\
		16 & 8.14$\times$10$^{10}$& 2.13$\times$10$^{9}$ & 46\\
		17 & 8.10$\times$10$^{10}$ & 1.75$\times$10$^{9}$ & 46\\
		18 & 8.05$\times$10$^{10}$ &2.51$\times$10$^{9}$ & 46\\
		19 & 7.79$\times$10$^{10}$ & 3.06$\times$10$^{9}$ & 45\\
		20 & 7.36$\times$10$^{10}$ & 2.09$\times$10$^{9}$ & 44\\		
		\hline
	\end{tabular}

		\begin{tabular}{lccr} 
		\hline
		\multicolumn{4}{c}{Effective model}\\
		\hline
		Obj. & M$_{h}$ & M$_\star$ & R$_{\rm vir}$\\
		     &  (M$_\odot$) & (M$_\odot$)& (phys. kpc)\\
		\hline
		1 & 5.04$\times$10$^{11}$ & 2.18$\times$10$^{10}$  & 84\\
		2 &  4.91$\times$10$^{11}$ &3.07$\times$10$^{10}$  & 84\\
		3 & 4.90$\times$10$^{11}$  & 3.73$\times$10$^{10}$  & 84\\
		4 & 4.11$\times$10$^{11}$  & 2.23$\times$10$^{10}$  & 81\\
		5 &  3.66$\times$10$^{11}$& 2.70$\times$10$^{10}$ & 74\\
		6 & 3.04$\times$10$^{11}$ & 1.66$\times$10$^{10}$ & 71\\
		7 &  3.00$\times$10$^{11}$& 1.31$\times$10$^{10}$ & 71\\
		8 & 2.39$\times$10$^{11}$ & 1.16$\times$10$^{10}$ & 66\\
		9 &  2.32$\times$10$^{11}$& 7.50$\times$10$^{9}$ & 65\\
		10 & 2.12$\times$10$^{11}$ & 8.22$\times$10$^{9}$ & 63\\
		11 &  9.91$\times$10$^{10}$&2.81$\times$10$^{9}$ & 49\\
		12 & 9.66$\times$10$^{10}$ &3.07$\times$10$^{9}$  & 48\\
		13 & 9.40$\times$10$^{10}$ & 3.48$\times$10$^{9}$ & 48\\		
		14 & 9.01$\times$10$^{10}$ &  4.99$\times$10$^{9}$& 47\\
		15 & 8.95$\times$10$^{10}$ &2.21$\times$10$^{9}$  & 47\\
		16 & 8.93$\times$10$^{10}$& 3.64$\times$10$^{9}$ & 47\\
		17 & 8.25$\times$10$^{10}$ & 2.21$\times$10$^{9}$ & 46\\
		18 & 7.78$\times$10$^{10}$ &2.30$\times$10$^{9}$ & 45\\
		19 & 7.77$\times$10$^{10}$ & 3.48$\times$10$^{9}$ & 45\\
		20 & 3.73$\times$10$^{10}$ & 3.64$\times$10$^{9}$ & 35\\	
		\hline
	\end{tabular}
	
\end{table}

In order to interpret the observed data, we have built a set of simulated data, which we have processed in a way as close as possible
to that adopted for observations. First of all, we have built simulated mock spectra and then we have constructed absorption
systems and computed thier column densities. We give details of those operations in Appendices \ref{ss:LOS} and \ref{s:Ncomparison}
and the following subsection.

%
\begin{figure*}
 \centering

\includegraphics[width=\textwidth]{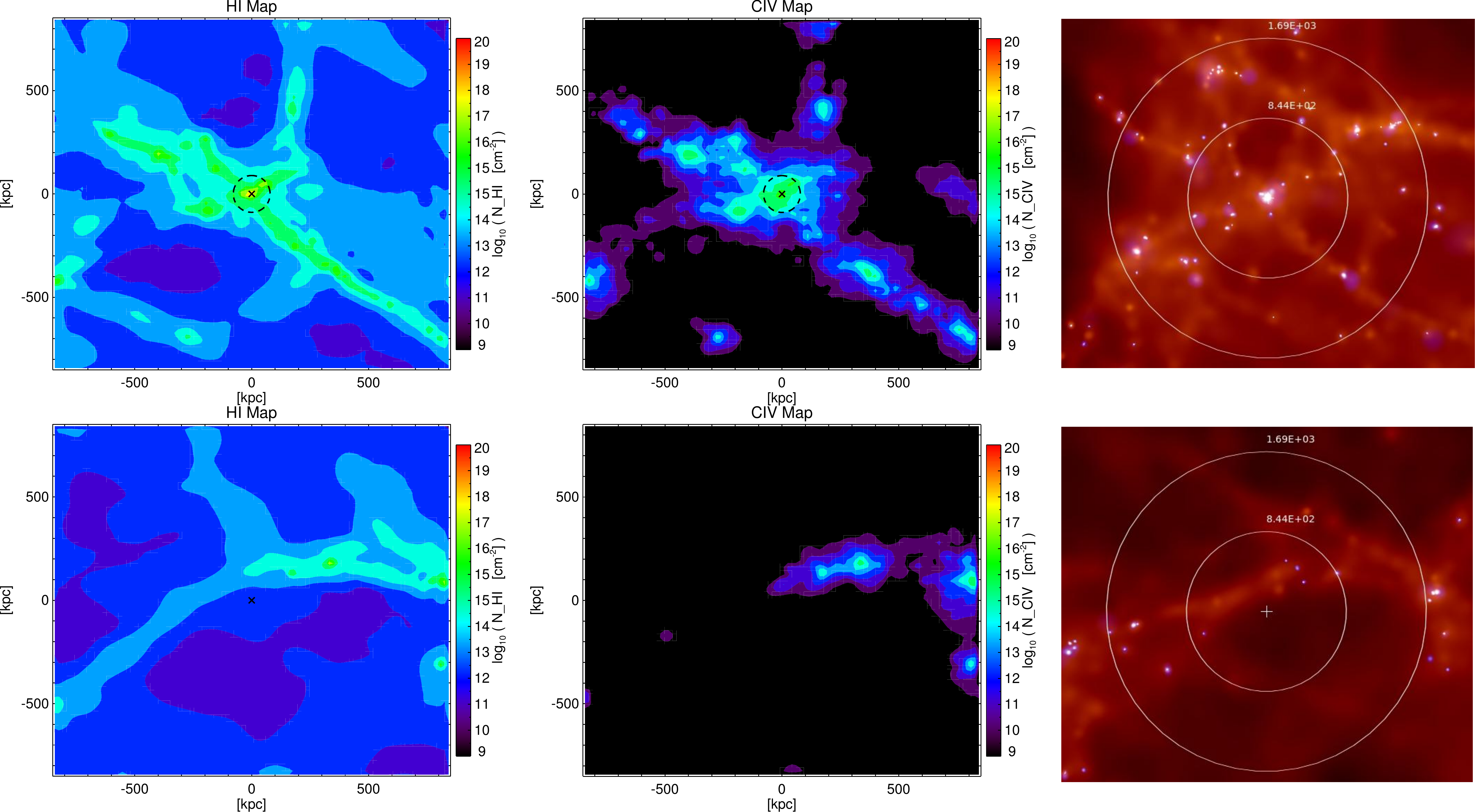}

  \caption{$Top$: projected maps of the \mbox{H\,{\sc i}} (left) and \mbox{C\,{\sc iv}} (centre) column density distribution around a given galaxy of the investigated
  sample (distances are in proper kpc). 
  Column densities have been computed by integrating the density profile of these sub-boxes along a perpendicular slice 
  of width $\pm$300 kpc around galaxy's position. Image of the chosen galaxy (right) obtained from \textsc{gadgetviewer} (a software for visualization 
  of N-body simulations. The image is colour coded according to particle properties: in red, it is represented as SPH smoothed density plot of gas particles 
  distribution, and in blue the same but for star particles): the chosen galaxy is in the centre of the image;
  the external circle has a radius of $1.6\times 10^3$ 
  comoving kpc, while the internal circle has a radius of $8.4 \times 10^2$ comoving kpc. $Bottom$: same as the $top$ panel but for
  a near-filament environment. The white cross here  represents the visually chosen ``centre of the object''.}
  \label{N_maps}
\end{figure*}

%

\subsection{Sample selection}
\label{ss:sampselect}
The extraction of line of sight (LOS) physical quantities follows the standard procedure of SPH schemes and is summarized in Appendix \ref{ss:LOS}.

In order to investigate the IGM/CGM of galaxies, we selected from the snapshot at $z=1.94$ of both cosmological runs 
two samples of 20 galaxies with a total halo mass $M_{\rm h}$ in the range $[\sim 10^{10}-10^{12}]$ M$_\odot$, identified with the 
SUBFIND algorithm  \citep{Springel2001} and stellar mass greater than $M_{\ast} \gtrsim 2 \times 10^9 \rm M_\odot$. 

The selection criteria were the following: the galaxies had not to be subject to major mergers  and to be the main 
halo of their friends-of-friends (FOF) group, so not a substructure, in order to compare similar conditions at a given distance from galaxy centre 
between the different galaxies of the sample. The first condition puts an upper limit on the chosen halo mass, as only six 
objects with $M_{\rm h} > 10^{12}$ M$_\odot$ are present in the two runs, but they are actually all formed by clusters of merging galaxies. 
The lower limit on the halo mass, instead, is related to the number of particles that sample the galaxy and its various 
components. We fixed a minimum number of $\sim100$ gas particles, in order to consider the object reliable. 

In Table~\ref{tab:Galaxies}, we report some properties of the two galaxy samples.
Virial radii are defined for each FOF halo as the radii of a sphere centred on the 
main halo of the FOF group and which contains an overdensity of 200 the critical density, so they are calculated with this 
formula:
\begin{equation}
 R_{\mathrm{vir}}= \left( \frac{M_h}{\frac{4}{3} \upi \cdot 200 \cdot \rho_{\mathrm{crit}}}\right)^{1/3}
\end{equation}
with $\rho_{\mathrm{crit}}=3$ H(z)$^2$/ $8\upi$ G.
We also visually selected from the two boxes, using the software 
\textsc{gadgetviewer\footnote{http://astro.dur.ac.uk/~jch/gadgetviewer/index.html}}, 10 near-filaments environments, in order to 
investigate the low-density IGM near galaxies. 

In the first row of Fig.~\ref{N_maps}, a galaxy from the MUPPI simulation is shown, while in the second row we plot an example of a
near-filament environment. In the left-hand panels the \mbox{H\,{\sc i}} column density maps around these two objects are reported, in the centre the 
\mbox{C\,{\sc iv}} column density maps and on the right we show the images of the two environments with \textsc{gadgetviewer}.

For near-filament environments, the ``centre'' is represented by the position of the white cross, as shown in the bottom left panel of Fig.~\ref{N_maps}.
\section{Results}
\label{s:results}
\subsection{The N$_{\rm{C \, \textsc{iv}}}$ versus N$_{\rm{H\,\sc{i}}}$ relation: piercing around objects}
\label{ss:Nrelation1}
The N$_{\rm{ C\, \textsc{iv}}}$ versus N$_{\rm{ H\, \textsc{i}}}$ relation, constructed here in order to compare with data, is 
the
observational equivalent of the theoretical overdensity-metallicity relation, as the overdensity is related to
N$_{\rm{ H\, \textsc{i}}}$ and
the metallicity can be traced by the presence of \mbox{C\,{\sc iv}}. The \mbox{C\,{\sc iv}} doublet, in fact, falls
in a region redward of the
Ly$\alpha$ forest, mostly free from line blendings, and observable from the ground in the visible band (3000-10000 \AA{}) at $0.9\lesssim z\lesssim 5.4$. 
Its ubiquitous presence in QSO 
spectra also indicates that it traces a gas phase, which is characteristic of many astrophysical  gas environments.

We pierced through the cosmological boxes random lines of sight around the 40 galaxies (20 per model) with impact parameters less than 800 
kpc, in order to characterize the environment around them. 

As the LOS is as long as the box side, we first identified the position of the centre of the galaxy along it. We, 
then, selected a region along the LOS $\pm$ 150 km s$^{-1}$  from this position and we considered the optical depth 
profile both for \mbox{H\,{\sc i}} and \mbox{C\,{\sc iv}} in this region (see 
Appendix~\ref{s:Ncomparison}). For each pixel of this region, we converted the value of the optical depth into a column density
adopting the apparent optical depth (AOD) method \citep[see Appendix~\ref{s:Ncomparison}, Eq.~\ref{eq:AOD}][]{SavageSembach1991}
and then we integrated in the considered velocity range. 
The same procedure is used for constructing \mbox{O\,{\sc vi}} and \mbox{Si\,{\sc iv}} systems, discussed in next sections.

In order to test the reliability of our simulation, we have computed the column density distribution functions (CDDFs) for a general sample 
of \mbox{C,\sc{iv}} and \mbox{H\sc{i}} lines. The CDDF is defined as the number of absobers in column density and redshift bins. The simulated CDDFs were then compared
with the observed ones computed with the sample of \citet{Kim2013}. The comparison is reported in Fig.~\ref{fig:MUPPI_cddf}
and Fig.~\ref{fig:EFF_cddf} for the MUPPI and the Effective models, respectively. In particular, the MUPPI model, which is our reference model, reproduced in a reasonable way
both the  \mbox{C,\sc{iv}} and \mbox{H\sc{i}} CDDFs in the range of interest.

\begin{figure*}
 \centering
  \makebox[\textwidth][c]{
  \subfloat{\includegraphics[width=0.4\textwidth]{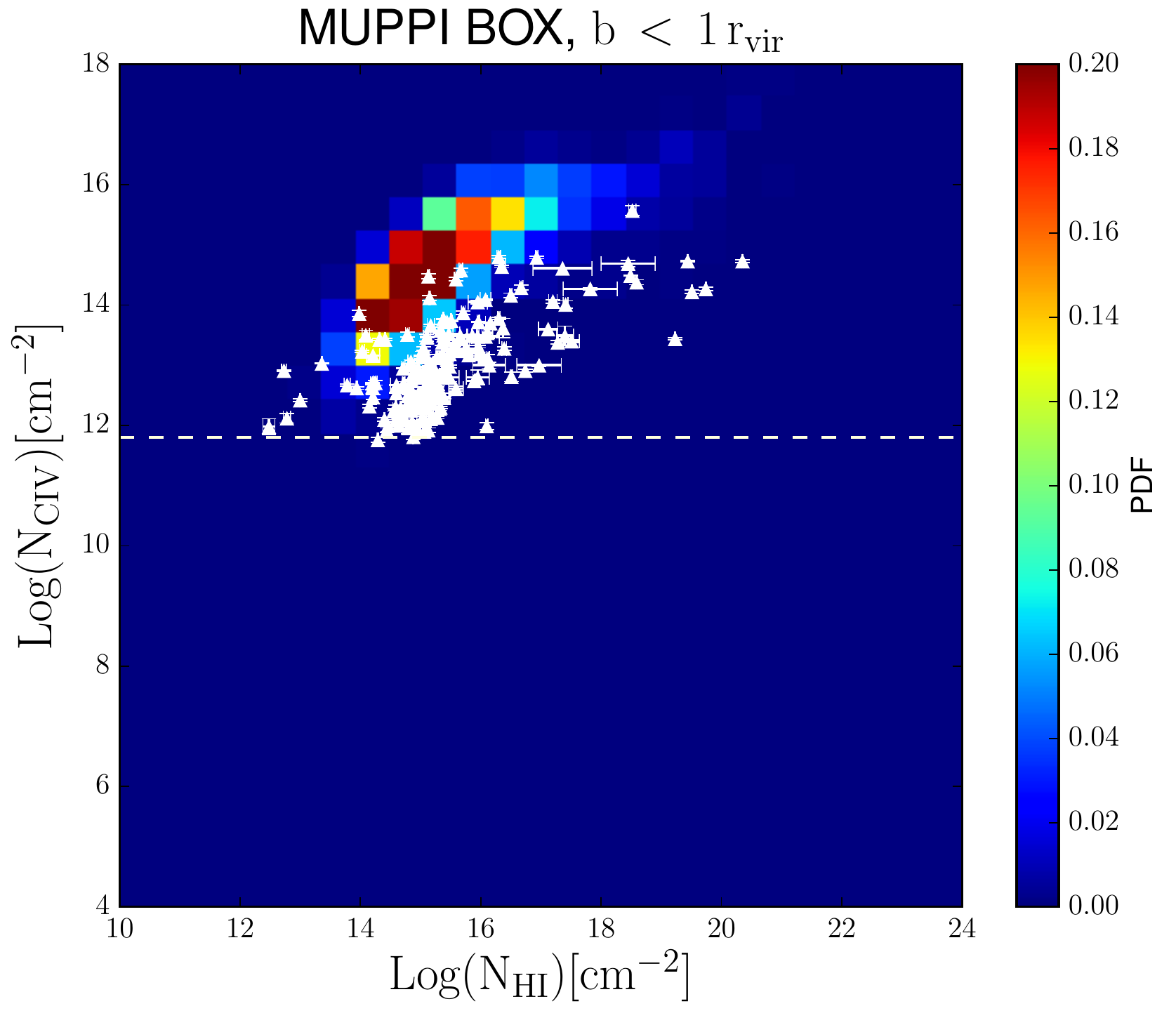}} 
  \hspace{0.5cm}
  \subfloat{\includegraphics[width=0.4\textwidth]{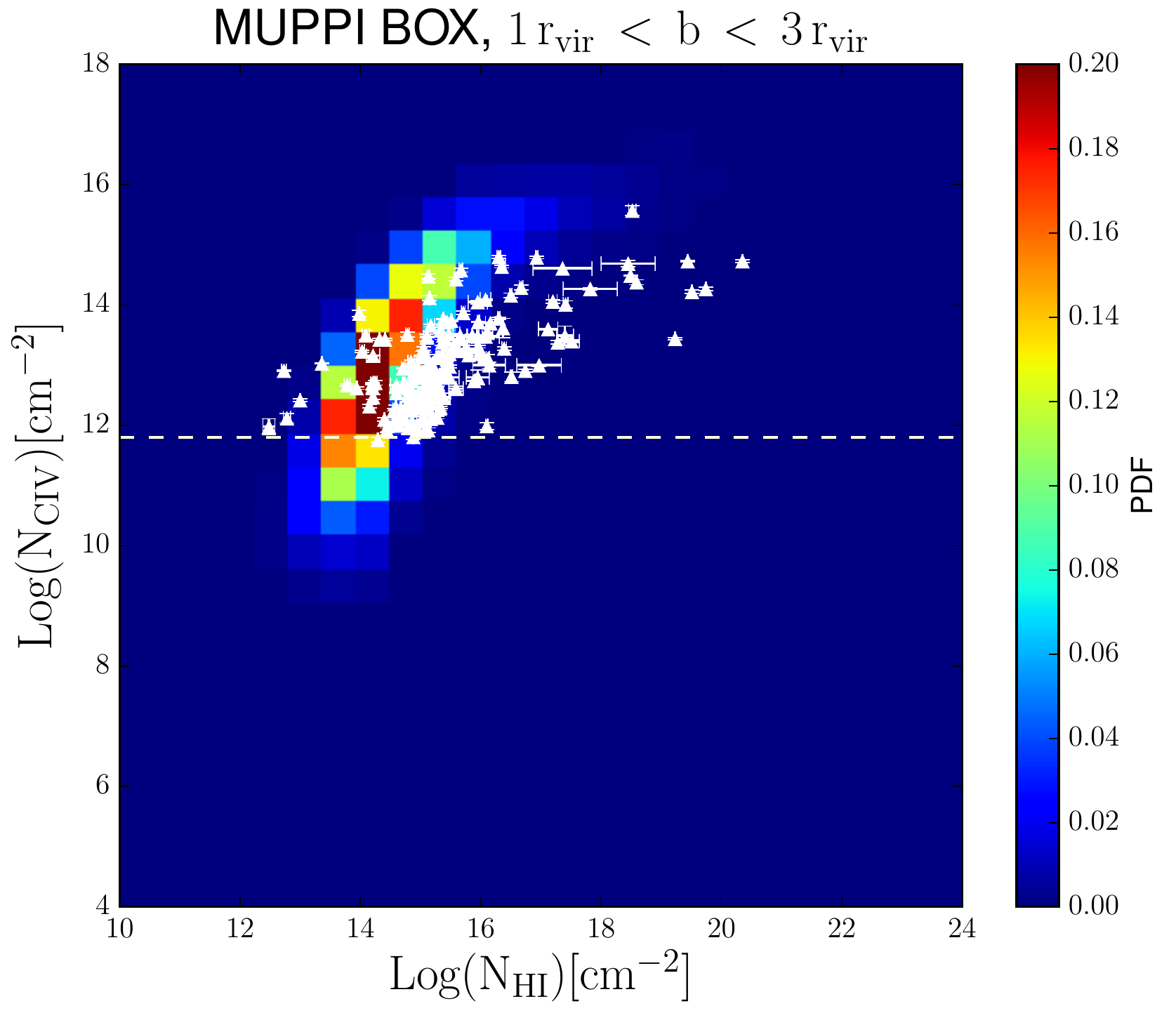}}
  
  }
   \vspace{0.1cm}
   \makebox[\textwidth][c]{
  \subfloat{\includegraphics[width=0.4\textwidth]{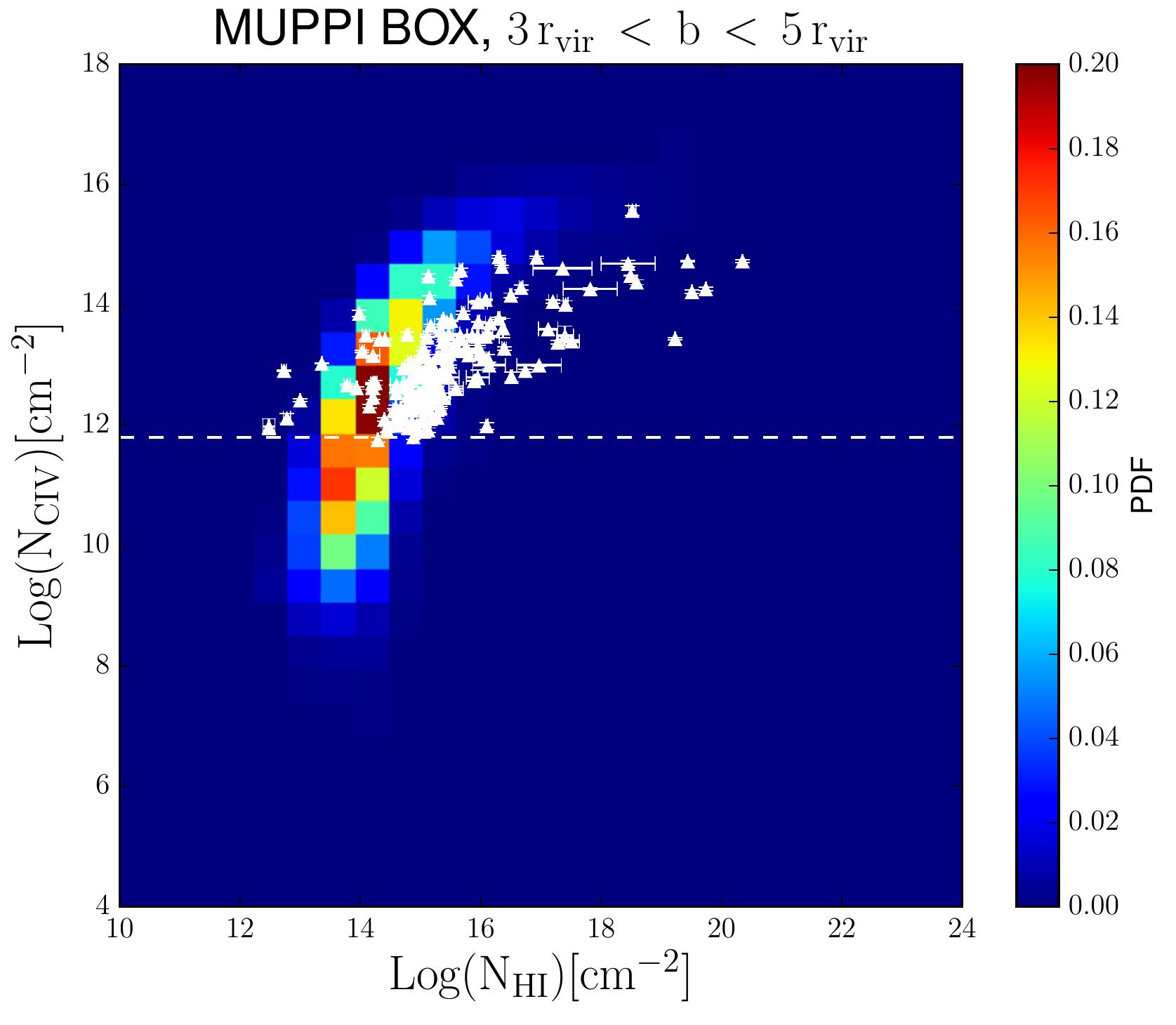}} 
  \hspace{0.5cm}
  \subfloat{\includegraphics[width=0.4\textwidth]{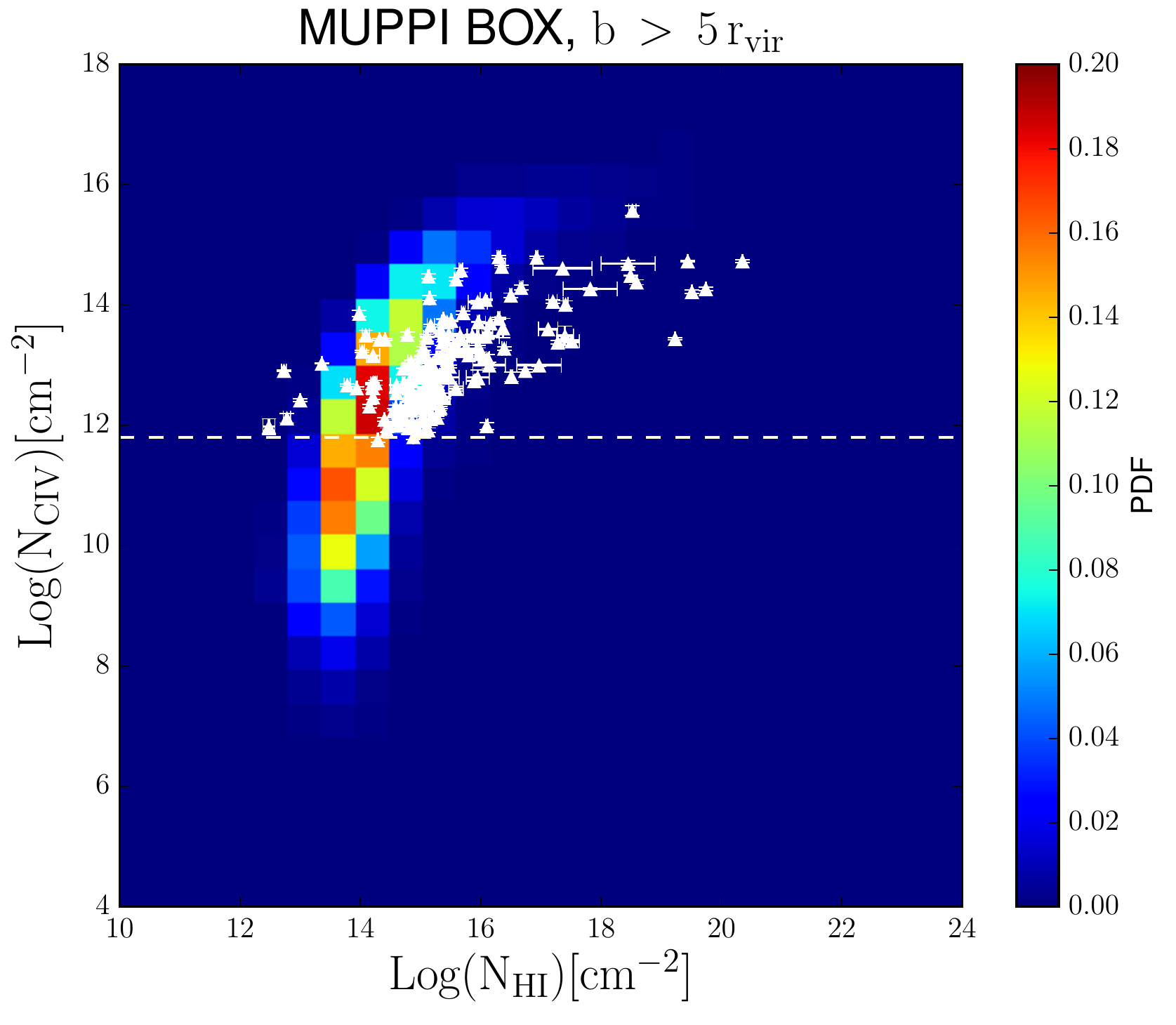}}
 
  }
  \caption{PDF of the simulated \mbox{C\,{\sc iv}} systems. {\it White triangles}: observational data from K16; and
  the same data are reported in each panel. {\it Dashed horizontal line}: observational detection limit.}
  \label{fig:PDFMUPPI}
\end{figure*}

\begin{figure*}
 \centering
  \makebox[\textwidth][c]{
  \subfloat{\includegraphics[width=0.4\textwidth]{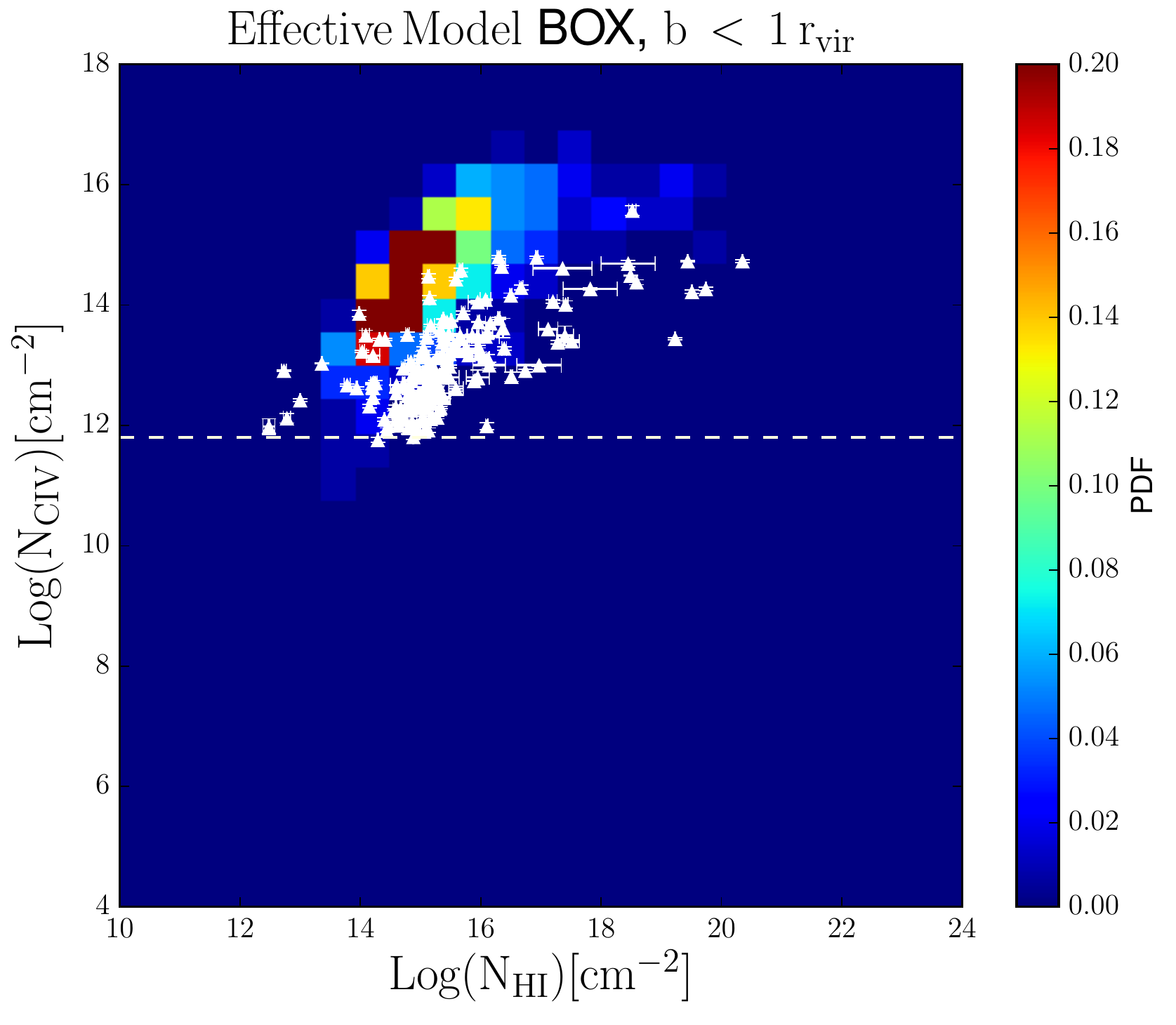}} 
  \hspace{0.5cm}
  \subfloat{\includegraphics[width=0.4\textwidth]{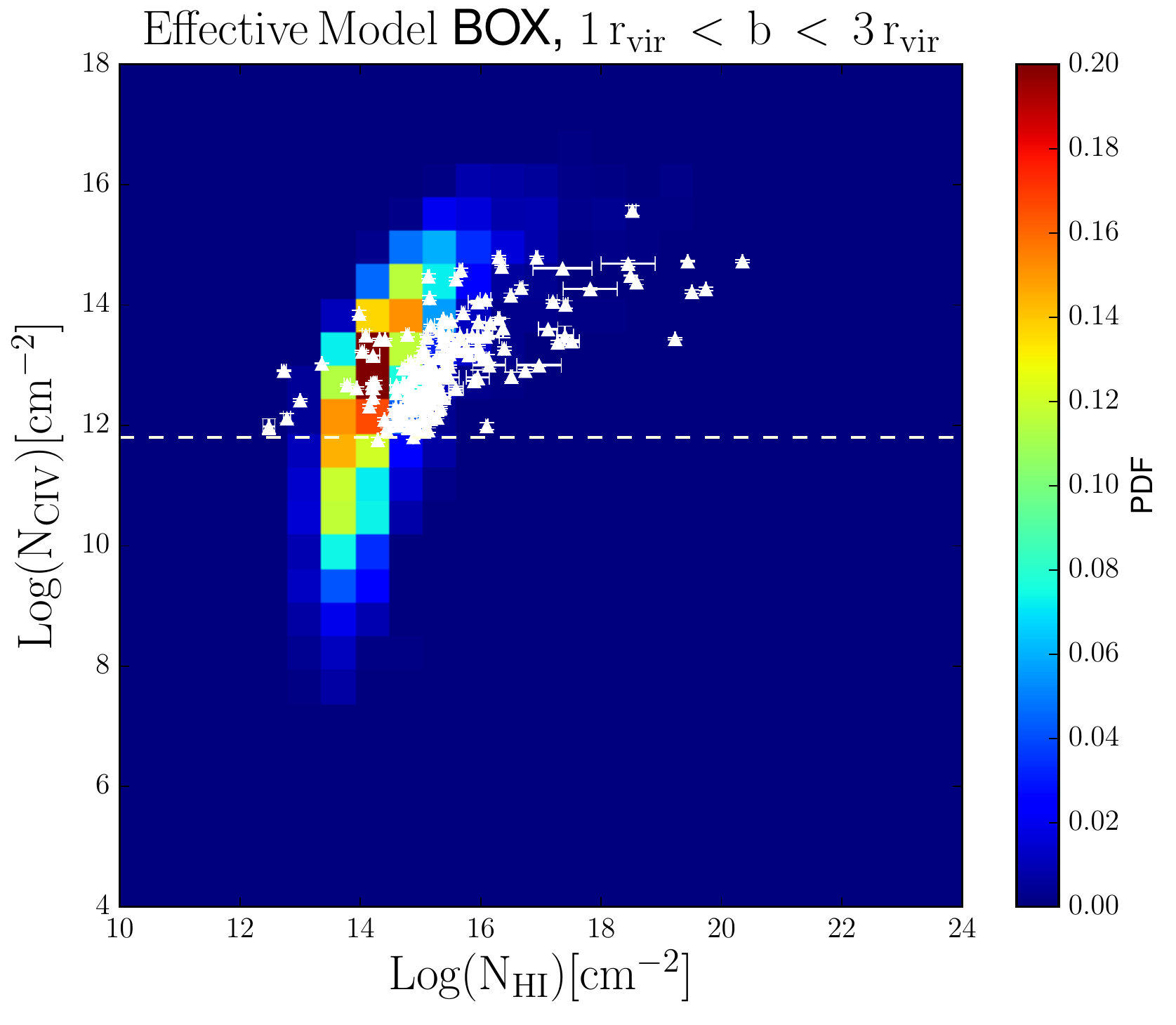}}
  
  }
   \vspace{0.1cm}
   \makebox[\textwidth][c]{
  \subfloat{\includegraphics[width=0.4\textwidth]{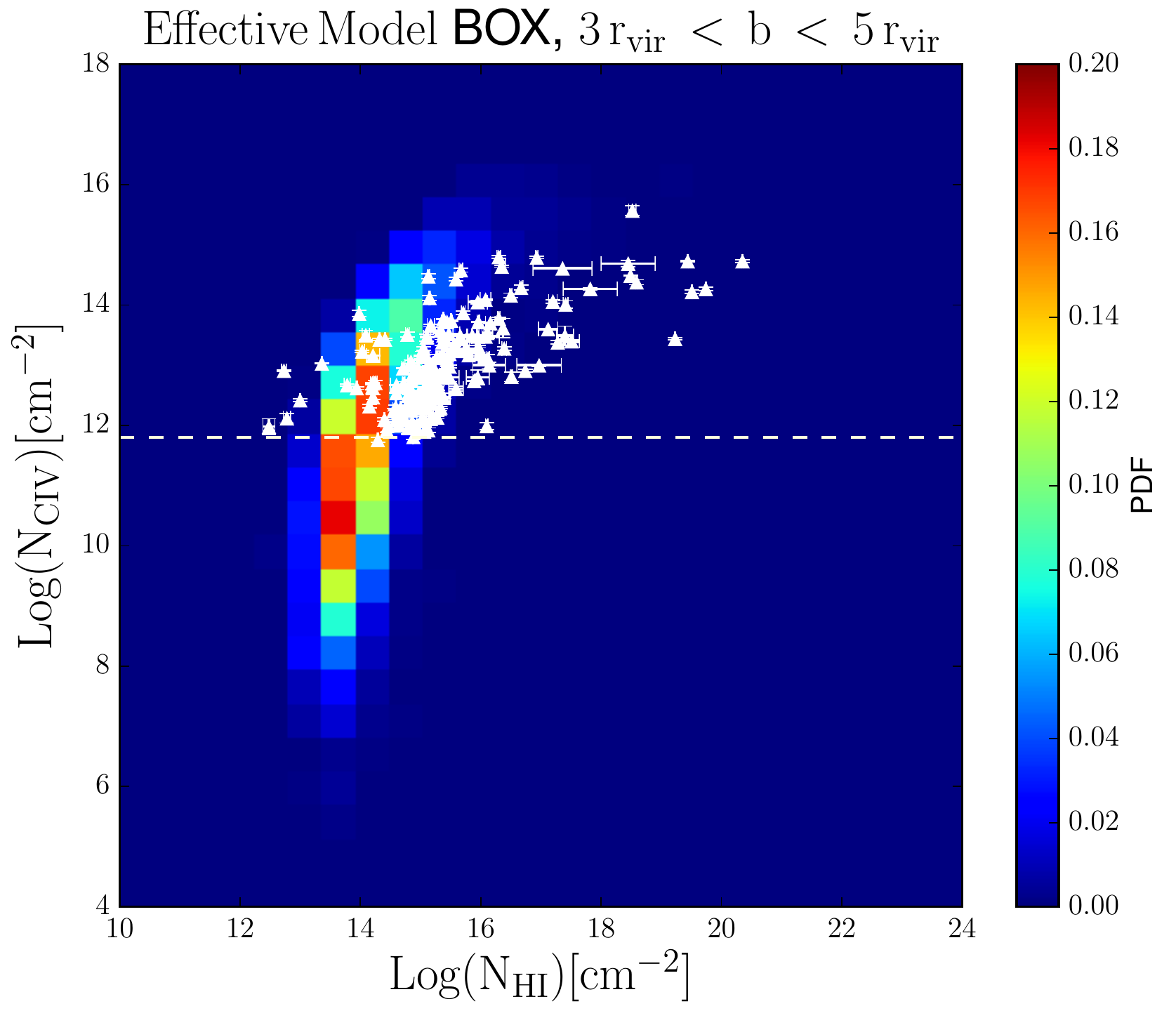}} 
  \hspace{0.5cm}
  \subfloat{\includegraphics[width=0.4\textwidth]{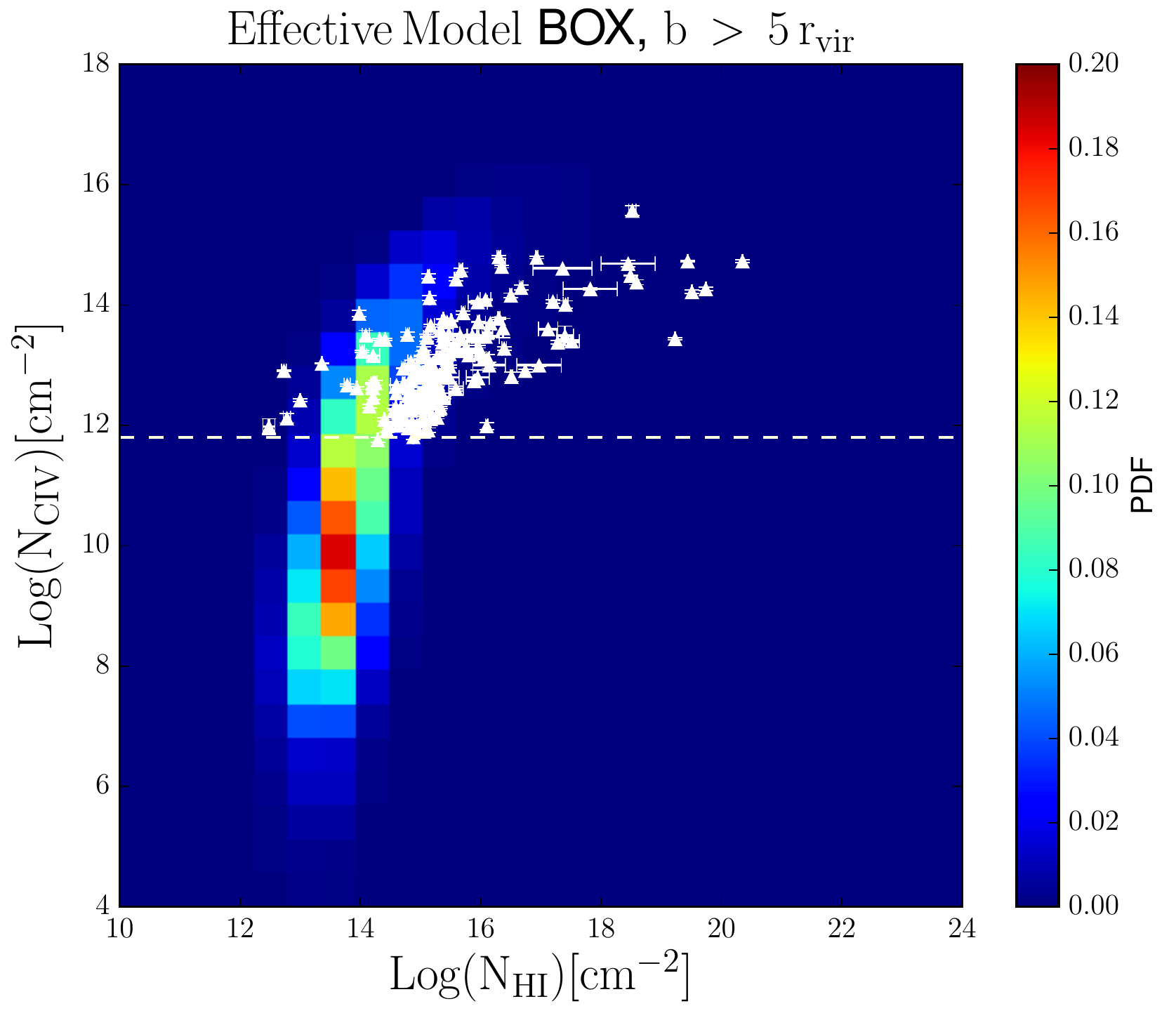}}
 
  }
  \caption{Same as Fig.~\ref{fig:PDFMUPPI}, but for the Effective model simulation.}
  \label{fig:PDFEFF}
\end{figure*}

In Figs.~\ref{fig:PDFMUPPI} and ~\ref{fig:PDFEFF}, we show the probability distribution functions (PDFs) of the N$_{\rm{ C\, \textsc{iv}}}$ versus N$_{\rm{ H\, \textsc{i}}}$ relation 
for the MUPPI and the Effective Models, respectively, compared to the observational sample by K16 (white triangles with error bars). The horizontal dashed lines in all 
plots represents the observational detection limit $\log\, \rm N_{\rm {C\, \textsc{iv}}} \simeq 11.8$.  
The simulated absorption systems are divided according to the distance of the corresponding LOS from the considered galaxy. From top-left to bottom-right, 
in each plot we show lines of sight with impact parameters $b$ between [0-1] $r_{\rm {vir}}$, [1-3] $r_{\rm {vir}}$, [3-5] $r_{\rm {vir}}$ and $b>5\,r_{\rm {vir}}$, where 
$r_{\rm {vir}}$ is the virial radius of the considered galaxy reported in Table~\ref{tab:Galaxies}. 

Going to larger distances, we can see that the probability distribution gradually shifts to lower values of \mbox{H\,{\sc i}} and 
\mbox{C\,{\sc iv}} column densities. Inside the virial radius, all the simulated lines of sight are above the observational 
limit. In this plot, the bulk of simulated points occupy a slightly different region than the data of K16: it looks like there is a shift toward larger \mbox{C\,{\sc iv}} 
column densities  with respect to observational data.  This could be due to either local radiation which is neglected in our models or to a wrong metallicity. In fact we 
are relying on an ultraviolet background (UVB) which is not accurate, as we are assuming that each gas particle is photoionized by Haardt $\&$ Madau UVB, neglecting 
that near galaxies there is also the contribution of the ionizing photons from the galaxy itself. It is however reasonable to assume that such effect is not large at least
for \mbox{H\, \sc{i}} absorbers at $\log (N_{\rm H\, \textsc{i}}/\rm{cm}^{-2}) < 17$ as suggested by \citet{Rahmati2013} and for \mbox{C\,\sc{iv}} could also be a minor effect due to the fact that radiation from galaxies falls sharply above 1 Ryd.
A more likely explanation is thus that the simulation produces too many metals, due to inefficient feedback, that in turn causes too much SF.



\begin{figure}
 \centering

\includegraphics[width=\columnwidth]{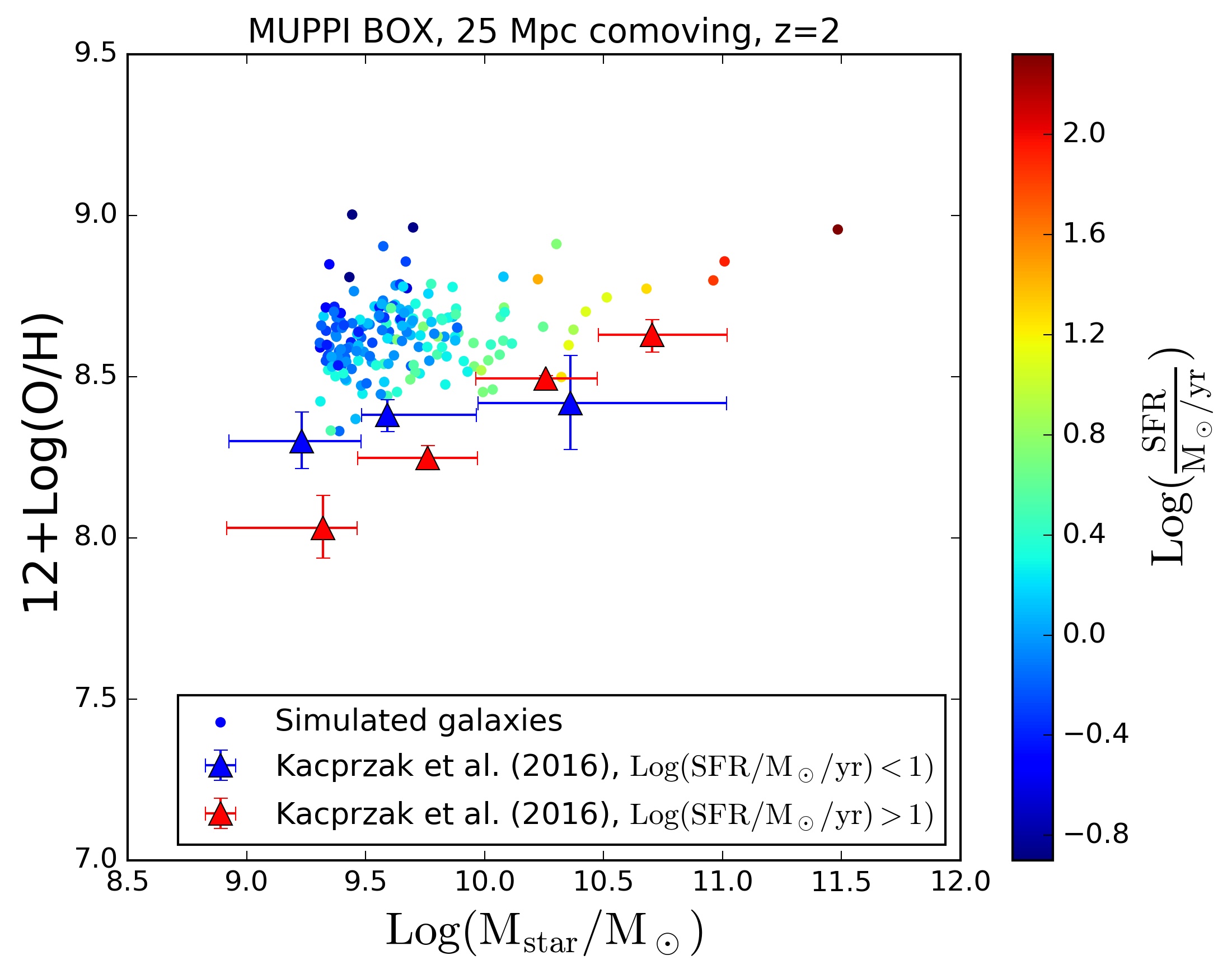}

  \caption{Mass-metallicity relation for all the MUPPI galaxies with $M_\ast > 2 \times 10^9$ M$_\odot$ compared to 
  the observed one by \citet{Kacprzak2016}. MUPPI galaxies are colour-coded according to their SF rate.}
  \label{fig:MassMet}
\end{figure}

To investigate this last issue, we constructed the mass-metallicity relation for all the galaxies 
of both simulations with stellar mass $M_\ast > 2 \times 10^9$ M$_\odot$. For simplicity, we show the results only for the 
MUPPI simulation. The Effective model provides similar results, as it shares the same chemical model and it does not include
AGN feedback as in MUPPI.
For each gas particle inside one-tenth of the virial radius of a galaxy, we 
computed the abundance O/H. For each galaxy, we then took the gas particle's mass-weighted average value of O/H.
Our result are shown in Fig.~\ref{fig:MassMet}. The simulation does actually produce too many metals with respect to the 
observed relation by \citet{Kacprzak2016} and this could explain the disagreement in the \mbox{C {\sc iv}} data. 

\citet{Goz2017} analysed the mass-metallicity relation of the MUPPI simulation at $z=0$. 
They compare the gas metallicity of their sample of MUPPI galaxies, taken from the same box
we used, as a function of stellar mass, with the observational relation by \citet{Tremonti2004}. They recover the same 
trend as the observed relation, but with a global offset of $\sim$ 0.1 dex, extending up to 0.2 dex in the 
low-mass end of the relation. They find that massive galaxies show a lack of proper quenching and small galaxies are too massive, passive, metallic and
with low atomic gas content. \citet{Barai2015}, using the same simulations, find an excess of SF rate density, in particular at high redshift. 
Both papers claim that these results could be due to the lack of AGN feedback in the simulations, capable of quenching the cooling and SF. 
In the case of our results, although AGNs are not dominant in the population of galaxies we have considered (the median halo mass is $\sim 10^{11}$ M$_\odot$),  
we believe the cumulative effect of their feedback during galaxy evolution could alleviate the excess of metallicity we measure at $z\simeq1.9$,
as its contribution at very high redshift prevents the formation at lower redshift of too many small galaxies, formed by too metal-enriched 
gas particles.


In Figs~\ref{fig:PDFMUPPI} and \ref{fig:PDFEFF}, we note that there are non-negligible probabilities of finding simulated systems with large values of 
N$_{\rm{ H\, \textsc{i}}}$ and N$_{\rm{ C\, \textsc{iv}}}$, even at large distances from the chosen galaxy. This is likely a consequence of the fact that galaxies are not isolated systems.

We investigated this issue by considering all the lines of sight with $b>$ 5 $r_{\rm {vir}}$
and with absorption systems with $14.0\leq$ $\log(N_{\rm C\, \textsc{iv}}/\mathrm{cm}^{-2}) \leq18.0$. 

\begin{figure}
 \centering

\includegraphics[width=\columnwidth]{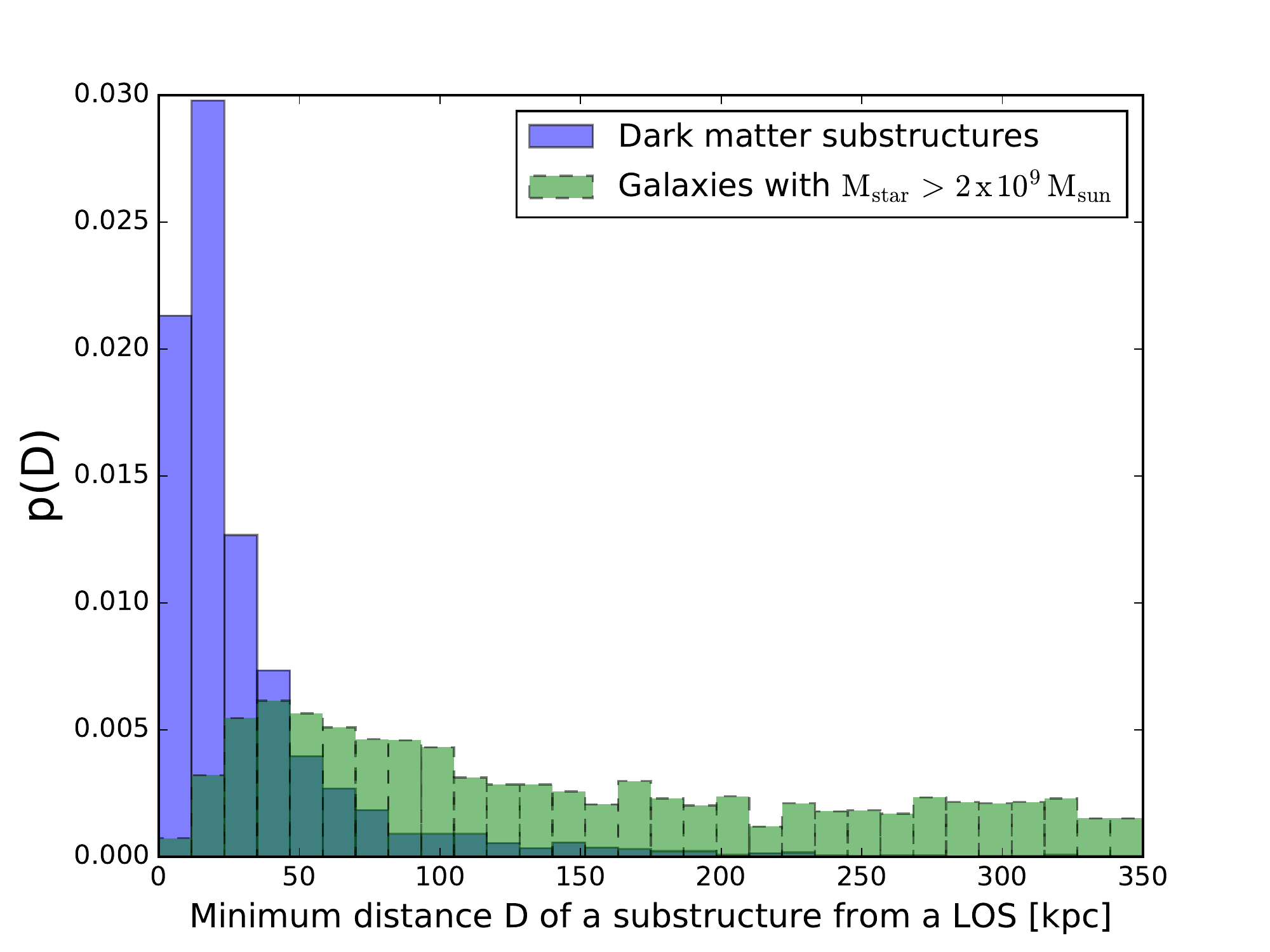}

  \caption{PDF of the minimum proper distance between an LOS with $b>$ 5 $r_{\rm {vir}}$
  and with an absorption 
  system with $14.0\leq$ $\log(N_{C\, \textsc{iv}}/{\mathrm cm}^{-2}) \leq 18.0$ and a DM substructure identified with 
  \textsc{subfind} (blue solid histogram) or a galaxy with stellar mass $M_\ast > 2 \times 10^9$ M$_\odot$ (green dashed histogram).}
  \label{fig:DistaSub}
\end{figure}

We searched for substructures 
near the lines of sight and we computed the minimum distance from an LOS to a substructure. The distribution of 
the distances is reported in Fig.~\ref{fig:DistaSub}. 
The blue solid histogram considers all the DM substructures identified by the \textsc{subfind} search algorithm  with no distinction in the mass. 
That includes also the smallest structures, simply formed by DM and gas clumps without SF. The green dashed histogram has a cut in the stellar 
mass of the substructure, as discussed in Section~\ref{ss:sampselect}. The blue solid histogram has a median value and a 1$\sigma$ error of 19$^{+29}_{-11} $ proper kpc,
while the green dashed one has a much broader distribution, with a median value and 1$\sigma$ error of 225$^{+200}_{-140} $ proper kpc. 
This result suggests that those lines of sight have a higher probability to be related to another nearby substructure. They could pass through a small DM
and gas clump or they could be near the halo of another galaxy, closer than the chosen one. In fact, even if the median value of 225 kpc is not a small distance, it is closer than the 
5 $r_{\rm {vir}}$ of the chosen galaxy.

For completeness, the median values with the 1$\sigma$ dispersion of the distance distributions in the ranges $b<$ 1 $r_{\rm {vir}}$,
1 $r_{\rm {vir}}$ $<b<$ 3 $r_{\rm {vir}}$ and 3 $r_{\rm {vir}}$ $< b <$ 5 $r_{\rm {vir}}$ are 15$^{+12}_{-8}$, 19$^{+18}_{-11} $ and 19$^{+21}_{-11} $ kpc respectively.
The total mass of the substructure is in the range $M_{\rm h}\sim 10^{8}-10^{12}$ M$_\odot$.
%


As imaging surveys are flux-limited, faint galaxies can be missed. For this reason, we can state from 
Figs.~\ref{fig:DistaSub}, \ref{fig:PDFMUPPI} and \ref{fig:PDFEFF} that finding a strong absorption system at a large distance from a 
galaxy is possible and the two can be related to each other. However, it is more probable that the absorption system is
related to a smaller galaxy, 
which is not detected in present observations.

\begin{figure*}
 \centering
  \makebox[\textwidth][c]{
  \subfloat{\includegraphics[width=0.4\textwidth]{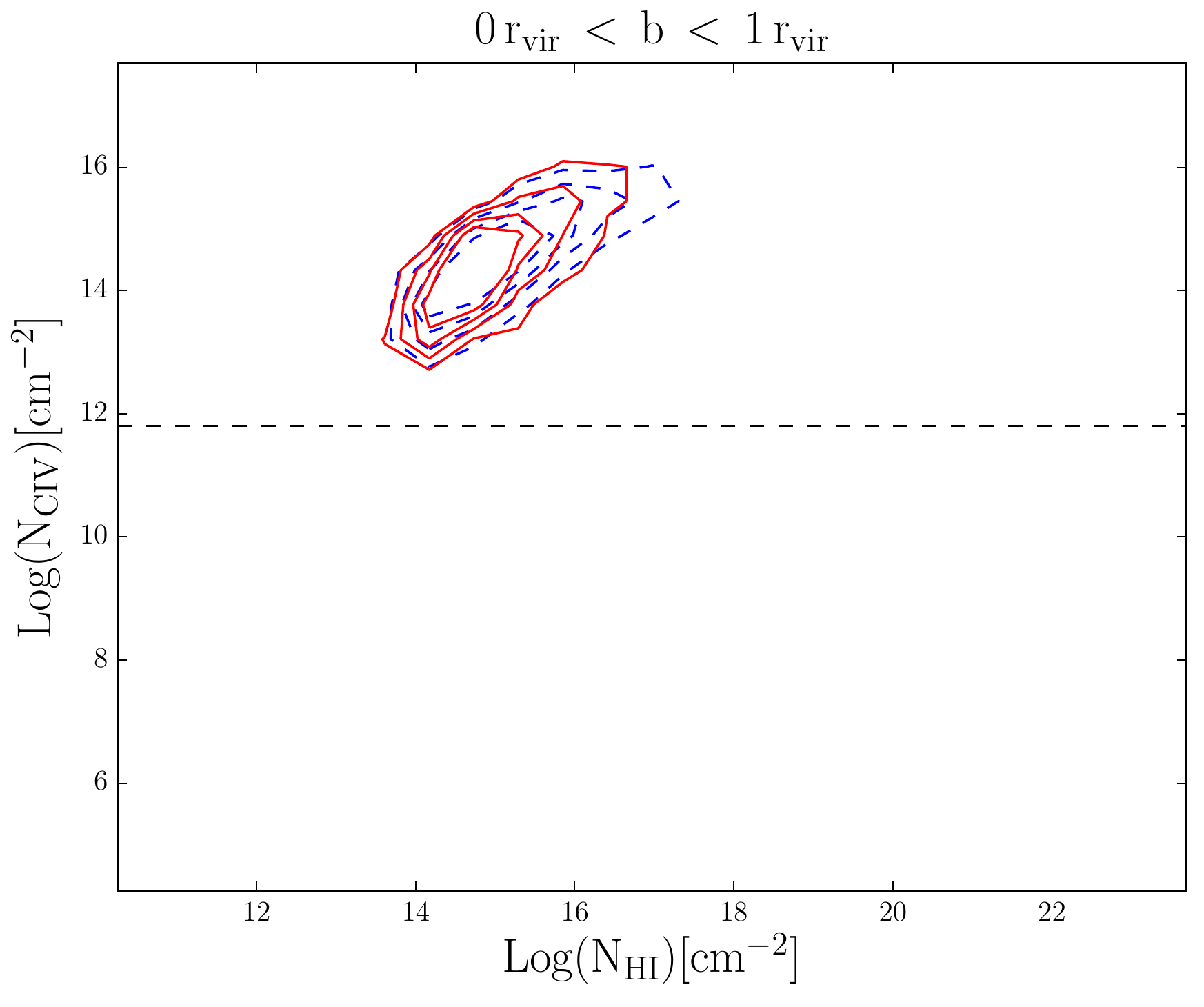}} 
  \hspace{0.5cm}
  \subfloat{\includegraphics[width=0.4\textwidth]{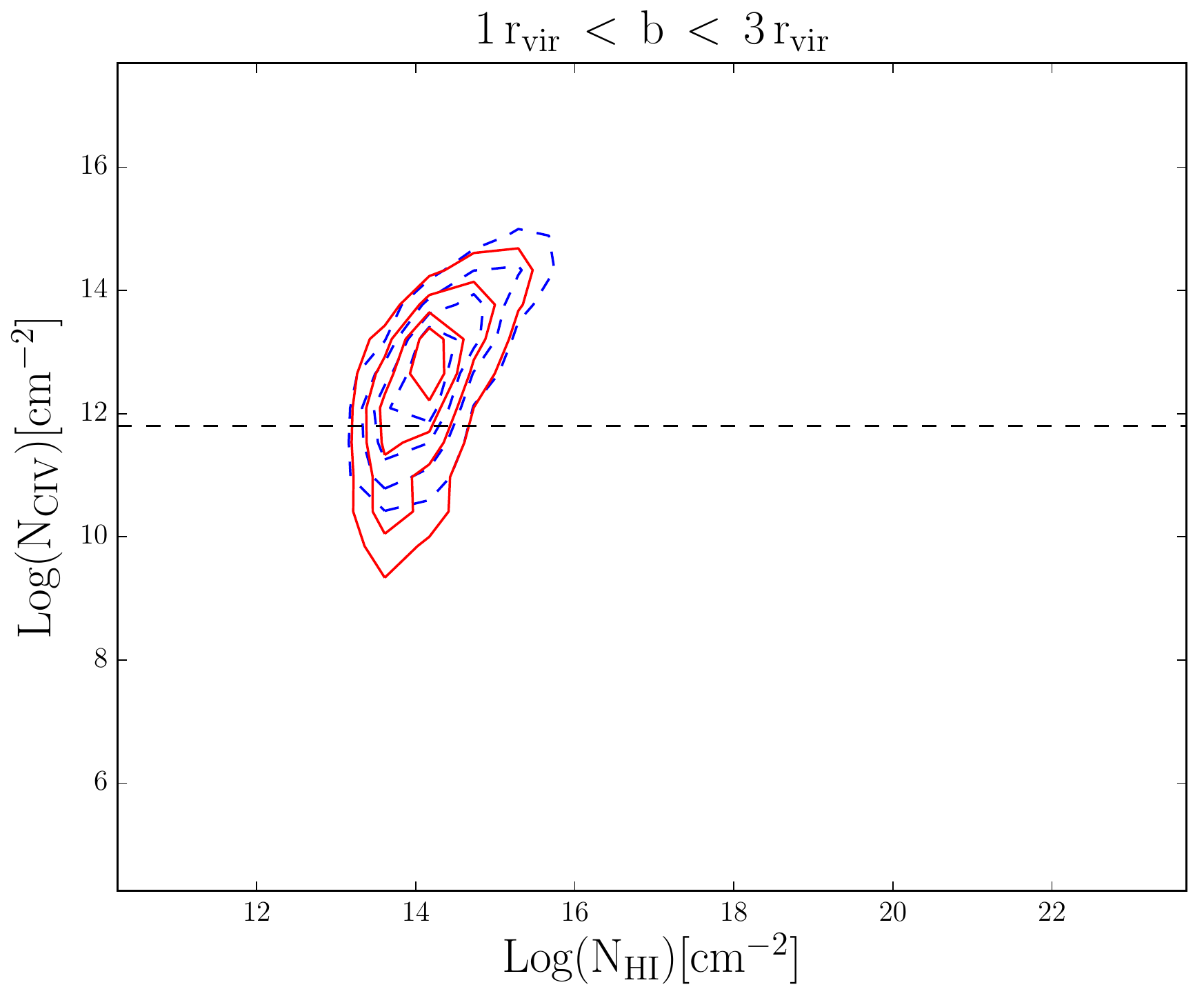}}
  
  }
   \vspace{0.1cm}
   \makebox[\textwidth][c]{
  \subfloat{\includegraphics[width=0.4\textwidth]{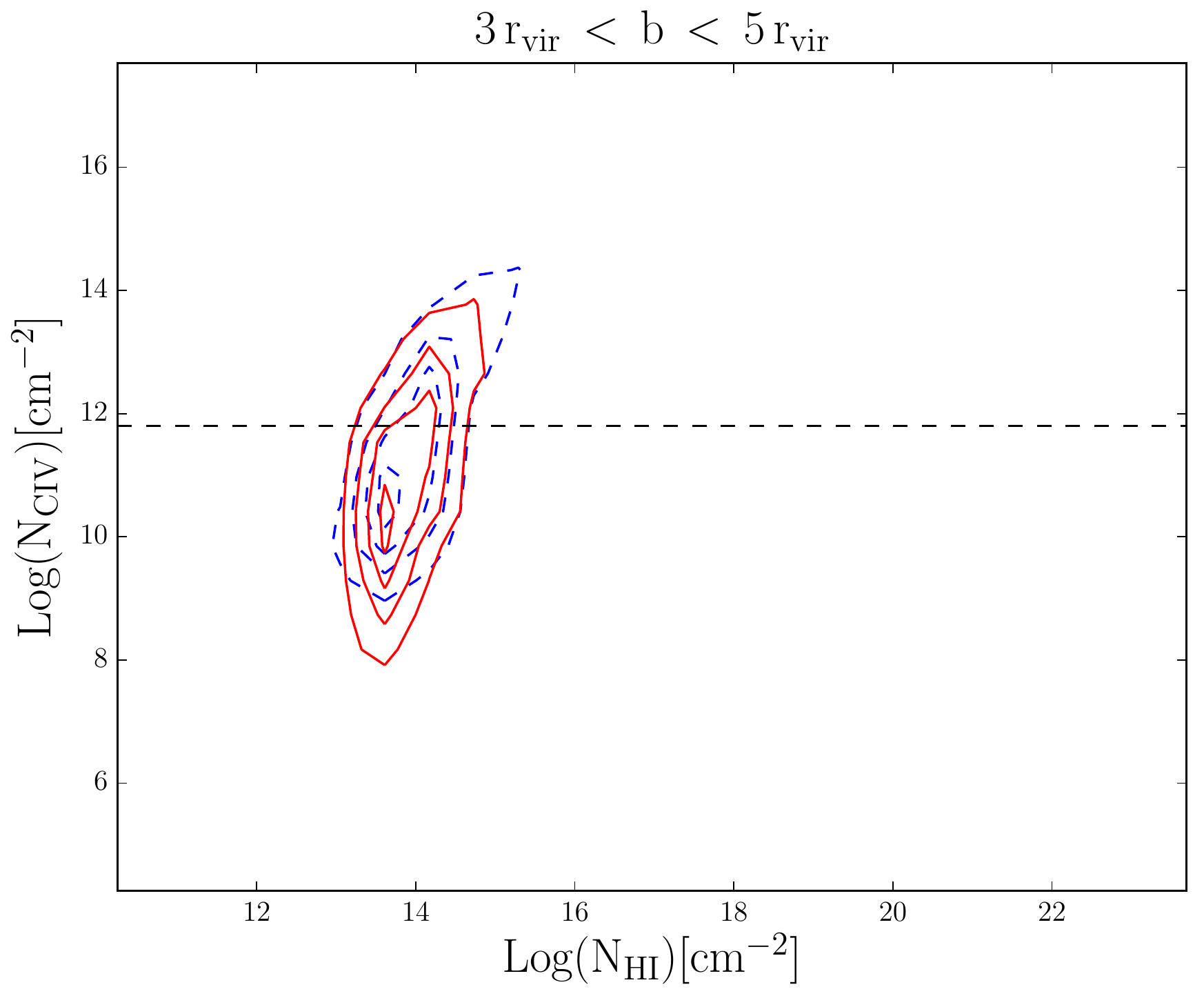}} 
  \hspace{0.5cm}
  \subfloat{\includegraphics[width=0.4\textwidth]{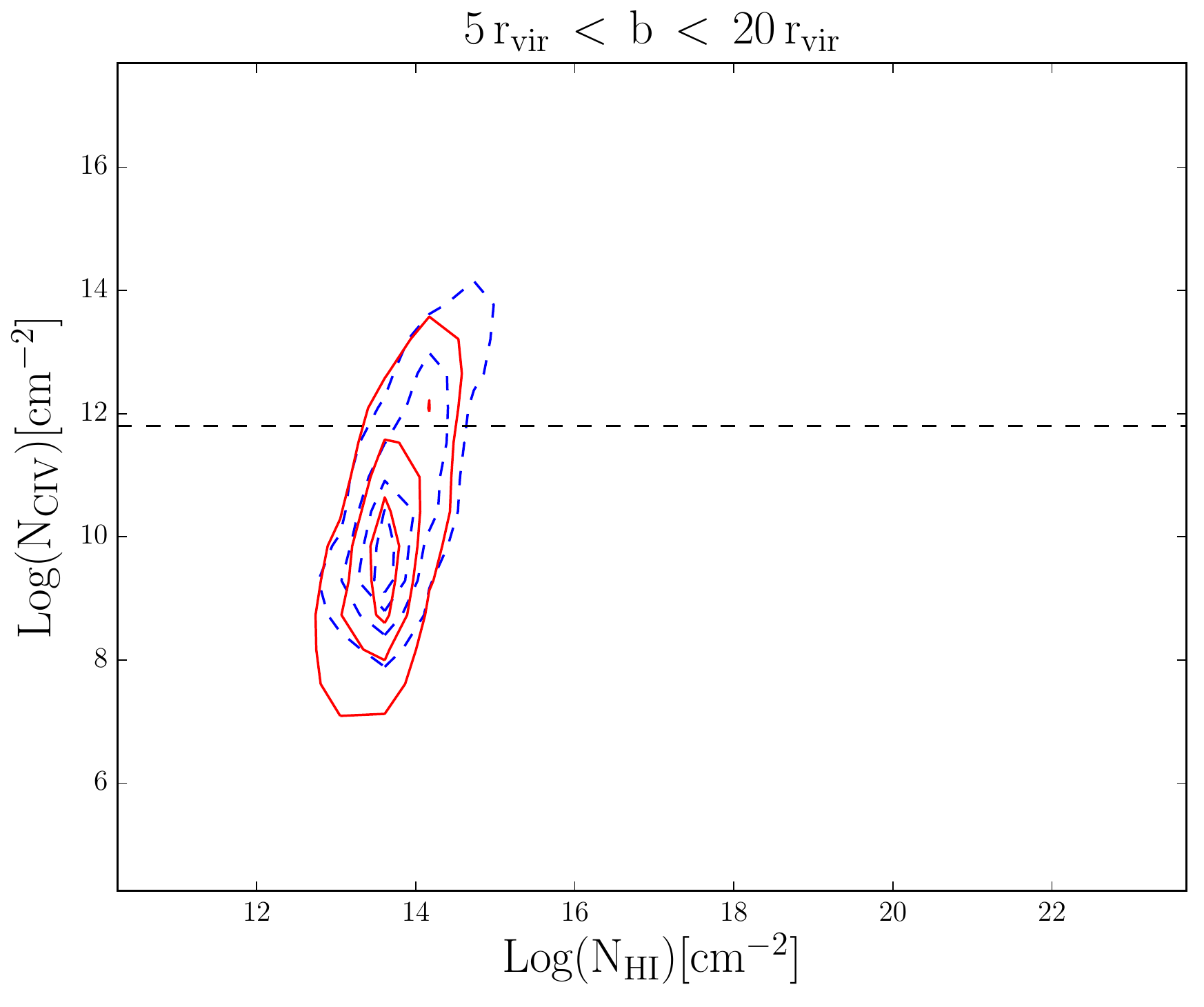}}
 
  }
  \caption{Comparison of the PDFs of the MUPPI (blue dashed) and the Effective (red solid) models for the considered distance ranges. Contour levels correspond to PDF
  values of [0.05,0.1,0.15,0.2].  {\it Dashed horizontal line}: observational 
  detection limit.  }
  \label{fig:ConfMUPPIEFF}
\end{figure*}


The plots in Figs.~\ref{fig:PDFMUPPI} and ~\ref{fig:PDFEFF} constitute the main result of this work: they give the probability to find an absorption system with a given 
\mbox{H\,{\sc i}} and \mbox{C\,{\sc iv}} column density at a certain distance from a galaxy. We can state that the observational data have the highest probability to originate
in a region within three virial radii from galaxies.
In the regime beyond 3 $r_{\rm {vir}}$, the most probable systems are below our observational detection limit. 

This result is strengthened by the fact that we do not recover any strong
difference between the MUPPI and the Effective models. 
In Fig.~\ref{fig:ConfMUPPIEFF}, we show the comparison between the two models (MUPPI in blue dashed and the Effective model in red
solid). The peaks of the probabilities are always coincident even though the Effective model shows a more extended tail for low 
values of \mbox{C\,{\sc iv}} column densities for all the radii above 1 $r_{\rm {vir}}$.
%
This could be due to the different mechanism of feedback of the Effective model, which is not capable to
spread metals with the same efficiency.



\begin{figure*}
 \centering
  \makebox[\textwidth][c]{
  \subfloat{\includegraphics[width=0.4\textwidth]{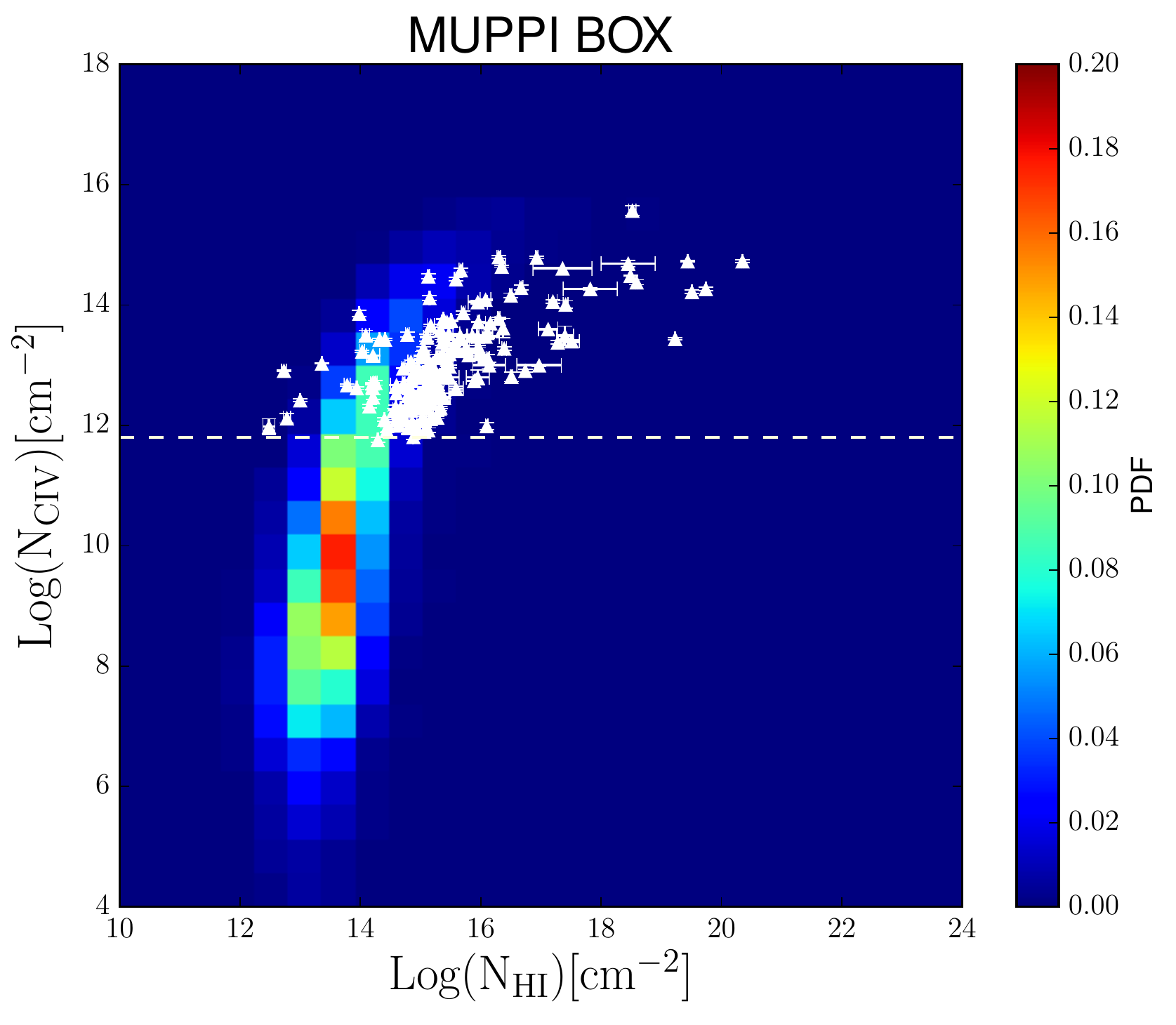}} 
  \hspace{0.5cm}
  \subfloat{\includegraphics[width=0.4\textwidth]{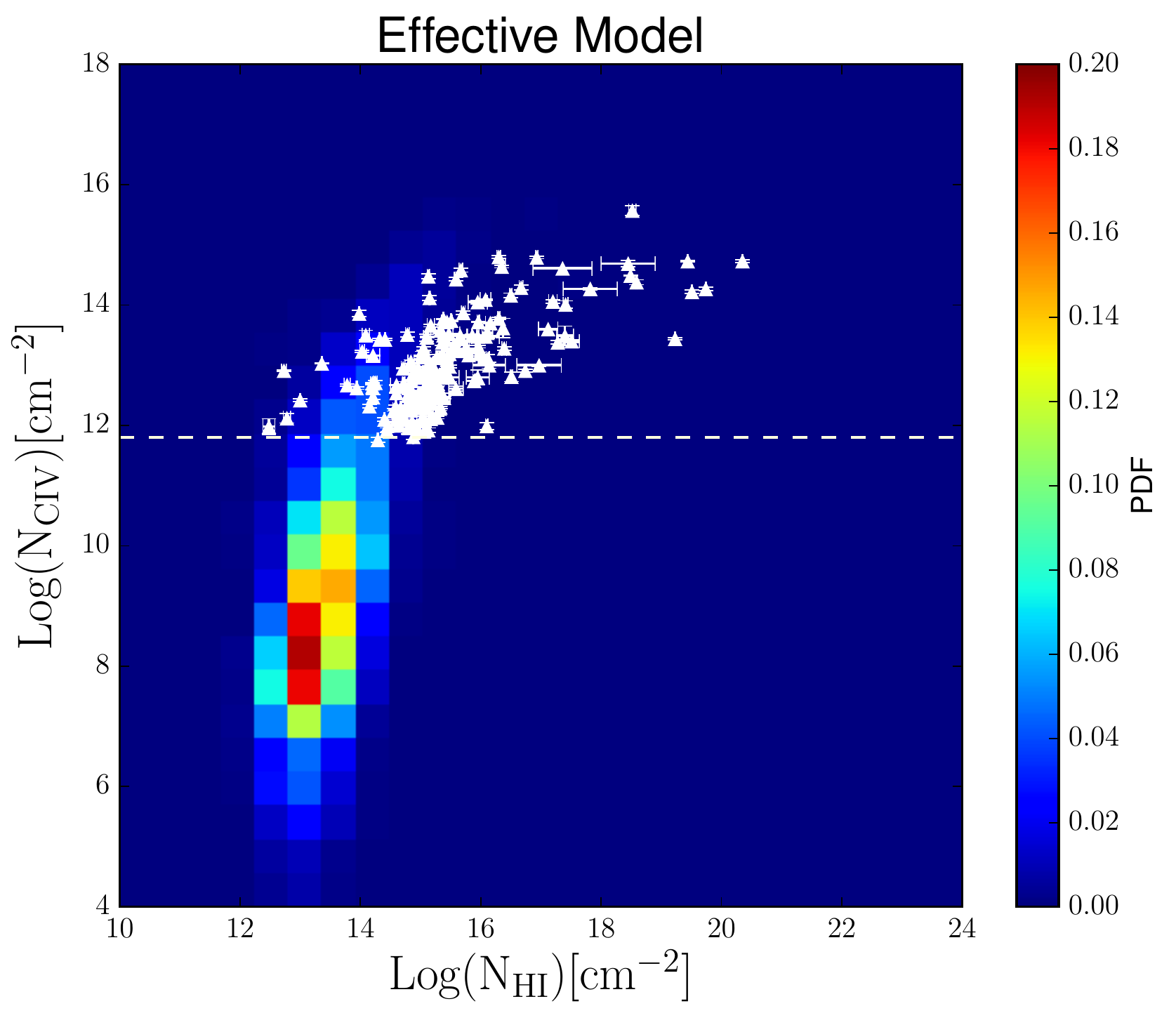}}
  
  }
%
  \caption{PDF for the N$_{\rm{ C\, \textsc{iv}}}$ versus N$_{\rm{ H\, \textsc{i}}}$ relation of lines of sight around 
  near-filament environments. All lines of sight with impact parameter less than 800 kpc are considered. {\it White triangles}: 
  observational data from K16. {\it Dashed horizontal line}: observational detection limit. {\it Left-hand panel}:  results for the MUPPI model.  {\it Right-hand panel}: 
  results for the Effective model. }
  \label{fig:PDFMUPPI_Fil}
\end{figure*}

We repeated the same procedure for near-filament environments, by piercing through the cosmological box random lines of sight 
around the chosen centre of near-filament environments with impact parameter less than 800 kpc.
Fig.~\ref{fig:PDFMUPPI_Fil} shows the PDFs of all lines of sight around near-filament 
environments for both simulations. 
Near-filament environments have the highest probability to have values of the \mbox{C\,{\sc iv}} and
\mbox{H \,{\sc i}} column densities smaller than what it is observed. The tail of the PDF at higher values of column densities
and slightly in correspondence with the observational sample could be due to the presence
of haloes near the chosen ``centre'' of a near-filament environment.

%

%
\subsection{The N$_{\rm{ C\, \textsc{iv}}}$ versus N$_{\rm{ H\, \textsc{i}}}$ relation: median values in concentric regions}
\label{ss:Nrelation2}

\begin{figure*}
 \centering
  \makebox[\textwidth][c]{
  \subfloat{\includegraphics[width=0.4\textwidth]{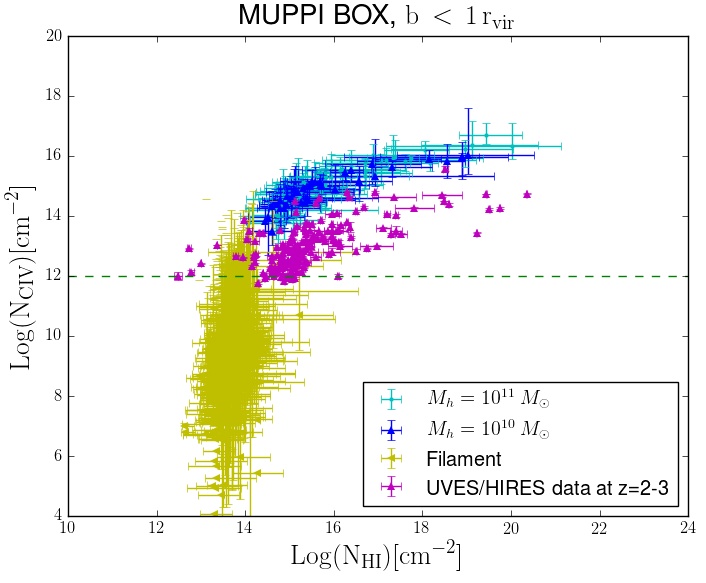}} 
  \hspace{0.5cm}
  \subfloat{\includegraphics[width=0.4\textwidth]{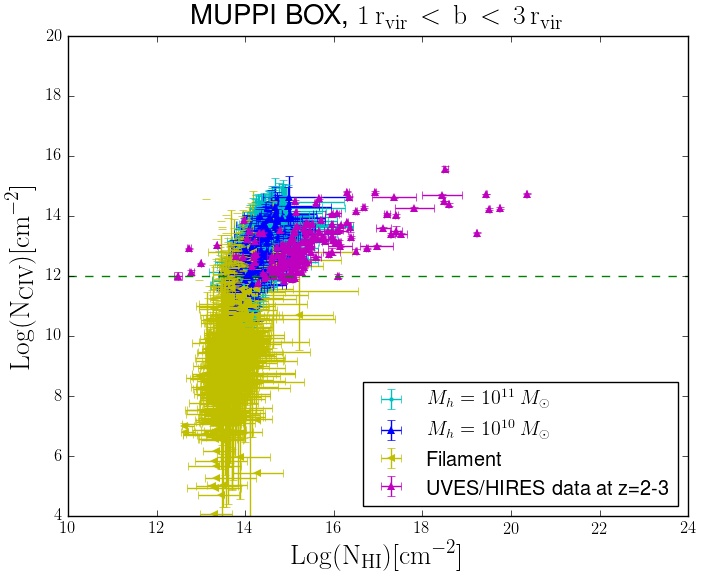}}
  
  }
   \vspace{0.1cm}
   \makebox[\textwidth][c]{
  \subfloat{\includegraphics[width=0.4\textwidth]{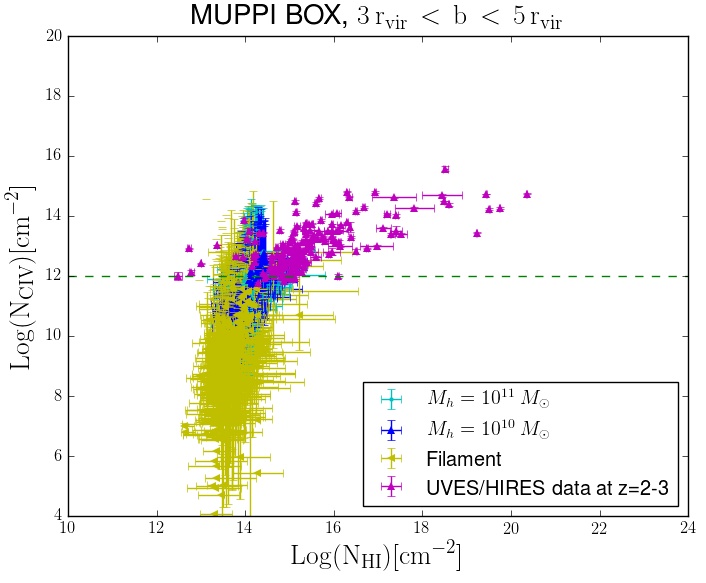}} 
  \hspace{0.5cm}
  \subfloat{\includegraphics[width=0.4\textwidth]{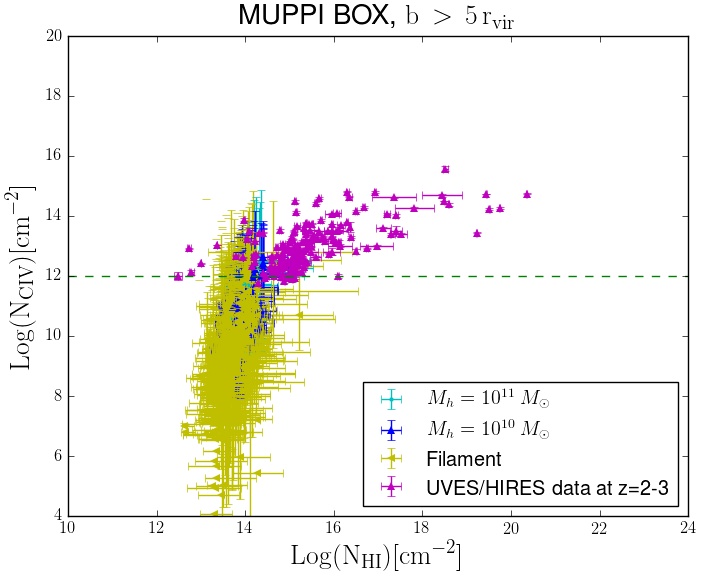}}
 
  }
  \caption{Each point is the spherical average value of \mbox{H\,{\sc i}} and \mbox{C\,{\sc iv}} column densities profiles in radial bins of width of $\sim$ 10 kpc 
  around the 30 objects (20 galaxies and 10 near-filament environments) for the MUPPI simulation. Error bars represent the 1$\sigma$ dispersion. $Cyan$ $points$: 
  mean values of lines of sights around haloes with $M_{\rm h}  \sim 10^{11}$ M$_\odot$. 
  $Blue$ $points$: mean values of lines of sights around haloes with $M_{\rm h}  \sim 10^{10}$ M$_\odot$. 
 $Yellow$ $points$: mean values of lines of sights around near-filaments points. $Magenta$ $points$: observational data from K16 as in previous plots (the same data are reported in each panel). 
 $Green$ $horizontal$ $dashed$ $line$: observational detection limit.
 Plots have been divided according to the distance of radial galaxy spherical averages from galaxy centre in units of virial radius. This
 division does not concern near-filaments points, whose all values are reported in each plot.}
  \label{fig:MedMUPPI}
\end{figure*}

\begin{figure*}
 \centering
 \makebox[\textwidth][c]{
  \subfloat{\includegraphics[width=0.4\textwidth]{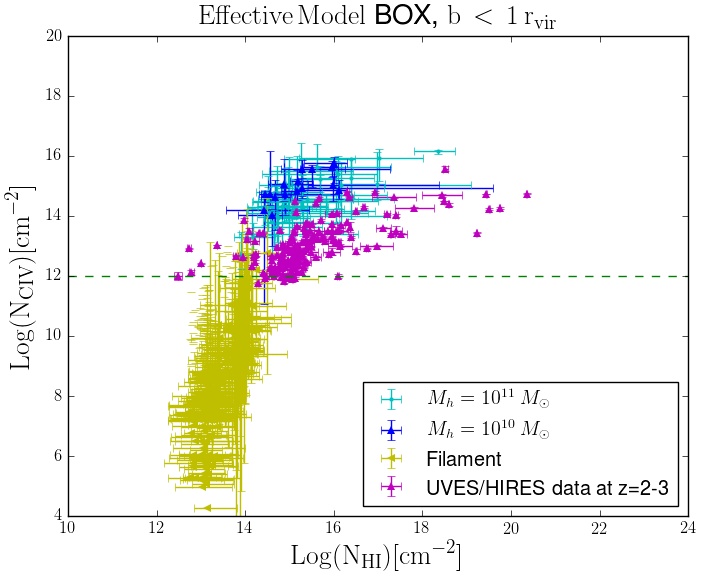}} 
  \hspace{0.5cm}
  \subfloat{\includegraphics[width=0.4\textwidth]{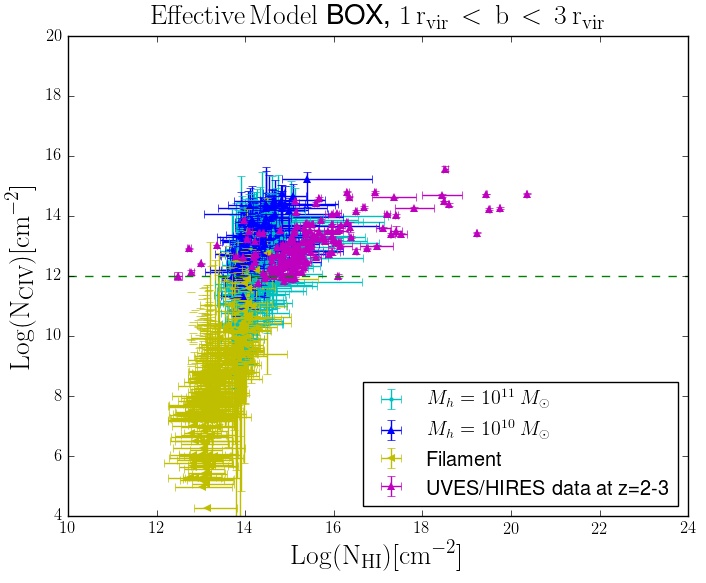}}
  
  }
   \vspace{0.1cm}
   \makebox[\textwidth][c]{
  \subfloat{\includegraphics[width=0.4\textwidth]{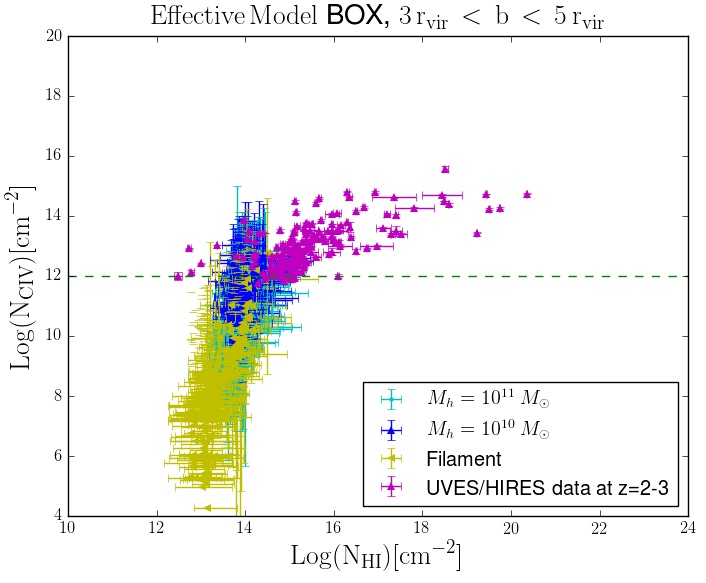}} 
  \hspace{0.5cm}
  \subfloat{\includegraphics[width=0.4\textwidth]{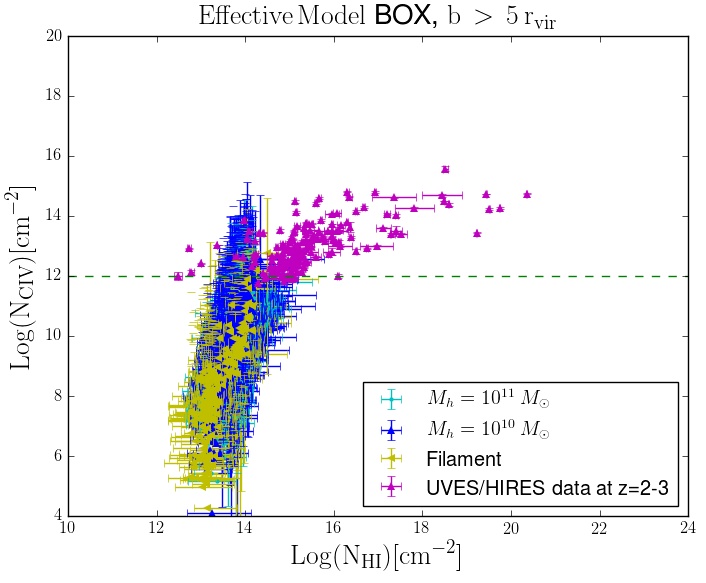}}
 
  }
  \caption{Same as Fig.~\ref{fig:MedMUPPI}, but for the Effective model simulation.}
  \label{fig:MedEFF}
\end{figure*}

We performed another analysis providing a more statistical picture of the distribution of metals in the 
CGM/IGM. 
We took the lines of sight of Section~\ref{ss:Nrelation1} and for each object, we divided them according to their distance 
from the centre of the object: going from 0 to 800 kpc, we divided the spatial region considered in radial bins of size 
$\sim$ 10 kpc and we grouped lines of sight in each bin according to their impact parameter.
For each object, we computed the mean and the 1$\sigma$ dispersion of the integrated N$_{\rm{ C\, \textsc{iv}}}$ and N$_{\rm{ H\, \textsc{i}}}$ values of all the 
lines of sight in each bin.
We report this modified N$_{\rm{ C\, \textsc{iv}}}$ versus N$_{\rm{ H\, \textsc{i}}}$ relation for the MUPPI and the Effective models in 
Figs.~\ref{fig:MedMUPPI} and \ref{fig:MedEFF}.


For each galaxy, we divided the mean values according to their distance from galaxy centre, so plotting mean values whose 
radial bin is at a distance from centre between [0-1] $r_{\rm {vir}}$, [1-3] $r_{\rm {vir}}$ , [3-5] $r_{\rm {vir}}$ and r $>$ 5 $r_{\rm {vir}}$. 
Each point refers to a single LOS. Blue points refer to lines of sight around galaxies with a total
halo mass between M$_{\rm h} \sim 10^{11}$ and 10$^{12}$ M$_\odot$, while cyan points refer to galaxies with a total halo mass in the range 
$M_{\rm h} \sim 10^{10}-10^{11}$ M$_\odot$. Yellow represents near-filament environments. Magenta points represent the observational sample by \citet{Kim2016} and the green
horizontal dashed line is the detection limit of this sample.

For near-filaments environments, it is obviously not possible to define a virial radius, so we plotted in each panel of 
Figs~\ref{fig:MedMUPPI} and \ref{fig:MedEFF}, the mean values of all the radial bins from 0 to 800 kpc.

Large error bars, representing the 1-$\sigma$ dispersion, are due to the fact that the medium is not homogeneous, so at same radial
distance we could find a low-density environment, having low column densities, or we could hit (or be close to) a substructure, 
having higher values of column densities.

As in the previous section, we can see that for galaxies median values shift to lower column densities as the distance 
increases and we do not find any correspondence between simulated and observed data above 3-5 r$_{\rm vir}$ except for the 
weakest systems of the sample of K16. Near-filaments points are confined to a region under the detection limit, 
suggesting that they have metallicities too low to be probed by present-day observations. 
First results in this sense have been obtained with extremely high S/N ratio spectra \citep{Dodorico2016,Ellison2000}.
However, to  collect statistical samples of low column density metal absorbers we will have to wait for next generation 
high-resolution spectrographs, like ESPRESSO (Echelle SPectrograph for Rocky Exoplanet and Stable Spectroscopic Observations) at the VLT \citep{Pepe2014e} or the
HIRES spectrograph at the E-ELT [European-Extremely Large Telescope, \citet{Marconi2016}].

\subsection{Comparison between MUPPI and the Effective model}
\label{ss:CompMUPPIEFF}
As already stated, the subresolution models do not show strong differences in the results for \mbox{C\,{\sc iv}}.

To better investigate this last issue, we performed the same overall analysis with two other ions: \mbox{O\,{\sc vi}} and \mbox{Si\,{\sc iv}}.
The results for the \mbox{O\,{\sc vi}} are discussed in Appendix~\ref{s:figures}, since the observational data were taken from the literature and the 
procedure to compute column densities is slightly different from our approach. In general, \mbox{O\,{\sc vi}} shows a similar behavior as \mbox{C\,{\sc iv}} without 
any strong differences between the two models, apart from the tail of the Effective model, which extends more to low 
values of \mbox{O\,{\sc vi}} column densities  in the range above 1 $r_{\rm {vir}}$.

\begin{figure*}
 \centering
  \makebox[\textwidth][c]{
  \subfloat{\includegraphics[width=0.31\textwidth]{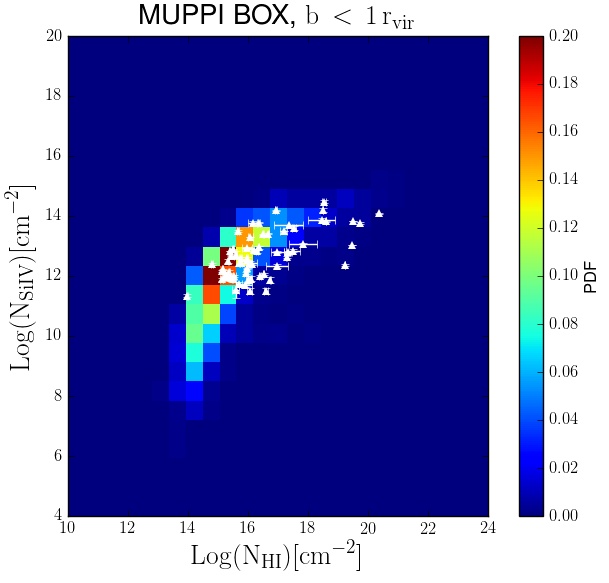}} 
  \hspace{0.5cm}
  \subfloat{\includegraphics[width=0.31\textwidth]{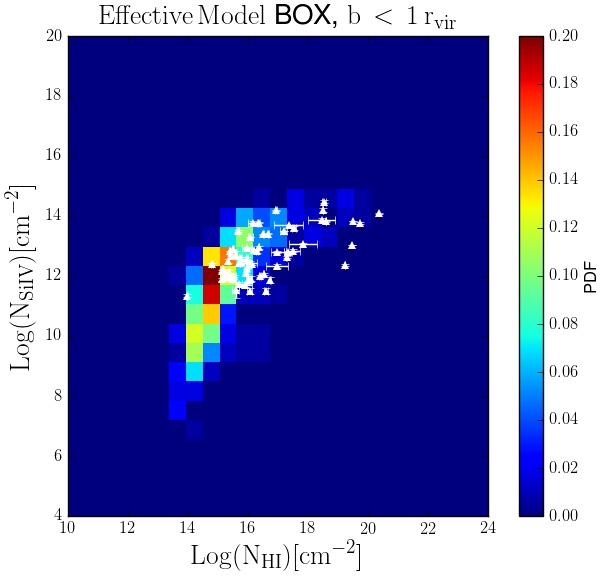}}
  
  }
   \vspace{0.1cm}
   \makebox[\textwidth][c]{
  \subfloat{\includegraphics[width=0.31\textwidth]{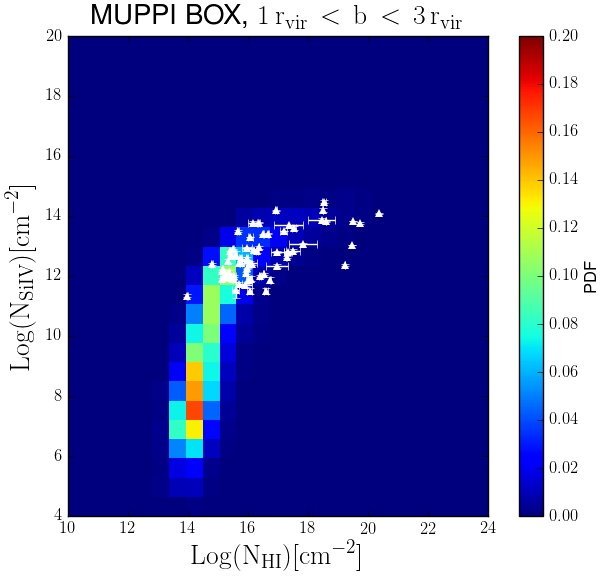}} 
  \hspace{0.5cm}
  \subfloat{\includegraphics[width=0.31\textwidth]{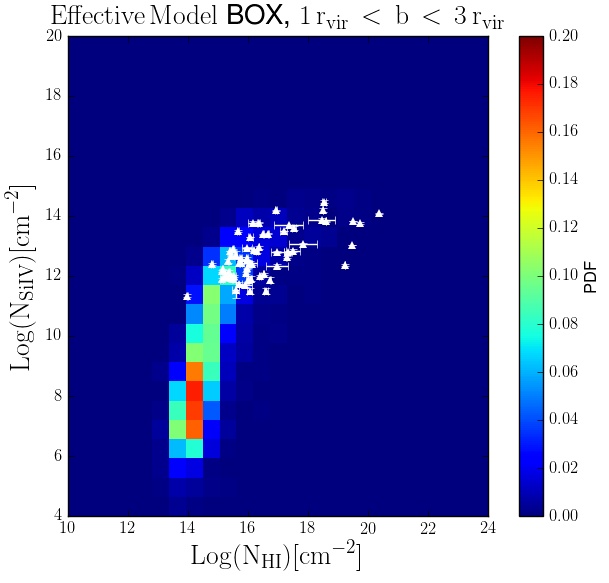}}
 
  }
      \vspace{0.1cm}
    \makebox[\textwidth][c]{
   \subfloat{\includegraphics[width=0.31\textwidth]{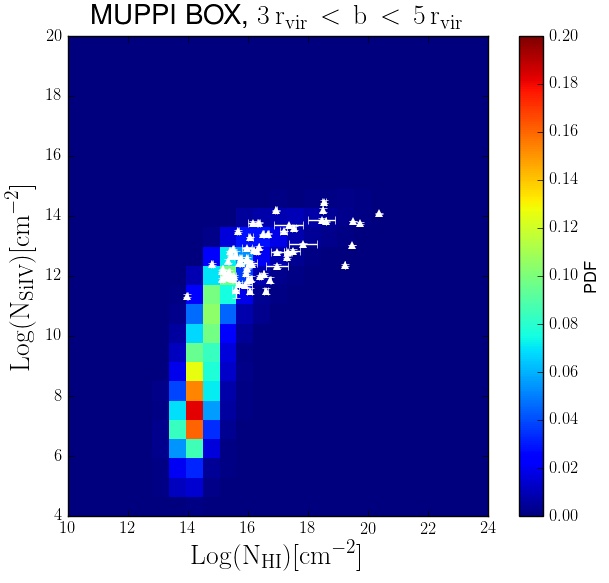}} 
  \hspace{0.5cm}
   \subfloat{\includegraphics[width=0.31\textwidth]{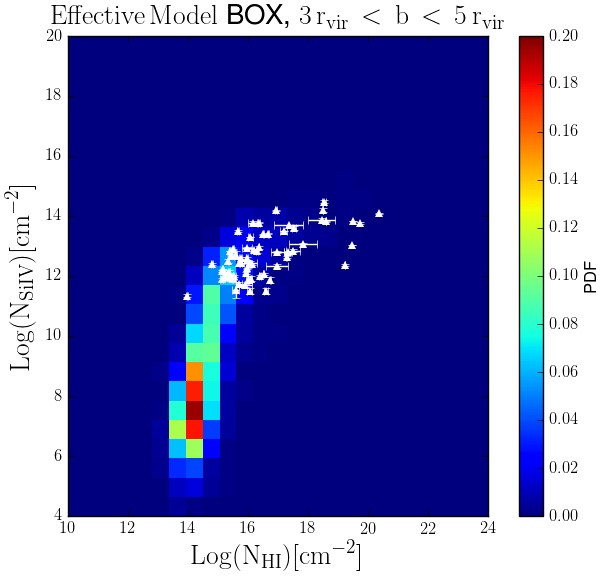}}
   }
     \vspace{0.1cm}
   \makebox[\textwidth][c]{
  \subfloat{\includegraphics[width=0.31\textwidth]{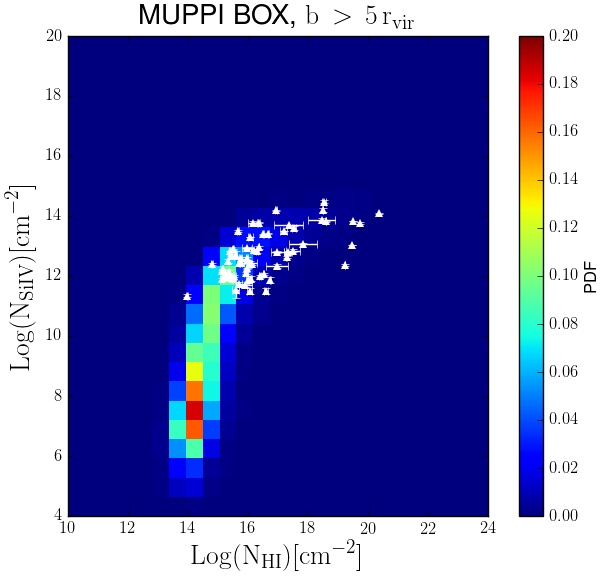}} 
  \hspace{0.5cm}
  \subfloat{\includegraphics[width=0.31\textwidth]{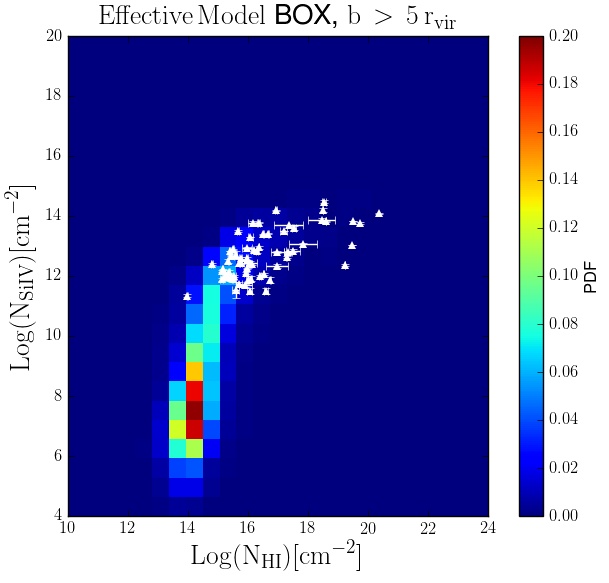}}
 
  } 
   \caption{PDF of N$_{\rm SiIV}$ versus N$_{\rm{ H\, \textsc{i}}}$ relation. $White$ $points$: observational data from K16 (the same data are reported in each panel).}.
  \label{fig:SiIVCompPDF}
\end{figure*}

\begin{figure*}
 \centering
  \makebox[\textwidth][c]{
  \subfloat{\includegraphics[width=0.34\textwidth]{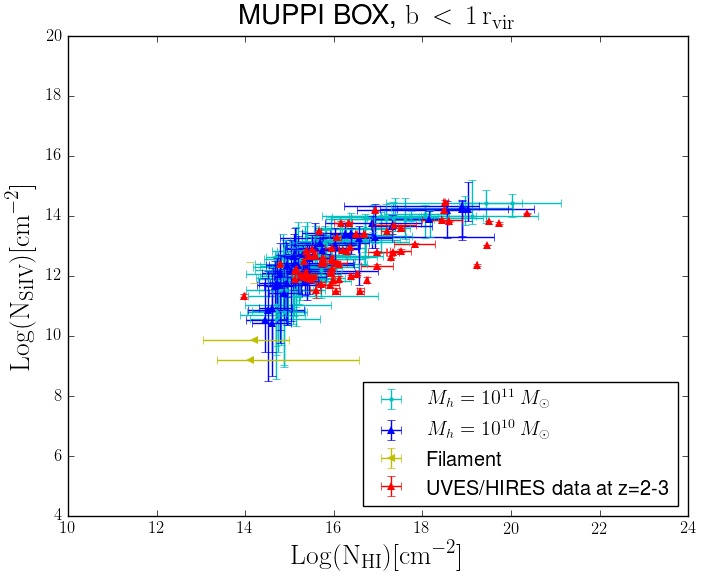}} 
  \hspace{0.5cm}
  \subfloat{\includegraphics[width=0.34\textwidth]{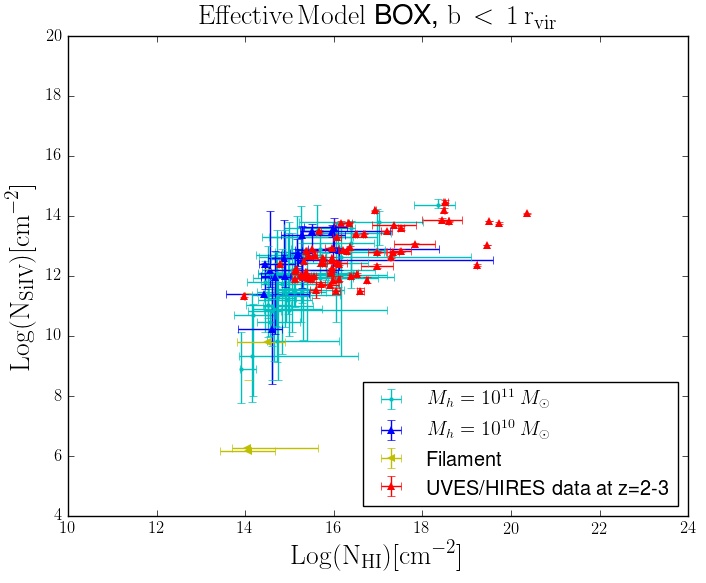}}
  
  }
   \vspace{0.1cm}
   \makebox[\textwidth][c]{
  \subfloat{\includegraphics[width=0.34\textwidth]{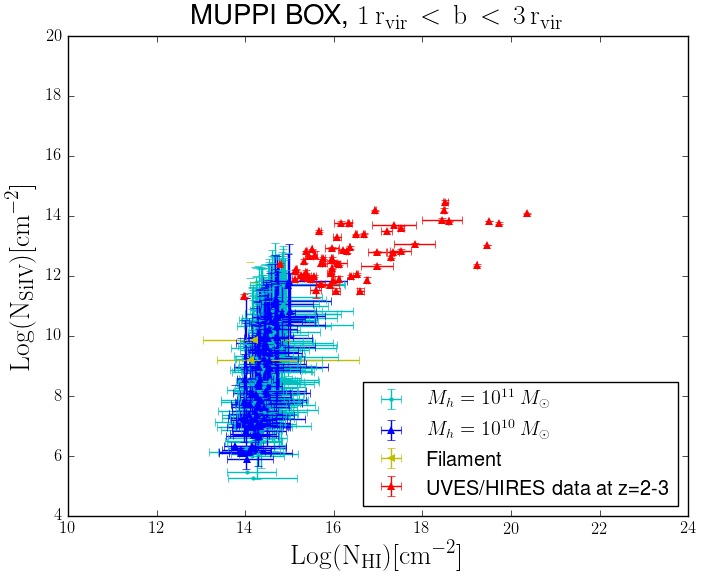}} 
  \hspace{0.5cm}
  \subfloat{\includegraphics[width=0.34\textwidth]{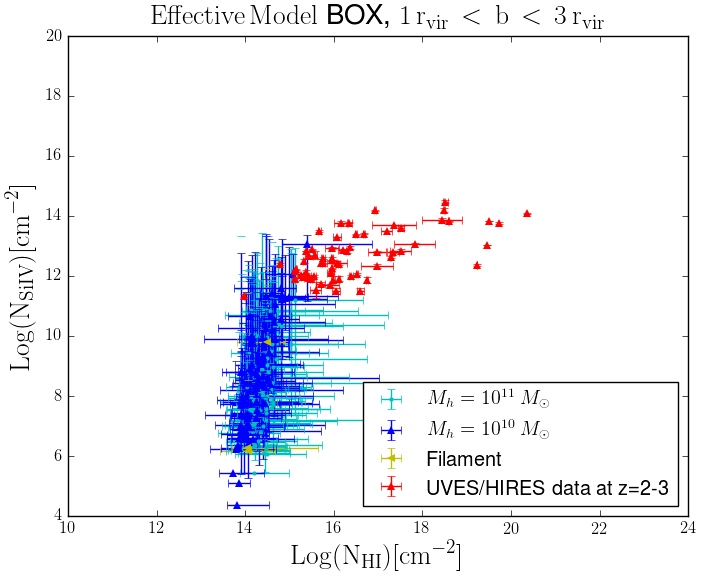}}
 
  }
      \vspace{0.1cm}
    \makebox[\textwidth][c]{
   \subfloat{\includegraphics[width=0.34\textwidth]{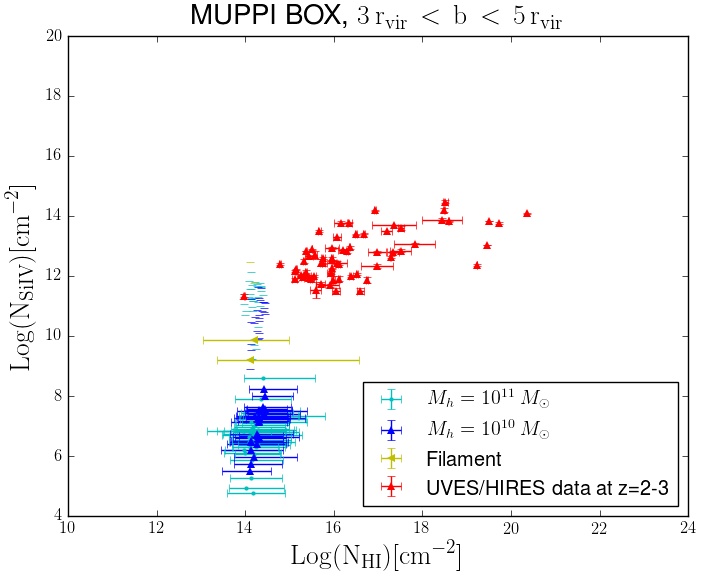}} 
  \hspace{0.5cm}
   \subfloat{\includegraphics[width=0.34\textwidth]{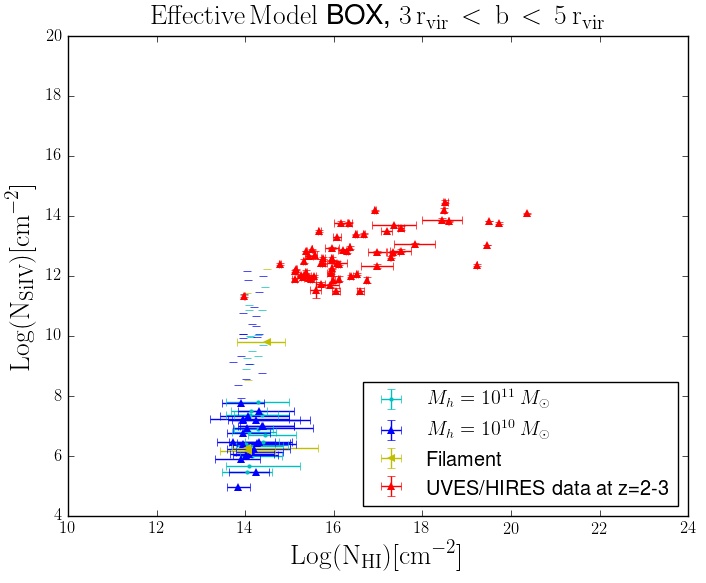}}
   }
     \vspace{0.1cm}
   \makebox[\textwidth][c]{
  \subfloat{\includegraphics[width=0.34\textwidth]{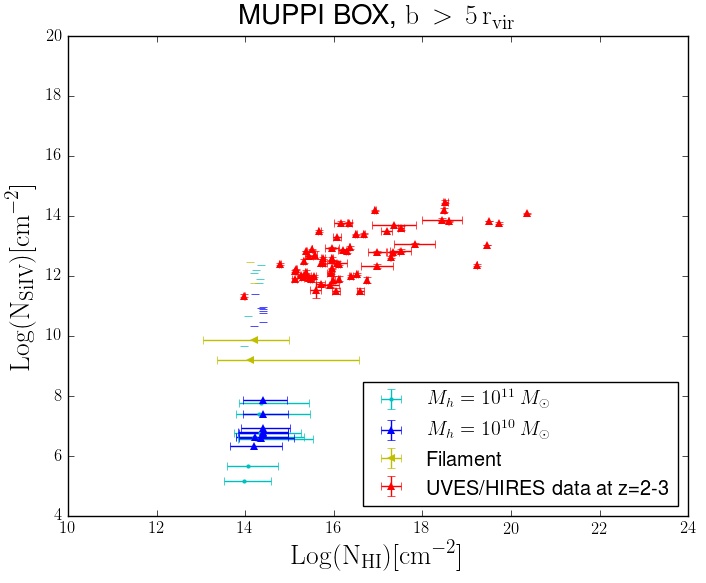}} 
  \hspace{0.5cm}
  \subfloat{\includegraphics[width=0.34\textwidth]{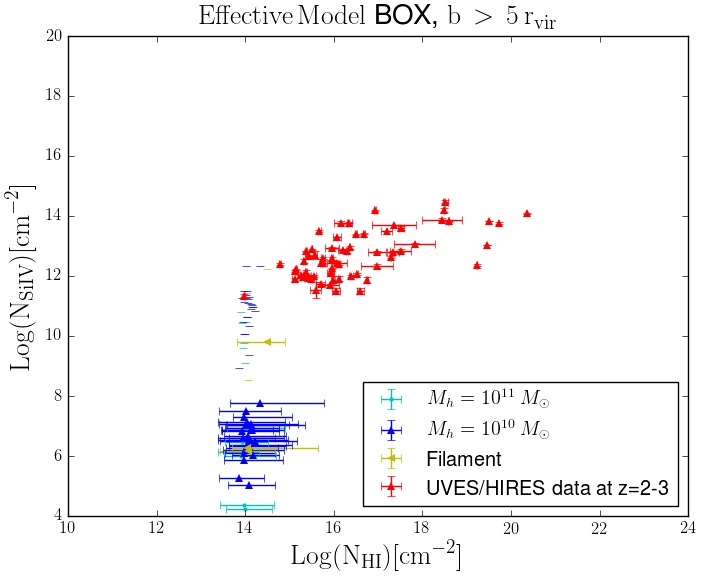}}
 
  } 
   \caption{Each point is the spherical average value of \mbox{H\,{\sc i}} and \mbox{Si\,{\sc iv}} column densities profiles in radial bins of width of $\sim$ 10 kpc 
  around the 30 objects (20 galaxies and 10 near-filament environments) for the MUPPI simulation. Error bars represent the 1$\sigma$ dispersion. $Cyan$ $points$: mean
  values of lines of sights around haloes with $M_{\rm h}  \sim 10^{11}-10^{12}$ M$_\odot$. 
  $Blue$ $points$: mean values of lines of sights around haloes with $M_{\rm h}  \sim 10^{10}-10^{11}$ M$_\odot$. 
 $Yellow$ $points$: mean values of lines of sights around near-filaments points. $Red$ $points$:
 observational data from table 4 of K16 (red points are equal to white points in Fig.~\ref{fig:SiIVCompPDF}); the same data are reported in each panel.
 Plots have been divided according to the distance of radial galaxy spherical averages from galaxy centre in units of virial radius. 
 Near-filament points are the same in each plot.}.
  \label{fig:SiIVCompMed}
\end{figure*}

The results for \mbox{Si\,{\sc iv}} are instead shown in Figs~\ref{fig:SiIVCompPDF} and \ref{fig:SiIVCompMed}.
%
%
For the \mbox{Si\,{\sc iv}}, we compared always with the observational sample by K16. 
Unlike \mbox{C\,{\sc iv}} and \mbox{O\,{\sc vi}}, we do not see any difference between the models in every radial bin. \mbox{Si\,{\sc iv}} traces a gas phase at higher 
N$_{\rm{ H\, \textsc{i}}}$, closer to galaxies. At these close distances to galaxies, the effects of the feedback could not be traced.

\begin{figure*}
  \centering
  \makebox[\textwidth][c]{
  \subfloat{\includegraphics[width=0.45\textwidth]{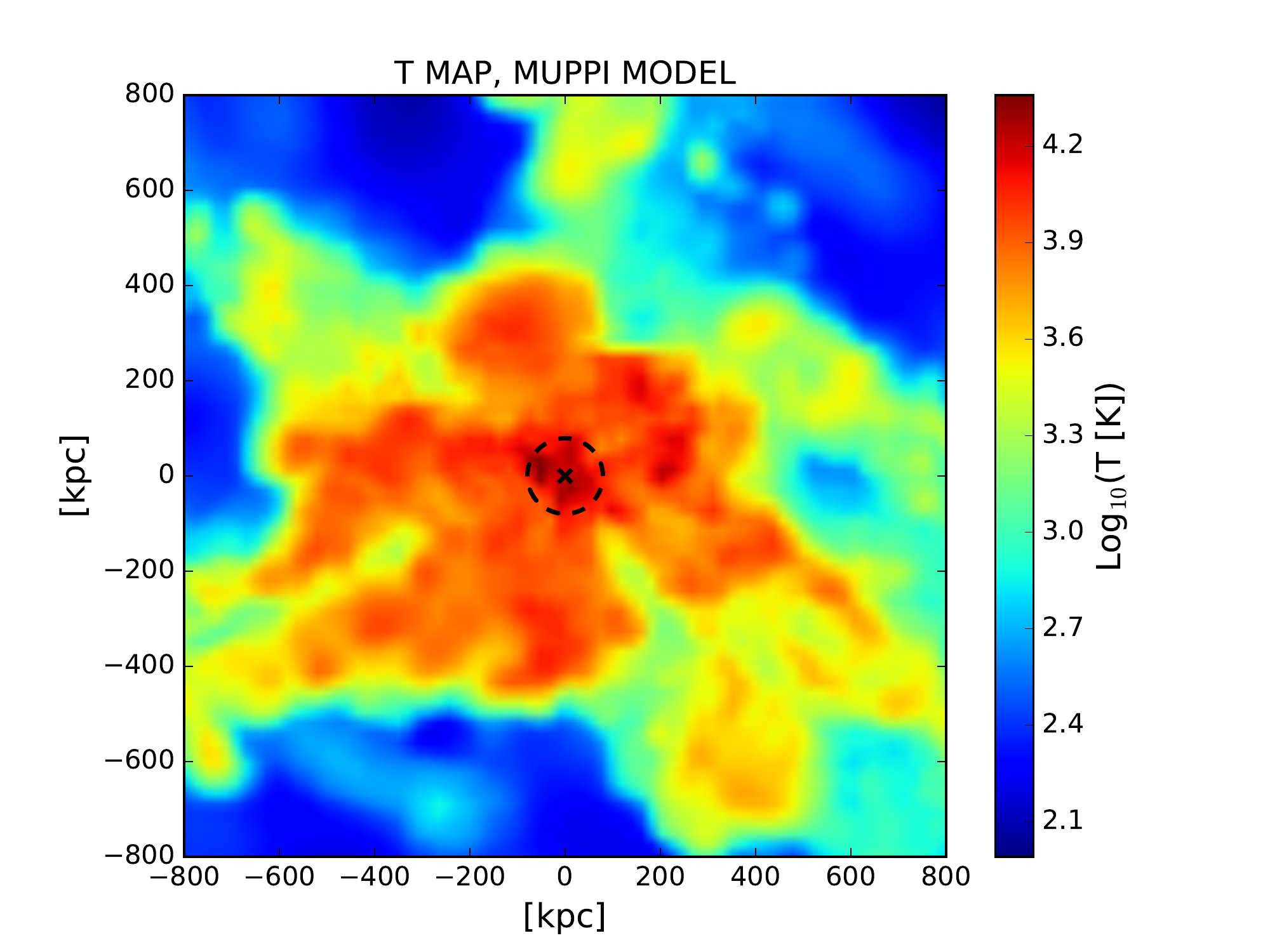}} 
  \hspace{0.5cm}
  \subfloat{\includegraphics[width=0.45\textwidth]{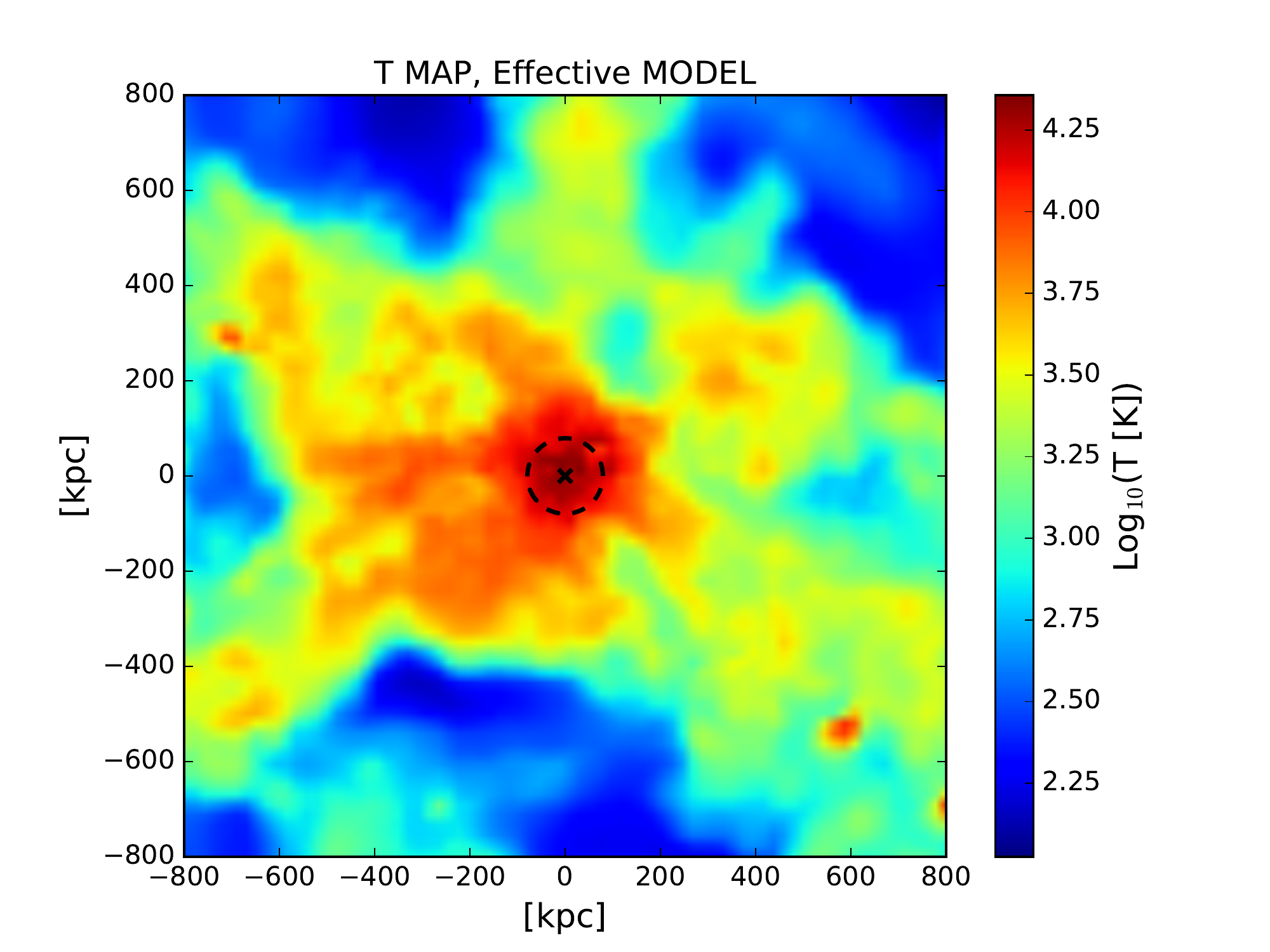}}
  
  }

  \caption{Comparison between temperature maps of the same galaxy between the MUPPI and the Effective
   models. Temperatures are computed by taking the mean value of the temperature profile along the LOS in a slice $\pm$ 300 kpc from galaxy's position.
   (Distances are in proper kpc.)}
  \label{fig:CompTempMAP}
\end{figure*}

One explanation of these results could be that the differences in the feedback prescriptions of the two models could be more 
related to the thermal feedback than to the kinetic one. Fig.~\ref{fig:CompTempMAP} shows a temperature map around
the same galaxy for both models, in which the mean value of the temperature profile along the LOS in a slice
$\pm$300 proper kpc is computed.  It is clear that the MUPPI model is more 
capable of heating the surrounding gas with respect to the Effective model, while the differences related to kinetic 
feedback, that is how metals are distributed outside galaxies, are less visible.

\citet{Barai2015} already studied different feedback schemes, among which the ones used
in this work. In their paper, they constructed radial profiles of the total gas metallicity around galaxy centres 
at $z\sim$2 and they inferred that the MUPPI model distributes metals more adequately than the Effective model.
We do also recover a slight difference between the two models in the range above 1 $r_{\rm {vir}}$, but when
comparing with observational data, these differences seem not to significantly impact on the IGM properties investigated here.
It is important to highlight, as already said in Section~\ref{s:sim}, that the MUPPI simulation, that we used, was run with slightly
different model parameters, due to small changes in the chemical sector with respect to the one by \citet{Barai2015}.

\begin{figure*}
 \centering
  \makebox[\textwidth][c]{
  \subfloat{\includegraphics[width=0.4\textwidth]{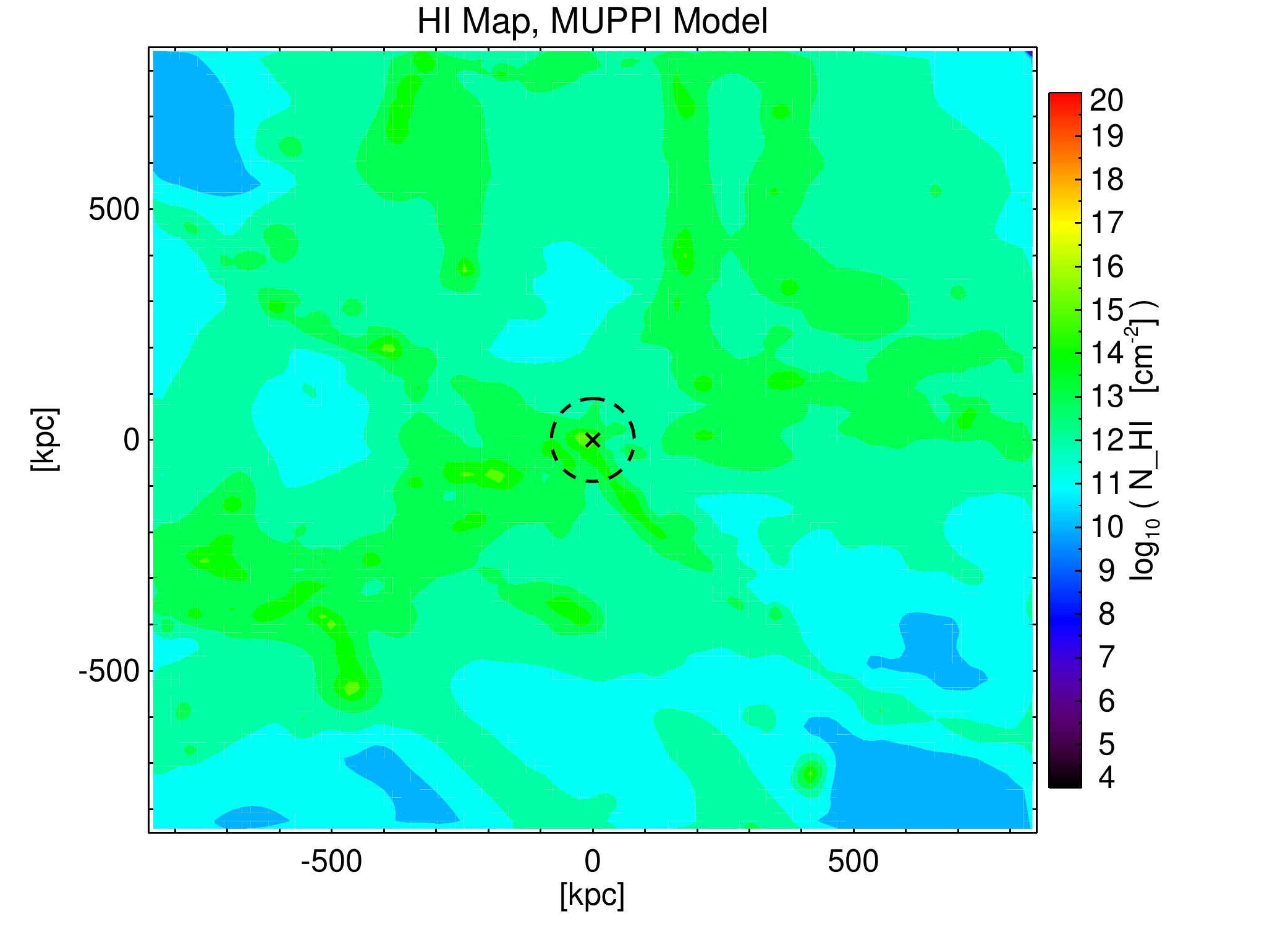}} 
  \hspace{0.5cm}
  \subfloat{\includegraphics[width=0.4\textwidth]{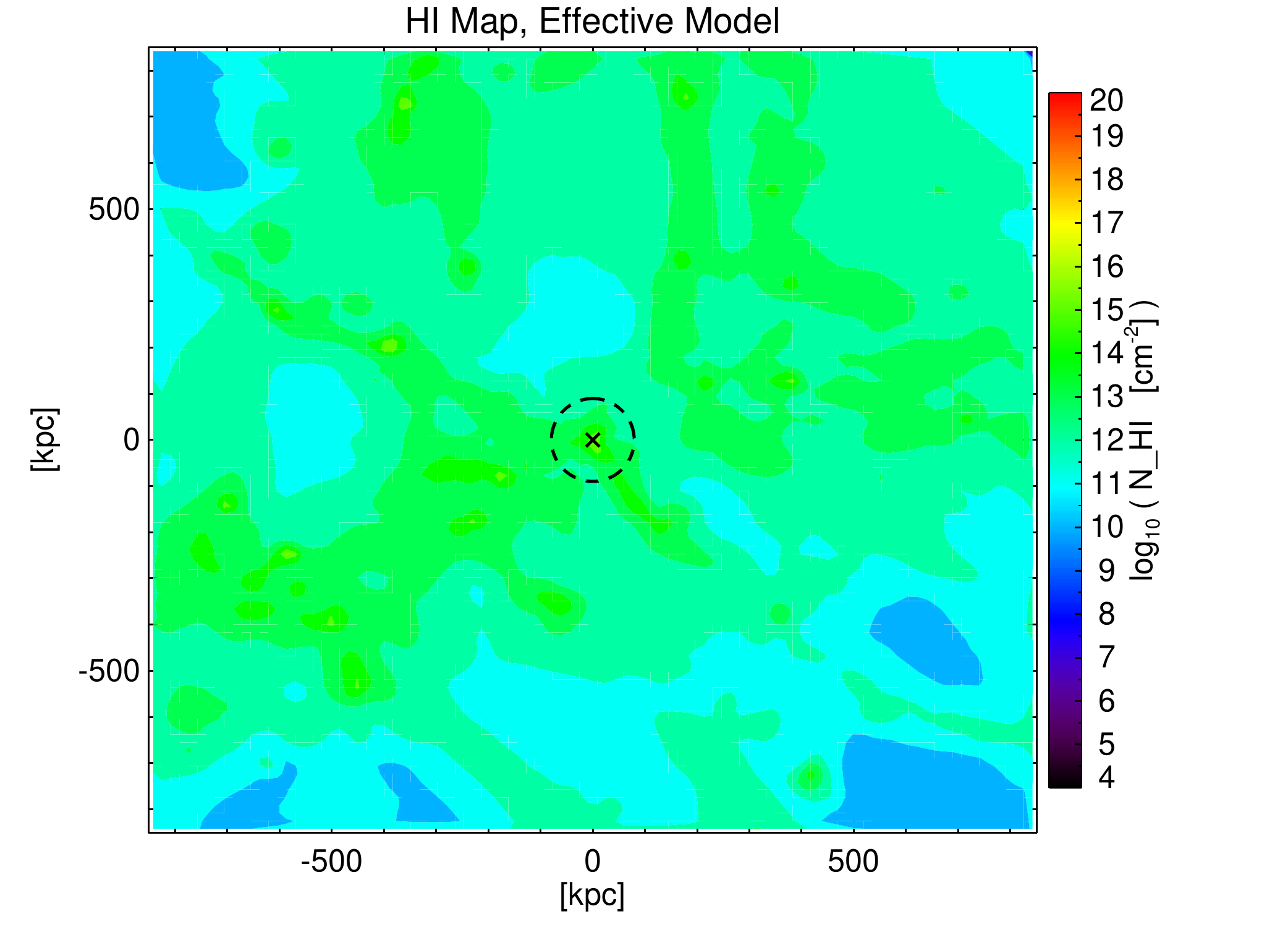}}
  
  }
   \vspace{0.1cm}
   \makebox[\textwidth][c]{
  \subfloat{\includegraphics[width=0.4\textwidth]{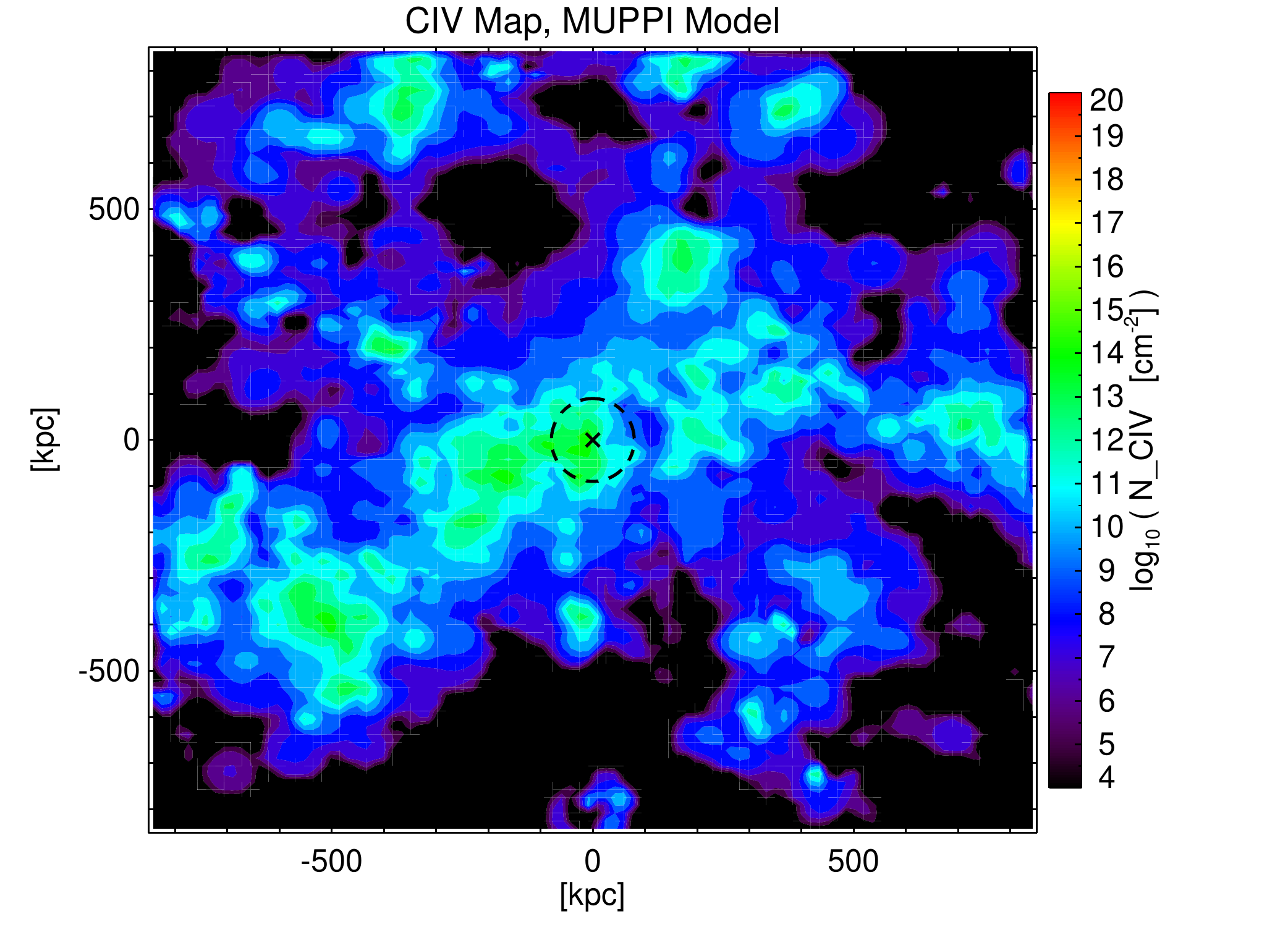}} 
  \hspace{0.5cm}
  \subfloat{\includegraphics[width=0.4\textwidth]{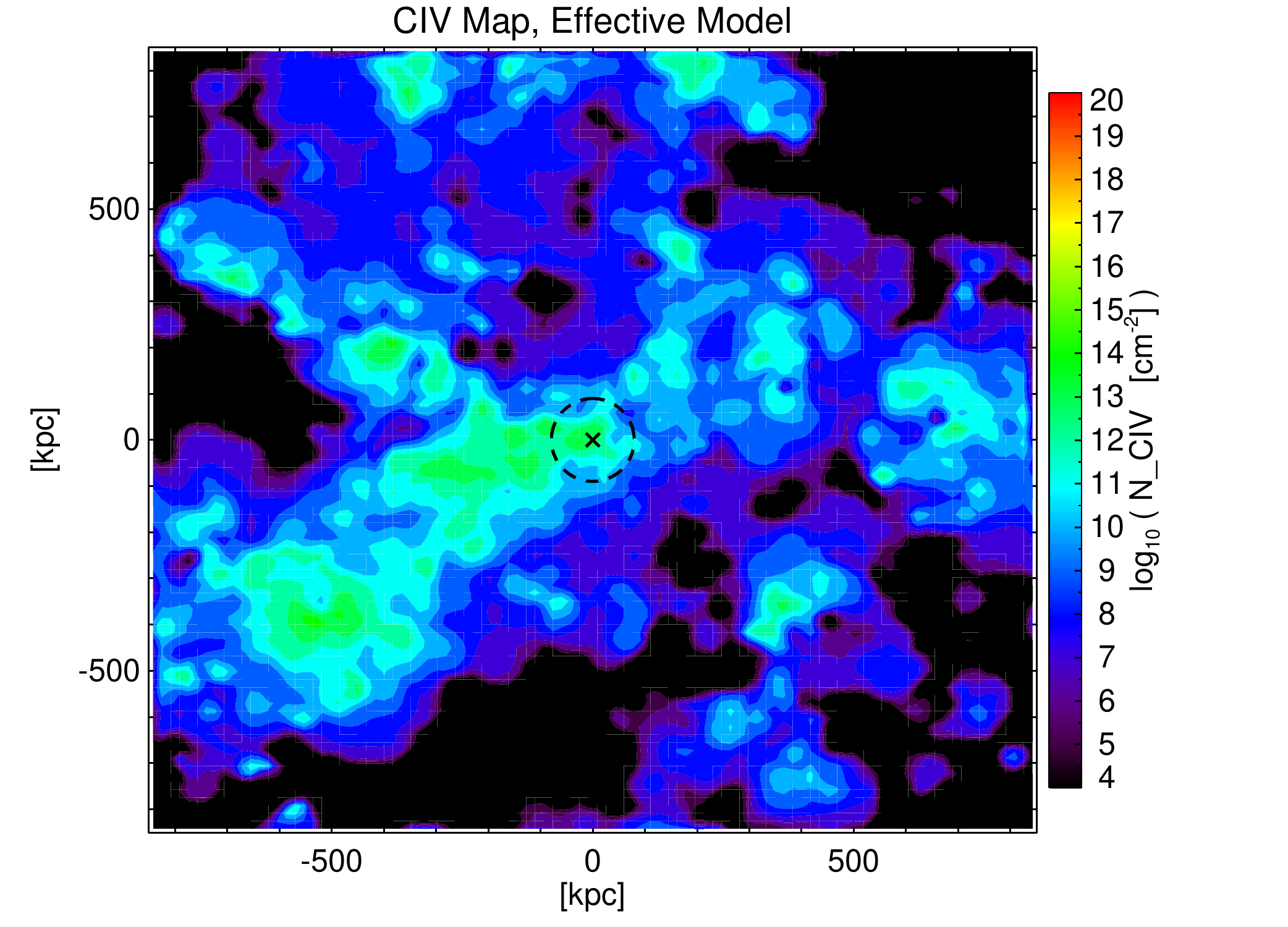}}
 
  }
      \vspace{0.1cm}
    \makebox[\textwidth][c]{
   \subfloat{\includegraphics[width=0.4\textwidth]{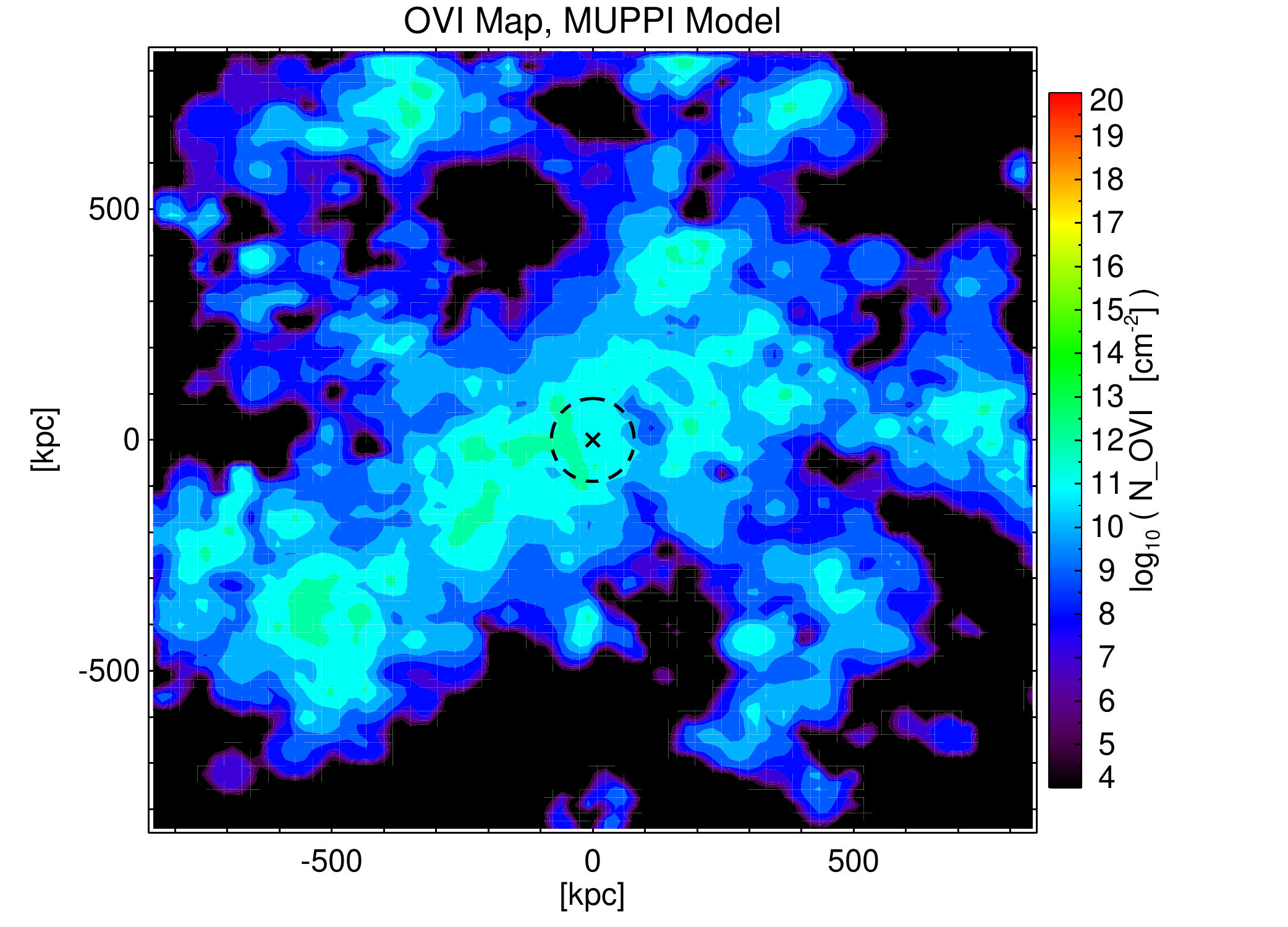}} 
  \hspace{0.5cm}
   \subfloat{\includegraphics[width=0.4\textwidth]{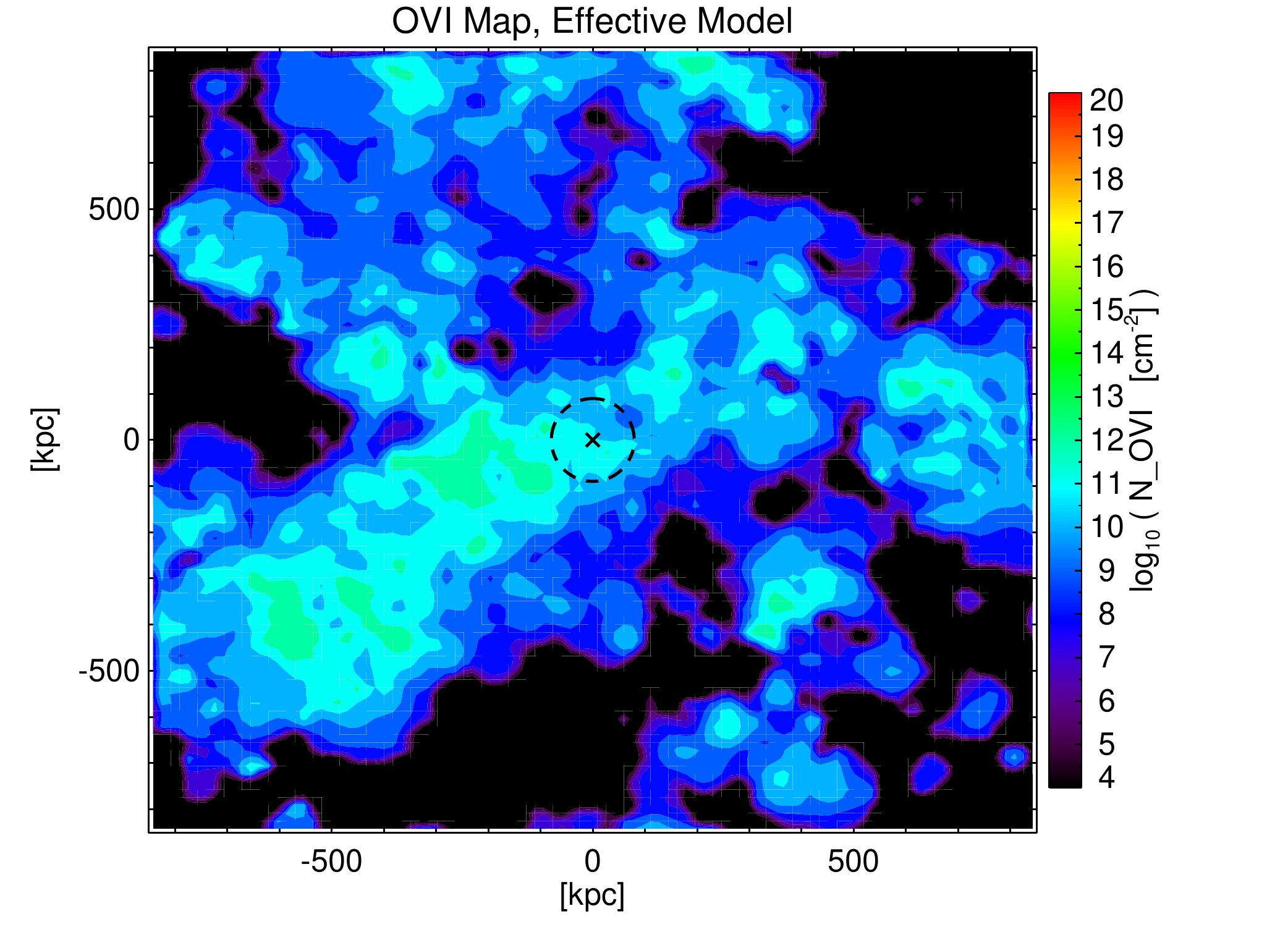}}
   }
     \vspace{0.1cm}
   \makebox[\textwidth][c]{
  \subfloat{\includegraphics[width=0.4\textwidth]{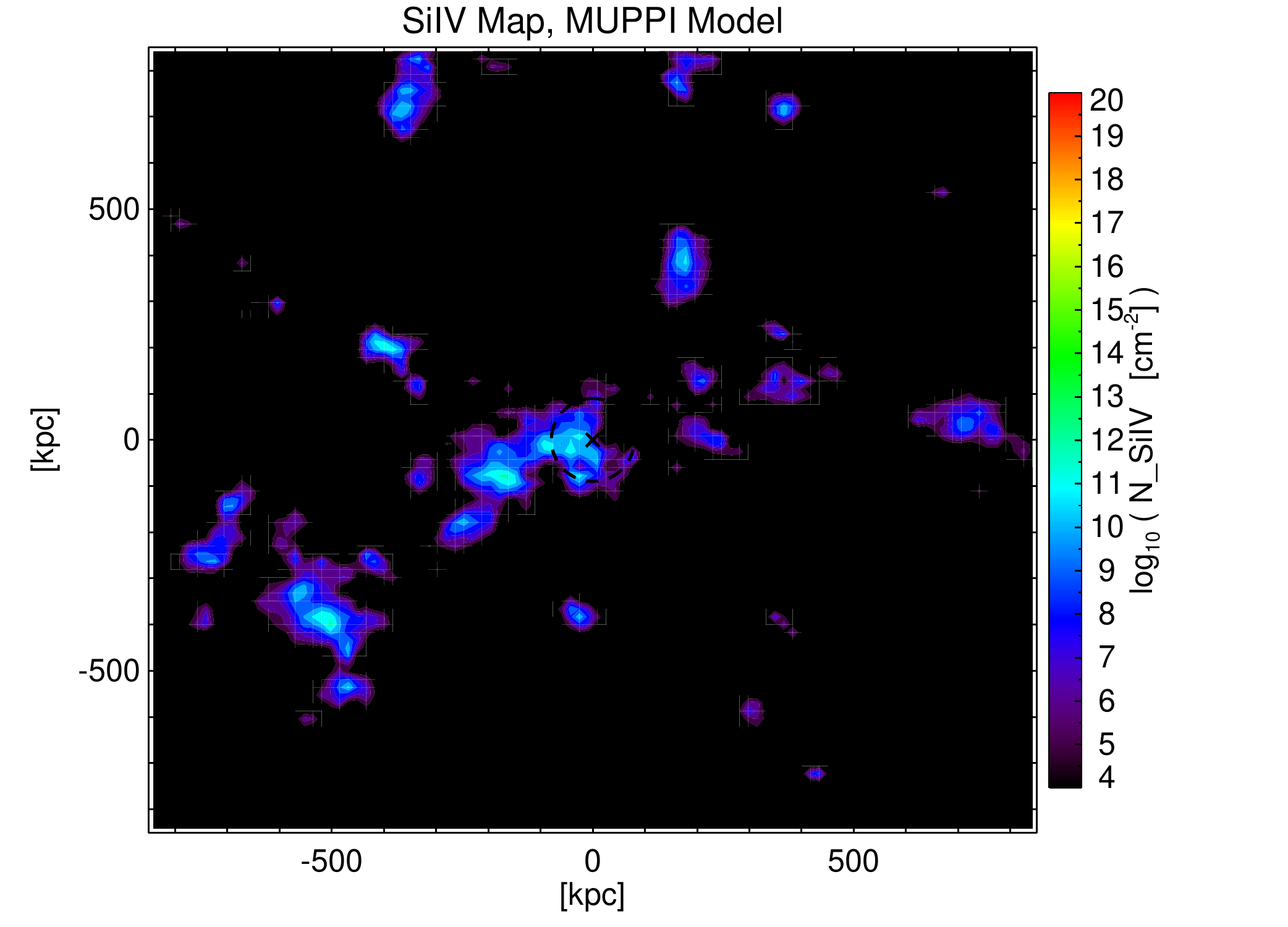}} 
  \hspace{0.5cm}
  \subfloat{\includegraphics[width=0.4\textwidth]{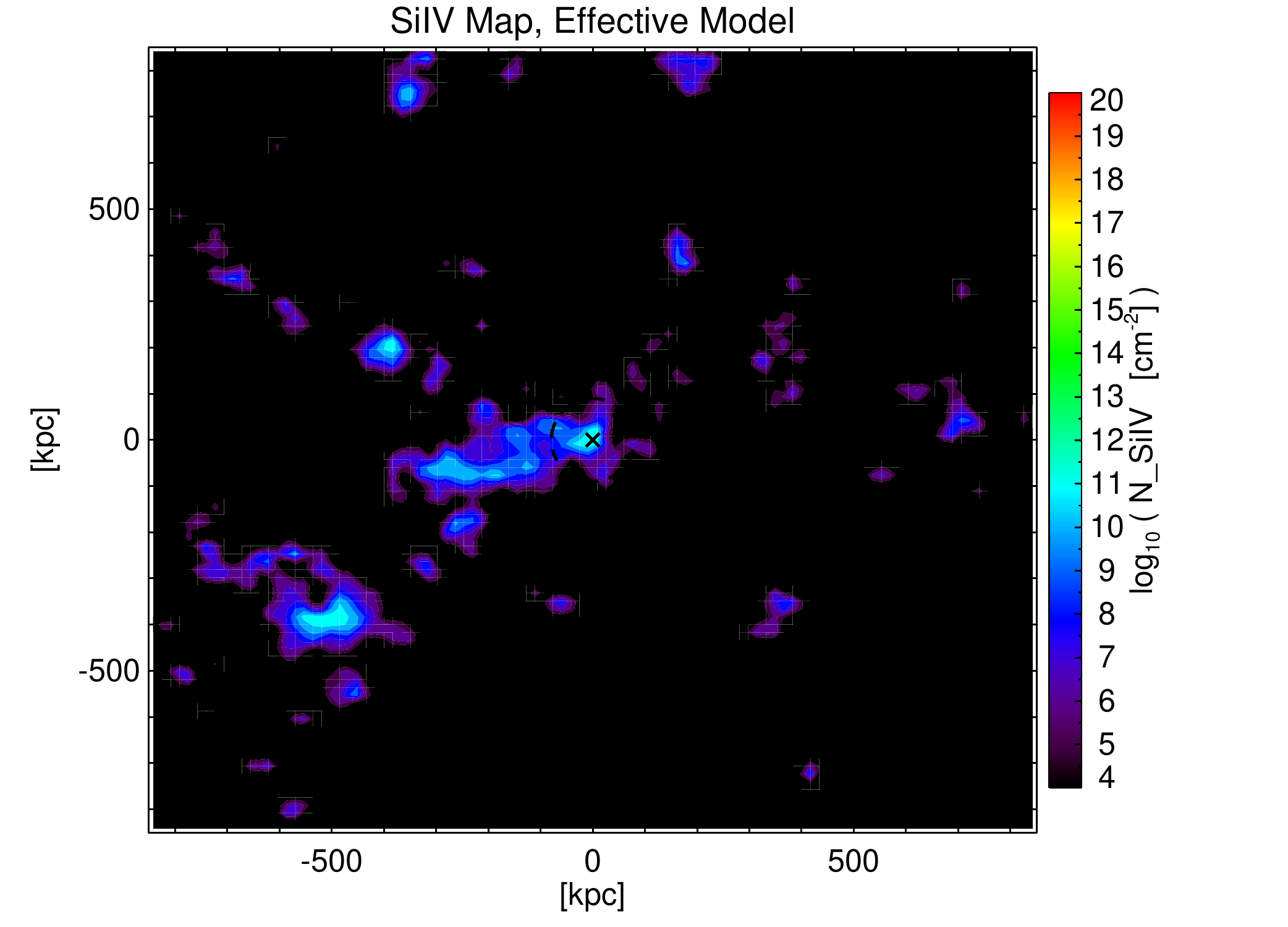}}
 
  } 
   \caption{Comparison between column density maps of different chemical elements between the MUPPI and the Effective
   models. Column densities are computed using the AOD method $\pm$ 150 km s$^{-1}$ from galaxy's position. (Distances are in proper kpc.)}.
  \label{fig:CompModel_AOD}
\end{figure*}

In Fig.~\ref{fig:CompModel_AOD}, we show a comparison for a particular galaxy between column density maps for 
\mbox{H \,{\sc i}}, \mbox{C \,{\sc iv}}, \mbox{O \,{\sc vi}} and \mbox{Si \,{\sc iv}} calculated with the AOD method 
as previously done, so taking into account gas peculiar motions.
Also by looking at the distributions of the gas around the galaxy for each element, we do not see any strong difference
between the two models. 

%
%
%
%

\subsection{Covering Fraction}
\label{ss:coveringfraction}

\begin{figure}
 \centering
 \includegraphics[width=\columnwidth]{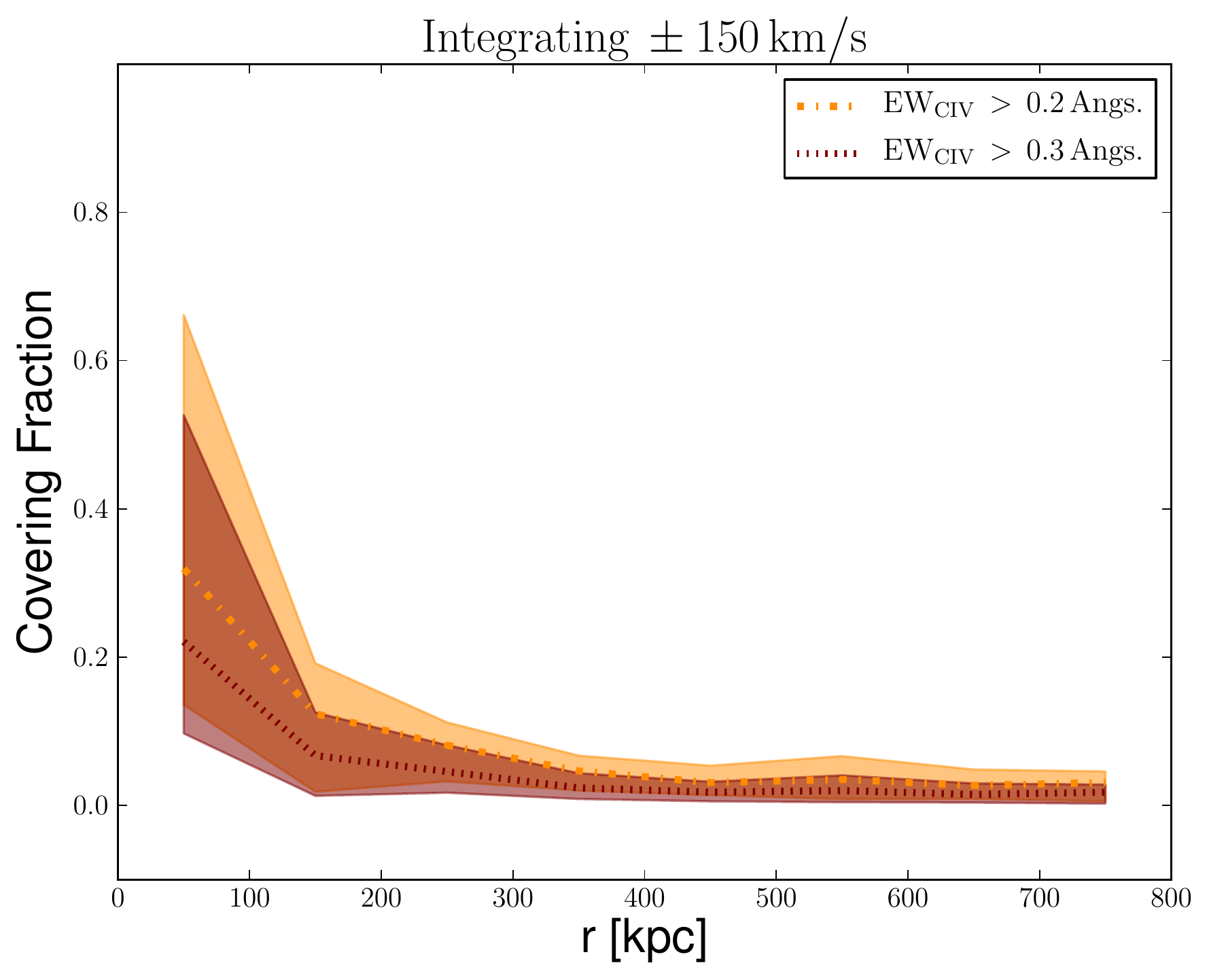}
  \caption{\mbox{C\,{\sc iv}} covering fraction of our sample of lines of sight, in which rest-frame EW have been integrated 
  $\pm$ 150 km s$^{-1}$ from galaxy position. Dotted and dot-dashed lines
  are the median values corresponding to two different EW thresholds (orange: EW$_{\rm {C\, \textsc{iv}}}>$0.2 \AA{}; red: EW$_{\rm {C\, \textsc{iv}}}>$0.3 \AA{}), while shaded areas are the
  1$\sigma$ confidence intervals. (Distances are in proper kpc.)}
  \label{fig:CovFracEWMIO}
\end{figure}

\begin{figure}
 \centering
  \includegraphics[width=\columnwidth]{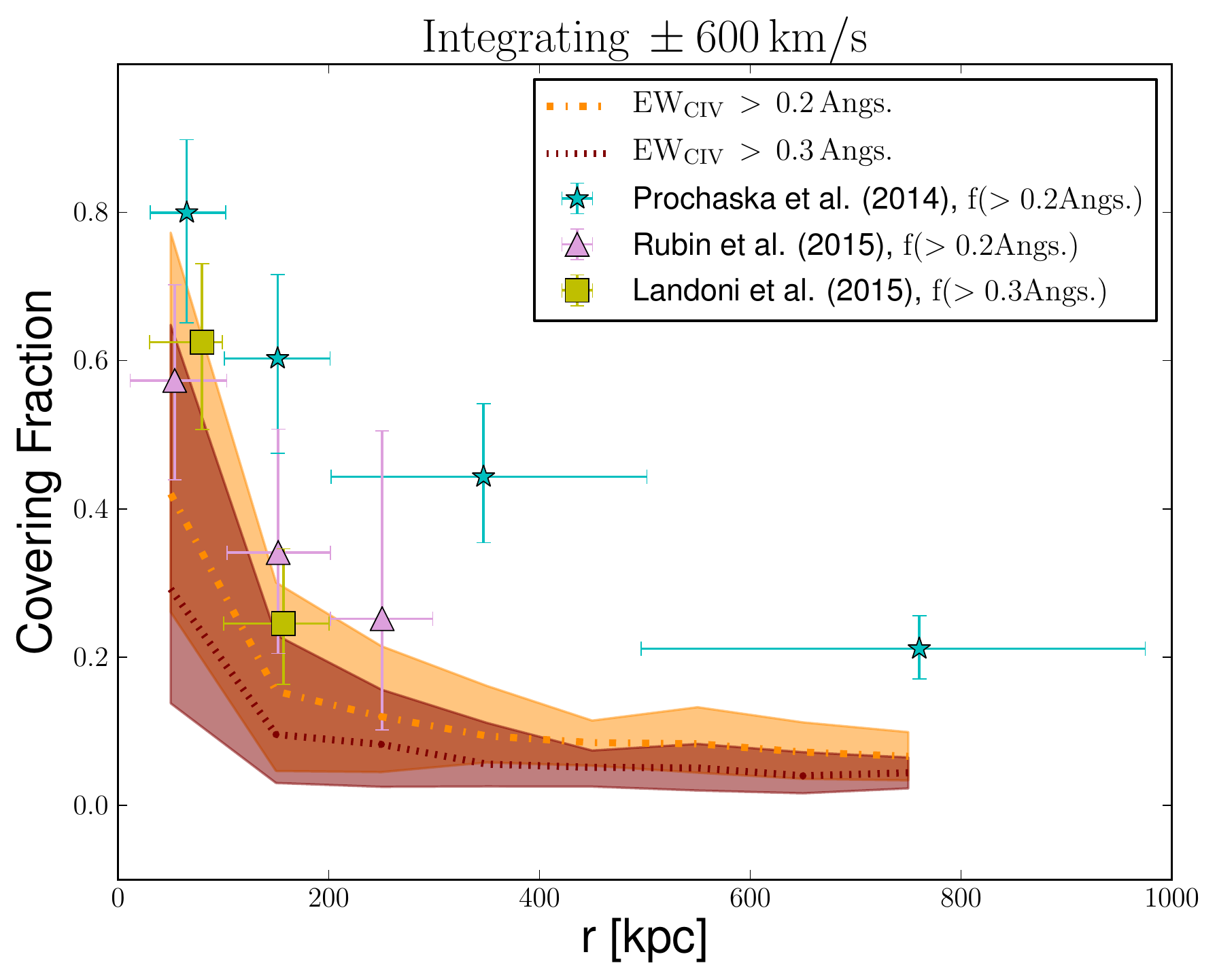}
  \caption{ \mbox{C\,{\sc iv}} covering 
  fraction of our sample of lines of sight, in which rest-frame EW have been integrated $\pm$ 600 km s$^{-1}$ , to be consistent with the observational sample of
  \citet{Rubin2015} and \citet{Landoni2016}. The integration window of \citet{Prochaska2014} is  $\pm$ 1500 km s$^{-1}$,
  a value larger than our box size. This explains the disagreement with the sample of \citet{Prochaska2014}. Dotted and dot-dashed lines
  are the median values corresponding to two different EW thresholds (orange: EW$_{\rm {C\, \textsc{iv}}}>$0.2 \AA{}; and red: EW$_{\rm {C\, \textsc{iv}}}>$0.3 \AA{}), while shaded areas are the
  1$\sigma$ confidence intervals. (Distances are in proper kpc.)}
  \label{fig:CovFracEW}
\end{figure}

The covering fraction of a given ion is by definition the ratio between the number of lines of sight showing an absorption 
system due to that ion with an 
equivalent width (EW) greater than a threshold value and the total number of lines of sight. In practice, it is a measure of the 
clumpiness of the medium (which depends also on the ionization of the medium itself). If the medium is perfectly homogeneous, the covering fraction is equal to 1. The clumpier the 
medium, the lower the covering fraction is. In  Fig.~\ref{fig:CovFracEWMIO}, we report the \mbox{C\,{\sc iv}} covering fractions for our sample 
of MUPPI data, calculated in bins of 100 kpc, in which we computed equivalent widths by always integrating $\pm$ 150 km 
s$^{-1}$ from galaxy position along the LOS. Dashed and dot-dashed lines are the median values of the \mbox{C\,{\sc iv}} covering fractions of our total sample of lines of sight for two 
different threshold values of the EW. Shaded areas are the 1$\sigma$ confidence intervals.
In  Fig.~\ref{fig:CovFracEW}, 
instead, we show the covering fractions obtained by integrating $\pm$ 600 km s$^{-1}$, in order to be consistent with the 
sample of \citet{Landoni2016} and \citet{Rubin2015}. 

Our calculated covering fraction (orange line with EW $>$ 0.3 \AA{}) is in good agreement with the sample of 
\citet{Landoni2016}. The fact that the data by \citet{Landoni2016} have a slightly higher normalization than our prediction 
could be explained
by larger masses of the galaxies they are probing (since they consider QSO host galaxies).
For this reason, it should be more correct to report the relations as a function of the distance
in units of virial radius. Another possible bias is related to the fact that they are observing all face-on galaxies.
 In other directions, their covering fraction could be lower, due to inhomogeneities of the medium.

The sample of \citet{Rubin2015} is actually formed by 40 DLAs, so a little bit different from
our work. The masses of the structures forming the DLAs are not known and this could explain why the sample of \citet{Rubin2015} has slightly higher values.

The sample of \citet{Prochaska2014} has a window of integration of $\pm$ 1500 km s$^{-1}$, 
due to uncertainties in the redshifts of their quasar sample. Unfortunately, our box size is not big enough, as
our maximum range of integration is $\pm$ 800 km s$^{-1}$. We tried to integrate along all the box size and we did recover 
higher values with respect to those shown in Fig.~\ref{fig:CovFracEW}, but still not in agreement with the \citet{Prochaska2014}
sample. The comparison with a bigger cosmological box can be done in a future work.
\begin{figure*}
 \centering
\includegraphics[width=0.9\textwidth]{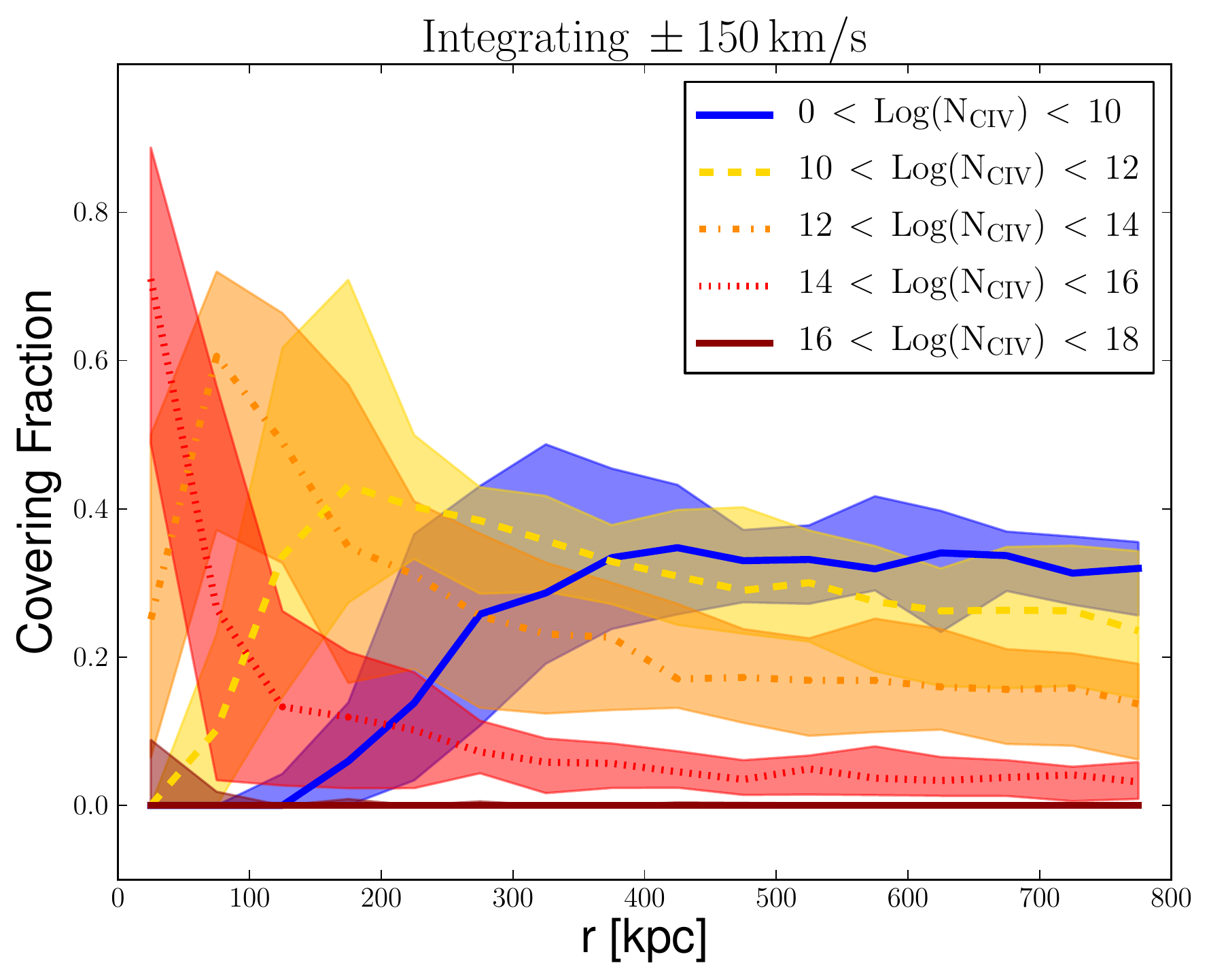}
  \caption{\mbox{C\,{\sc iv}} covering fraction of our sample of lines of sight as a function of the column density. Column densities are integrated $\pm$ 150 km s$^{-1}$ from galaxy position. Dotted and dot-dashed lines
  are the median values, while shaded areas are the 1$\sigma$ confidence intervals. (Distances are in proper kpc.)
}
  \label{fig:CovFracN}
\end{figure*}
Computing the covering fraction as a function of the EW is what it has been done so far in the literature, as
measuring the EW of an absorption system is the simplest thing that can be done, especially with low-resolution
data. 
With the advent of high-resolution spectroscopy, it has become possible to measure column densities of absorption 
systems, using Voigt Profile Fitting codes, which give a more reliable and complete information on the physical state of the gas
that produced the observed absorption.

For comparison with future samples of high-resolution data,
we decided to compute the covering fraction as a function of the column density, as it can be seen in 
Fig.~\ref{fig:CovFracN}.
We divided the covering fractions in different column densities ranges: 0 $\leq\log(\rm N_{\rm {C\, \textsc{iv}}}/{\mathrm {cm}}^{-2})\leq$ 10.0,
10.0 $\leq\log(\rm N_{\rm {C\, \textsc{iv}}}/{\mathrm {cm}}^{-2}) \leq$ 12.0, 12.0 $\leq\log(\rm N_{\rm {C\, \textsc{iv}}}/{\mathrm {cm}}^{-2}) \leq$ 14.0,
14.0 $\leq\log(\rm N_{\rm {C\, \textsc{iv}}}/{\mathrm {cm}}^{-2}) \leq$ 16.0 and  16.0 $\leq\log(\rm N_{\rm {C\, \textsc{iv}}}/{\mathrm {cm}}^{-2}) \leq$ 18.0.
We can see that \mbox{C\,{\sc iv}} absorption systems with $\log(\rm N_{\rm {C\, \textsc{iv}}}/{\mathrm {cm}}^{-2}) >$ 14.0 dominate in the haloes of galaxies 
(d $<$ 100 kpc), even if their covering fraction does not drop to zero at greater distances, due to the presence of 
substructures.
At higher distances, weaker systems start to dominate, with their covering fraction increasing.

\section{Conclusions}
\label{s:conc}

We analysed the output of high-resolution hydrodynamical simulations
with SN feedback implemented both in the thermal and kinetic forms. In
particular, two different subresolution models were considered: the
MUPPI model \citep{Murante2010,Murante2015} and the Effective model
\citep{SpringelHernquist2003}. These two sets share the same large-scale structure evolution, but they are decoupled in the
hydrodynamical part, as they have different SF and
feedback prescriptions.

The main findings of this work can be summarized as follows.
\begin{itemize}

\item We performed a state-of-the-art post-processing analysis and
  produced a set of mock galaxies and mock spectra.  Spectra are
  constructed using an SPH formulation and profiles of different
  physical quantities can be reproduced along the LOS. A
  sample of 40 galaxies with halo mass in the range $M_{\rm h} \sim
  10^{10}-10^{12}$ M$_\odot$, half of which has been selected from the
  MUPPI simulation and the other half from the Effective model
  box. They have to satisfy the following criteria: galaxies do not
  have to be on a major merger, they have to be the main halo of the
  FOF group, in order to analyse the same conditions at the same
  distance and be sampled by a sufficient number of particles. We also
  considered 20 near-filaments environments from the two runs, by
  visually centring our volume as close as possible to a filamentary
  structure.

\item We identified an optimal analysis of mock QSO spectra based on
  the AOD method and compared to the results
  obtained with Voigt Profile Fitting methods, as described in
  Appendix~\ref{s:Ncomparison}.  We performed a comparison between
  column densities derived with the \textsc{vpfit} code and the ones directly
  derived from the simulation. We found the best and faster method to
  use on mock QSO spectra when comparing with observational data is
  the AOD method. With this method, gas peculiar motions are taken
  into
  account and, at the time, a reliable measure of the column density is obtained.

\item We performed a full environmental characterization of the CGM
  and IGM for absorption systems. In particular, we pierced 4000 lines
  of sight around each selected object in the simulation with impact
  parameters less than 800 kpc. Using the \mbox{C\,{\sc iv}} as a
  tracer of the metallicity, we constructed the N$_{\rm{ C\, \textsc{iv}}}$ versus
  N$_{\rm{ H\, \textsc{i}}}$ relation in different radial bins from the object's
  centre, which we compared with observational data by
  K16.  We found that observational data have the highest
  probability to be confined in a region up to 3-5 virial radii from
  galaxies, which correspond (at this redshift) to a physical distance
  of $\sim$ 150 - 400 kpc. Near-filament points are instead confined
  to a region under the detection limit, suggesting that they have
  metallicities too low to be probed by present-day
  observations. These results are validated by the fact that we do not
  recover any strong difference between the
  MUPPI and the Effective models, which display similar trends in the results.

\item We presented a probabilistic approach to the galaxy/IGM
  interplay which carefully quantifies the probability to find an
  absorption system with a given column density at a certain distance
  from a galaxy. For example, if a system with a column density
  N$_{\rm{ C\, \textsc{iv}}}$ of $\sim$10$^{15}$ cm$^{-2}$ and N$_{\rm{ H\, \textsc{i}}}$ of
  $\sim$10$^{16}$ cm$^{-2}$ is found, it has a probability five times
  higher to be located inside the virial radius of a galaxy of mass
  $\sim$10$^{10-11}$ M$_\odot$ than at a distance of 3-5 virial radii
  from such a galaxy. If such a system is observed at this great
  distance from a luminous galaxy, it could be actually related to
  another smaller and less luminous galaxy.

\item We quantified the distribution of ionization species around
  galaxies in terms of covering factors and compared to data taken
  from \citet{Landoni2016}, \citet{Rubin2015} and
  \citet{Prochaska2014}. Considering also the 1$\sigma$ dispersion, our
  data are in good agreement with the observed one, even if they show
  a somewhat lower normalization. This could be due to the fact that
  the observational samples probe galaxies with slightly higher masses
  than those in our simulation.  Covering fractions are computed as a
  function of the EW of the absorption system.
  Furthermore, we constructed covering fractions depending on the
  column density of absorption systems, which can be used to compare
  with future data. Higher column density systems ($\log(N_{\rm{C\, \textsc{iv}}}/{\mathrm{cm}}^{-2})\geq 14$) dominate at impact parameters $\lesssim$ 100 kpc,
  while lower column density systems become prevalent at larger
  distances. The contribution of higher column density systems does
  not drop to zero with increasing separation due to the fact that
  galaxies are not
  isolated systems, but they are surrounded by many other structures.

\item We extended the work also to other two chemical elements,
  \mbox{O\,{\sc vi}} and \mbox{Si \,{\sc iv}}. Simulated \mbox{O\,{\sc vi}}
  absorption systems show a very similar behaviour as \mbox{C\,{\sc iv}}. This
  is an indication that these two elements trace a similar gas phase,
  having both the highest probability to be confined up to a few
  virial radii from galaxies, characterizing a region typical of the
  CGM. The comparison with the observational sample suffers
  from the limitations of the simulation, which is not capable to properly treat 
  ionization. \mbox{Si\,{\sc iv}} traces, instead, more
  internal regions, as it has the highest probability
  to be observed inside the virial radius, as the comparison with data shows.

\item The two simulations which we analysed in this work give similar
  results apart from the comparison in the radial bins $b >$ 1 $r_{\rm {vir}}$, for \mbox{C \,{\sc iv}} and
  \mbox{O \,{\sc vi}}. The Effective Model has a higher probability to
  have absorption systems with smaller values of \mbox{H \,{\sc i}}-\mbox{C \,{\sc iv}}-\mbox{O \,{\sc vi}} column density and this
  could be due to a less efficient feedback with respect to the MUPPI
  model, which is not capable to spread metals with the same strength.
  For \mbox{Si \,{\sc iv}} no difference is observed, but as it is
  confined in a smaller region, the effects of the feedback could be
  less visible.  This is a quite surprising and unexpected result from
  the simulation point of view, as the two models are significantly
  different, both in the SF and feedback prescriptions.
  It suggests that galactic feedback and different SF
  processes are not strongly impacting on the IGM properties
  investigated here.

\end{itemize}

In conclusion, we have shown that observed integrated N$_{\rm{ C\, \textsc{iv}}}$ versus N$_{\rm{ H\, \textsc{i}}}$ relation
likely arises in a region around the galactic halo, more typical of a
CGM environment (meaning the region surrounding the galactic halo up to 
$\sim$ 300-400 kpc) than of the IGM . This could be a hint that physical
processes that are responsible for the metal enrichment did not affect
significantly regions at very low density.

The fact that we obtain similar results in two different numerical models used to interpret the data sets suggests
that these results are qualitatively correct.

Similar studies on the importance of absorption lines systems
in the proximity of galaxies as derived from hydrodynamical simulations
have been presented in several works [e.g. \cite{Hummels2013,Suresh2015,Bordoloi2014,Turner2016}]. However,
it is quite difficult to compare with these findings given the different problems addressed and also the physical implementations
of feedback models and the setup of the cosmological simulations (in terms of considered redshift, box sizes, objects studied
and numerical resolution). Our work relies on a subresolution model (the MUPPI one), which succeeded in reproducing realistic disc galaxies
with new implementations of the SF and feedback subresolution recipes.

In the future, it would be interesting to better investigate the
comparison between the different subresolution models, in order to
understand which physical processes affect more deeply the IGM
properties.

From the observational point of view the comprehension of the
enrichment mechanism could be enlarged thanks to the new facilities
that will come online in the next decade.  The very weak metal
absorptions associated with the IGM will be revealed, if present, by
the next generation of spectrographs like ESPRESSO at the VLT, which
will see the first light in 2017, and even more with the
high-resolution spectrograph for the 39 m European ELT telescope. On
the other hand, the relation between the level of enrichment of the
IGM studied with metal absorbers and the distribution of galaxies in
the same field will be better quantified when also the very faint
objects will be identified, in particular with the JWST (James Webb Space Telescope).

\section*{Acknowledgments}
MV is supported by INFN/PD51 Indark, PRIN-INAF, PRIN-MIUR, and ERC-StG (European Research Council-Starting Grant)
``cosmoIGM'' grants under ERC Grant agreement 257670-cosmoIGM. TSK is
supported by ERC-StG ``cosmoIGM''. TSK also acknowledges the support from "NSF-AST118913". CM wishes to thank D. Goz, 
E. Munari, and A. Zoldan for useful discussion.
We are grateful to an anonymous referee who provided very useful comments that allowed us to clarify several points in the paper.





\bsp	
\label{lastpage}

 \bibliographystyle{mnras}

\bibliography{biblio}

\appendix

\section{Constructing lines of sight}
\label{ss:LOS}

Simulated data are obtained by piercing through the cosmological box with lines of sight, either randomly or forcing them to 
have impact parameters smaller than a certain value 
around a given galaxy or position in the simulation. 

Lines of sight are as long as the box side and are always parallel to it along one of the three directions $x-y-z$.

Each LOS is divided in 2048 pixels of $\Delta L\sim 10$ proper kpc each.
Along each LOS we can compute different quantities, such as the density profile of the total gas or of a particular
chemical element, the temperature profile, the gas peculiar velocity or the optical depth profile of a given chemical element.

Profiles are computed following the prescriptions by \citet{Theuns1998}: the total gas density $\rho_i$ of each pixel $i$ is 
the sum of the 
densities of all gas particles $\rho_{\rm part}$ (multiphase or single phase) convolved with their SPH kernel functions $W$, that means that all the gas particles
$j$ contribute to the pixel's density; these particles have a smoothing length $h$ which intersect the LOS. The density of a gas particle is defined as $\rho_{\rm part}=M_p/h^3$. The same 
procedure is applied to compute the other physical 
quantities, such as the gas peculiar velocity or the temperature.
\begin{equation}
\begin{split}
 & \rho_i = \sum\limits_{j}  W_{ij} \\
 & v_i = \sum\limits_{j} v_{part,j} W_{ij} \\
 & T_i = \sum\limits_{j} T_{part,j} W_{ij}
 \end{split}
 \label{eq:pixelcomput}
\end{equation}
Here, $W_{ij}$ is:
\begin{equation}
\begin{split}
  W_{ij} = \rho_{part} \cdot W(q_{ij}) & = \rho_{part} \cdot \biggl[ \frac{1}{4\upi}(4-6q_{ij}^2+3q_{ij}^3) \biggr] \quad , {\mathrm{if}} \quad q_{ij}\leq1 \\
                    & = \rho_{part} \cdot \biggl[ \frac{1}{4\upi}(2-q_{ij})^3 \biggr] \quad , {\mathrm{if}}\quad  1 \leq q_{ij} \leq 2 \\
                    & = 0 \quad {\mathrm{otherwise}}\\     
\end{split}
\label{eq:kernel}
\end{equation}
with
\begin{equation}
 q_{ij}=\frac{\textbar x(i)-x(j)\textbar}{h}
 \end{equation}
$x(i)$ is the pixel's position and $x(j)$ is the particle's position.

We can also compute density profiles of a given chemical species in a given ionization state. 

The metallicity of a specific chemical element in a single gas particle is defined as the ratio between the mass of that element in the gas particle 
and 
the total mass of the particle, such as: 
\begin{equation}
 Z_X=m_X/M_p
\end{equation}
with X=[H, He, C, Ca, O, N, Ne, Mg, S, Si, Fe].
In order to have the fraction of that element in a particular ionization state $k$ for a specific gas particle, we used the 
formula:
\begin{equation}
 f_{X,k}=f_{X,k,\odot}\cdot Z_X
\end{equation}
$f_{X,k,\odot}$ is the fraction of an element X in the ionization state $k$ per solar metallicity. Given the density, the temperature 
and the redshift of the particle, $f_{X,k,\odot}$ is computed using
the \textsc{cloudy} code\footnote{http://www.nublado.org/} \citep{Ferland2013} with the UVB by Haardt $\&$ Madau (2005)\footnote{The UV backgrounds could be 
obtained from http://www.ucolick.org/$\sim$pmadau/CUBA/HOME.html}.
For the \mbox{H\,{\sc i}}, $f_{\rm {H\, \textsc{i}}}$ is an output of the simulation and, as
described in Section \ref{s:sim}, it is computed considering a
UVB by \citet{HaardtMadau2001}, which is not significantly different from Haardt $\&$ Madau (2005) at $z\sim2$. 
Defining $\rho_{X,k}=f_{X,k} \cdot \rho_{\rm part}$ and using this quantity in place of $\rho_{\rm part}$ in equation~(\ref{eq:kernel}), we can 
compute all the profiles of equation~(\ref{eq:pixelcomput}) but for specific ions.

Optical depths along lines of sight were computed by considering their density, temperature, and velocity in each pixel.
We consider 
a pixel $i$ and first calculate the central optical depth of the
line, which falls in that pixel, using its density and temperature with this formula:
\begin{equation}
 \tau_{0,i}=\frac{\sigma_\alpha c}{\sqrt{\upi}}\frac{\rho_i \Delta L}{b_i}
\end{equation}
with $b_\mathrm{i}=\sqrt{\frac{2K_\mathrm{B} T_\mathrm{i}}{m_{\mathrm{atom}}}}$ and $\sigma_\alpha=(3\upi\sigma_T/8)^{1/2}f\lambda_0$, where
$\sigma_T=6.625\times10^{25}$ 
cm$^{-2}$ is the Thomson cross-section, $f$ is the transition oscillator
strength and $\lambda_0$ is the rest-frame transition wavelength.

According to the Voigt profile line shape, neighbouring pixels $q$ will suffer absorption from pixel $i$ by an amount 
$e^{-\tau_q}$, where:

\begin{equation}
 \tau_q=\tau_{0,i} \cdot exp\biggl\{ - \biggl[\frac{v_q-v_i}{b_i}\biggr]^2\biggr\}
 \label{eq:taucomputation}
\end{equation}
The number of pixels $q$ that will suffer absorption from pixel $i$ is set to be equal to $3 b_i \Delta v_L$, where $\Delta v_L$ is the 
length of a pixel in velocity, so depending 
on temperature of pixel $i$.

Shifting from pixel to pixel, we can repeat the same operation along the LOS, summing each new contribution to the optical 
depth in a pixel to what was previously computed
a number $l$ of times, depending on the temperature of neighbouring   pixels.
In this way, the optical depth in one pixel is the sum of its own optical depth plus the contribution  from  $l$ pixels, where $l$ depends
on their temperature and density.

We can also construct $\tau$-weighted density or temperature profiles. Focusing only on density, we define:

\begin{equation}
 \rho_{\tau,q}=\frac{\sum\limits_l\rho_i(l) \cdot \tau_i(l)}{\sum\limits_l \tau_i(l)}
\end{equation}

where $\rho_{\tau,q}$ is the $\tau$-weighted density in a given pixel $q$, $\rho_i(l)$ is the density of an adjacent  pixel $i$,
considering only its contribution and not by its own adjacent  pixels,  and 
$\tau_i(l)$ the optical depth contribution to pixel
$q$ by pixel $i$ given by the Eq. \ref{eq:taucomputation}, that is the contribution to pixel $q$ is given by the wings
of the Gaussian profile whose central contribution is given by pixel $i$. The sum is over a number $l$ of pixels, as defined as before.

\section{Column densities of simulated absorption systems}

\subsection{Comparison among different methods to derive the column density}
\label{s:Ncomparison}
In the regime of high-resolution spectroscopy, observed parameters of absorption systems, such as the $b$-parameter or the column 
density, can be derived using Voigt Profile Fitting codes, such as \textsc{vpfit}\footnote{http://www.ast.cam.ac.uk/~rfc/vpfit.html}
\citep{Carswell2014}.
Observed absorption lines contain the information of gas peculiar motions and thermal broadening, so, in order to 
compare with them, it is important to take into account redshift space distortions also in the simulation. 
Using these fitting codes when dealing with simulated spectra is a delicate procedure and difficult to implement in an automatic way for thousands of spectra.
For this reason, we searched for alternative methods to compute column densities of simulated absorption systems, suitable for comparison with our sample of observational data.

Since we cannot just integrate the real gas density profile along the simulated lines of sight, because in this way we neglect gas peculiar motions, we considered two 
different methods that we compared with the \textsc{vpfit} output: 
\begin{itemize}
 \item the integration along the LOS of the optical depth weighted density profile;
 \item the AOD method.
\end{itemize}
%
%
We first computed our sample of simulated \textsc{vpfit} outputs to compare with the two methods. We therefore constructed simulated \mbox{C\,{\sc iv}}
spectra by piercing through the cosmological box 1000 random lines of sight, which we post-processed by adding instrumental noise (to reach S/N=100) and rebinning to 
the spectral resolution of 
the observational sample of K16 ($\Delta v=6.7$ km s$^{-1}$). We also rescaled all the fluxes by a factor $\approx$0.44 in order to match the effective optical depth 
by \citet{Kim2007}. We chose to perform the comparison with the \mbox{C\,{\sc iv}}, as \mbox{H\,{\sc i}} absorption lines 
are more easily saturated and \textsc{vpfit} is no longer reliable in estimating column densities without higher order transitions than Ly$\alpha$, which are not present in 
our simulated lines of sight. 
%
%

The produced spectra were then fitted by automatized \textsc{vpfit}. We discarded all the spectra having fitted single components with column densities
smaller than $\log(\rm N/\mathrm{cm}^{-2})=12.06$  and errors on $\rm b$ and $\rm N$ greater than 50\%. The first condition is due to the fact that many spectra show
no \mbox{C\,{\sc iv}} absorption, but \textsc{vpfit} attempts to fit the spectra anyway by putting a small component. The threshold of 12.06 is our chosen value
to discard all false detections. 369 spectra satisfy these conditions.

We then constructed absorption systems trying to be as consistent as possible with the procedure by K16.
With an algorithm similar to the ``FOF'', we identify 
an absorption system as a group of single \textsc{vpfit} components that lie within $\pm$ 150 km s$^{-1}$ of each other along the LOS.
The condition ``any friend of a friend is a friend'' is valid, that is, starting from one component, we identified all the components lying within 
$\pm$ 150 km s$^{-1}$ of this first one. Then we moved to the second closest component
and we repeated the operation. If new friends (new with respect to the first considered component) are identified, they are 
added to form a unique system with the first one, and so forth with all other components satisfying this algorithm.
This scheme was chosen in order to translate in an automatic way the procedure adopted by K16. 

The column density of the group is the sum of the column densities of all the single components that form the system.
We then determined a mean redshift of the group by weighting the redshifts of single components with their column density. 
Then, for every \mbox{C\,{\sc iv}} system, we considered its mean redshift and we defined the velocity range of 
integration by 
extending $\pm$ 150 km s$^{-1}$ from this mean redshift. 

At this point, we performed a comparison between column densities of simulated absorption systems derived with the \textsc{vpfit} code and the ones 
derived with the other two methods. 

\begin{figure*}
  \centering
  \makebox[\textwidth][c]{
  \subfloat{\includegraphics[width=0.5\textwidth]{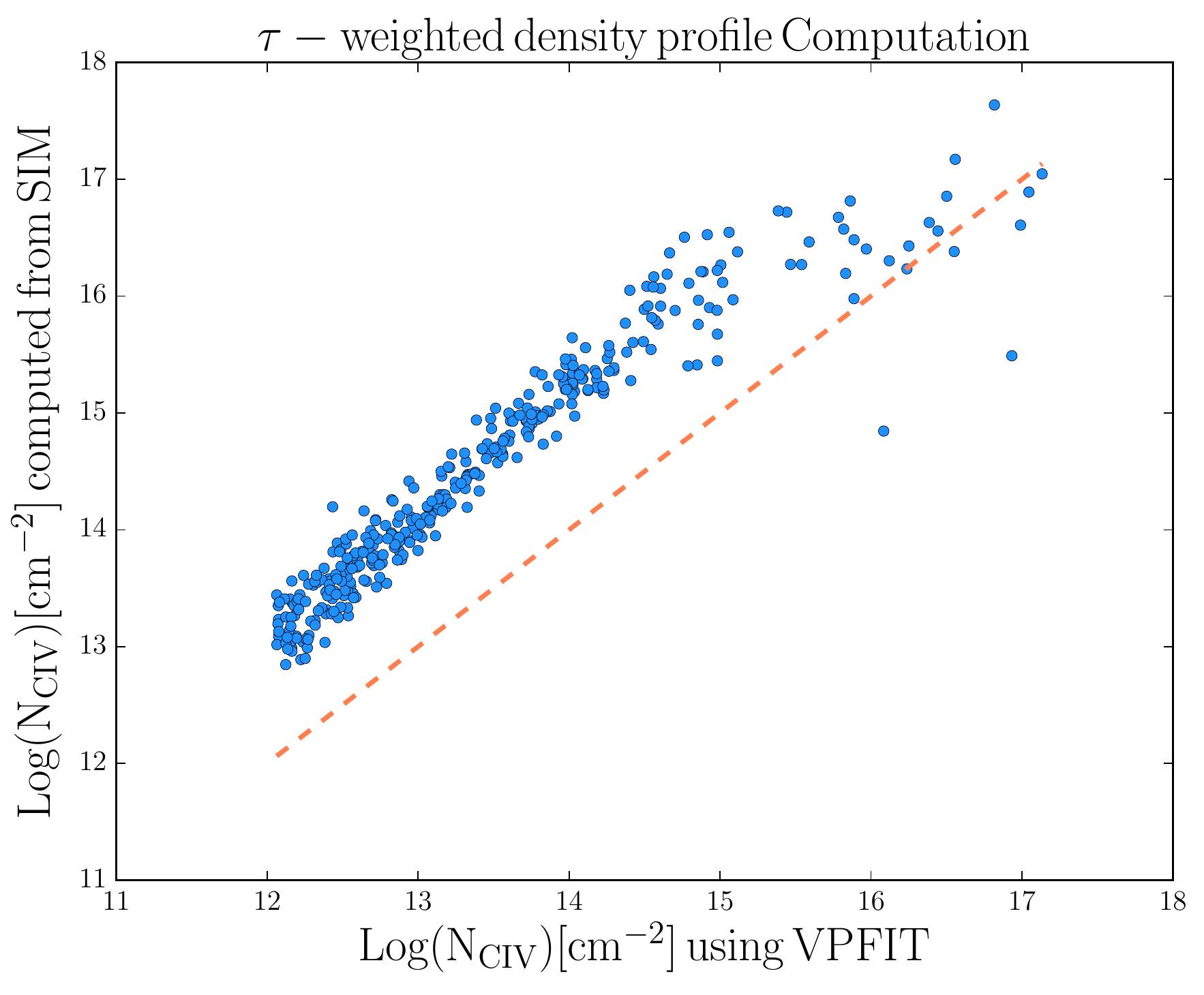}} 
  \hspace{0.5cm}
  \subfloat{\includegraphics[width=0.5\textwidth]{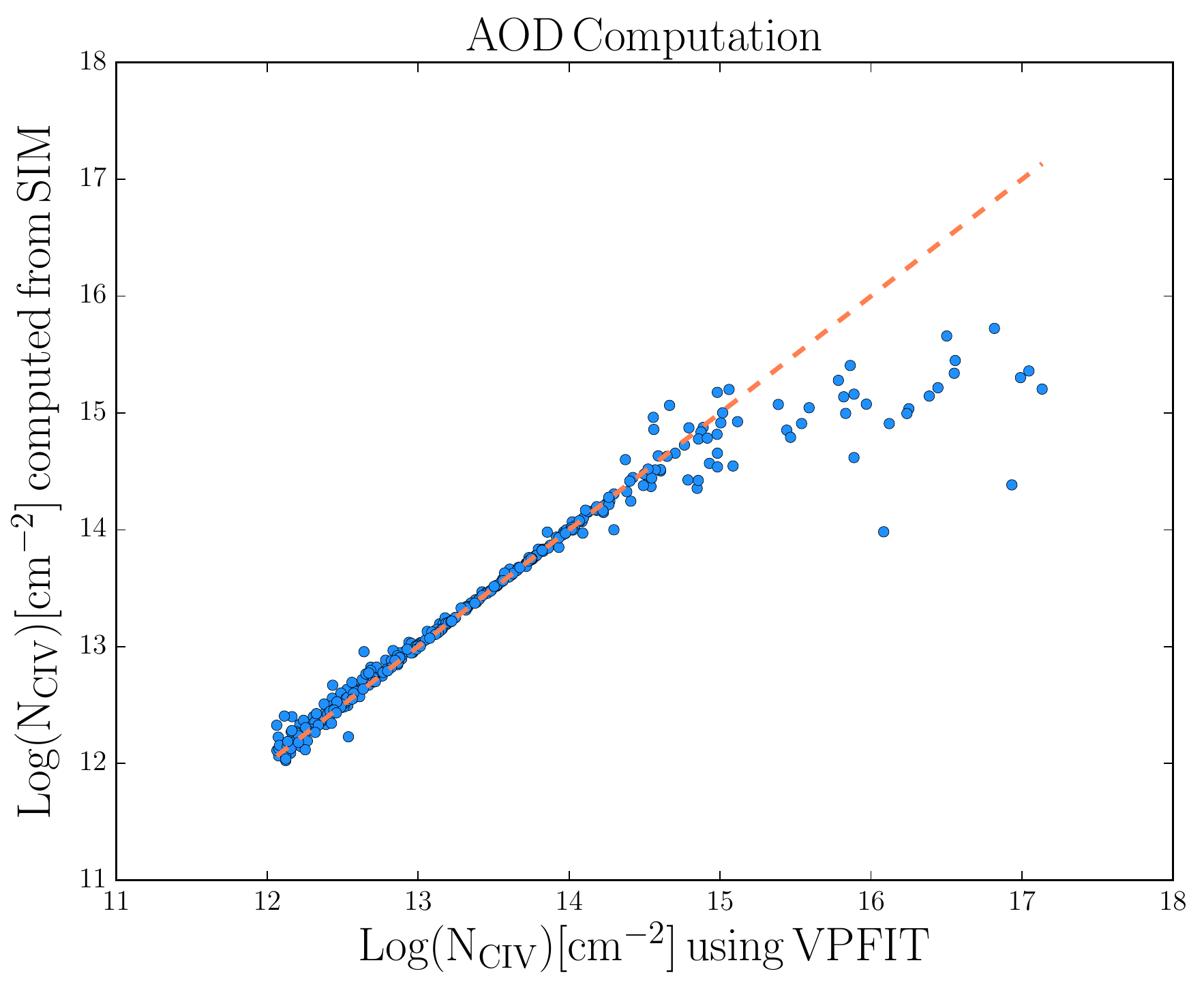}}
  
  }

  \caption{$Left:$ comparison between column densities of simulated absorption systems derived with the \textsc{vpfit} code and those derived by integrating the 
  $\tau$-weighted 
  density profile of the same spectra and in the velocity range $\pm$ 150 km s$^{-1}$ from \textsc{vpfit} mean group redshift. The orange line is the 
  1:1 ratio. A disagreement of one order of 
  magnitude is observed. $Right:$ comparison between column densities of simulated absorption systems derived with the \textsc{vpfit} code and those of the same spectra 
  derived using the AOD method.
  The two values are the same. In both panels the disagreement at high column densities ($\log(\rm N_{\rm{C\, \textsc{iv}}}/\mathrm{cm}^{-2}) \gtrsim15$) is due to
  the fact that when lines become saturated, \textsc{vpfit} is no more reliable in calculating column densities without higher order transitions.}
  \label{fig:VPFITvstau}
\end{figure*}

\citet{Schaye1999} showed that absorption line minima correspond to peaks in the $\tau$-weighted density profile and the $\tau$-weighted density measured in the line 
centre is proportional to the column density measured by Voigt Profile Fitting. However, this does not imply that integrating this quantity along the LOS 
it is possible to recover the correct column density of the absorption features, as we demonstrate in Fig.~\ref{fig:VPFITvstau} (left-hand panel).

In this comparison, we took the 369 simulated lines of sight in our sample, not post-processed for \textsc{vpfit}, and we computed the column density by integrating in the previously
defined velocity range the $\tau$-weighted density profile. These values are shown in the $y$-axis of Fig.~\ref{fig:VPFITvstau} (left-hand panel), while 
the \textsc{vpfit} total column density of simulated absorption systems, as defined above, is shown on the $x$-axis.
The orange line is the 1:1 ratio.
The plot shows clearly that there is a 
disagreement of one order of magnitude between the two determinations. This is due to the fact that the process of $\tau$-weighting does not 
preserve the integral of the density function along the LOS. 

We then tested the AOD method.
This method is often used dealing with observational data, because it provides a quick and reliable measure of the column density of 
unsaturated lines in a regime of high-resolution spectroscopy. According to this method, an apparent optical depth can be converted into
an apparent column density using the formula:
\begin{equation}
 \log[N_a(v)]= \log\, \tau_a(v)-log(f\lambda)+14.576
 \label{eq:AOD}
\end{equation}
Here, $\tau_a(v)$ is the apparent optical depth of a velocity pixel, that is the true optical depth $\tau_{true}(v)$ convolved with the
instrumental response, while $f$ is the transition oscillator strength and $\lambda$ the transition wavelength. 
In a regime of high-resolution spectroscopy, $\tau_a(v)\approx \tau_{true}(v)$, so for this reason, we used the simulated true optical depth
profile along the 
LOS not convolved with the instrumental response.

We performed the same comparison with the simulated outputs of the \textsc{vpfit} code: for every spectrum, we integrated column densities derived with Eq.~\ref{eq:AOD}
in the velocity range defined by the mean redshift of the \textsc{vpfit} system of the same spectra, as explained before. 
In this case, before the computation, we rescaled all the optical depth values
in each pixel by the same factor $\approx$ 0.44, in order to be consistent with \textsc{vpfit}.
The comparison is shown in the right-hand panel of Fig.~\ref{fig:VPFITvstau}. The AOD method predicts the same column densities as 
\textsc{vpfit}.

From an observational point of view, the AOD method does not give reliable column densities for saturated lines, because, in this case it is 
 impossible to recover the real optical depth that produced a flux equal to zero. On the other hand, in the simulations we know the value of the true optical depth 
 in each pixel, thus this method always gives the correct result. The disagreement at high column densities ( $\log(\rm N_{\rm{C\, \textsc{iv}}}/\mathrm{cm}^{-2})\gtrsim 15$) 
 shown in Fig.~\ref{fig:VPFITvstau}, is due to the difficulty of \textsc{vpfit} to derive reliable column densities for saturated lines as already described above. 
 
On the base of the carried out tests, in our study we have adopted the AOD method for the computation of column densities in the simulated spectra. 

\subsection{Comparison of simulated and observed CDDF}
\label{ss:CDDF}
We report here the results of our simulated \mbox{H\,{\sc i}} and \mbox{C\,{\sc iv}} CDDFs for both the two models, compared 
with the observational ones by \citet{Kim2013}. 

Simulated spectra were treated as described in Appendix~\ref{s:Ncomparison} with the same post-processing and absorption lines were fitted with the \textsc{vpfit} code.
The cut at 10$^{12}$ cm$^{-2}$ is artificially introduced as explained in Appendix~\ref{s:Ncomparison}.
In particular, the MUPPI model reproduced quite well both the \mbox{C\sc{iv}} and \mbox{H\sc{i}} CDDFs (up to 10$^{14}$ cm$^{-2}$). 
The Effective model shows slight worse agreement, in particular for \mbox{H \sc{i}} lines at column densities larger than 10$^{15}$ cm$^{-2}$.
In general, underproduction of \mbox{H \sc{i}} could be ascribed to several reasons, among which the large uncertainties in the column densities determinations with
the \textsc{vpfit} code without higher order transitions.
The overproduction of simulated \mbox{C \sc {iv}} can be instead ascribed to the fact discussed in Section~\ref{ss:Nrelation1} about metal production in the simulation.

%
%
%

\begin{figure*}
 \centering
  \makebox[\textwidth][c]{
  \subfloat{\includegraphics[width=0.45\textwidth]{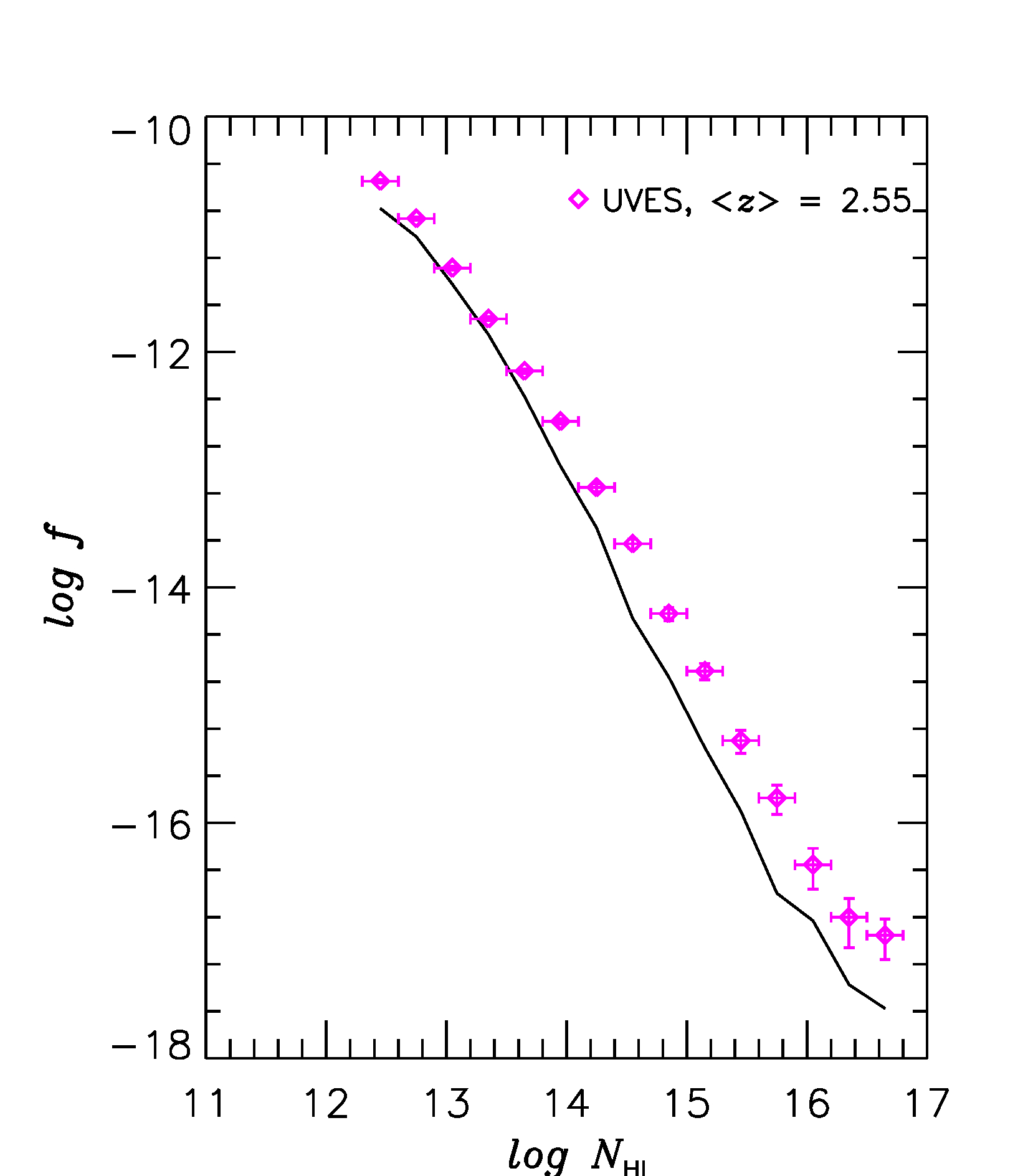}} 
  \hspace{0.5cm}
  \subfloat{\includegraphics[width=0.45\textwidth]{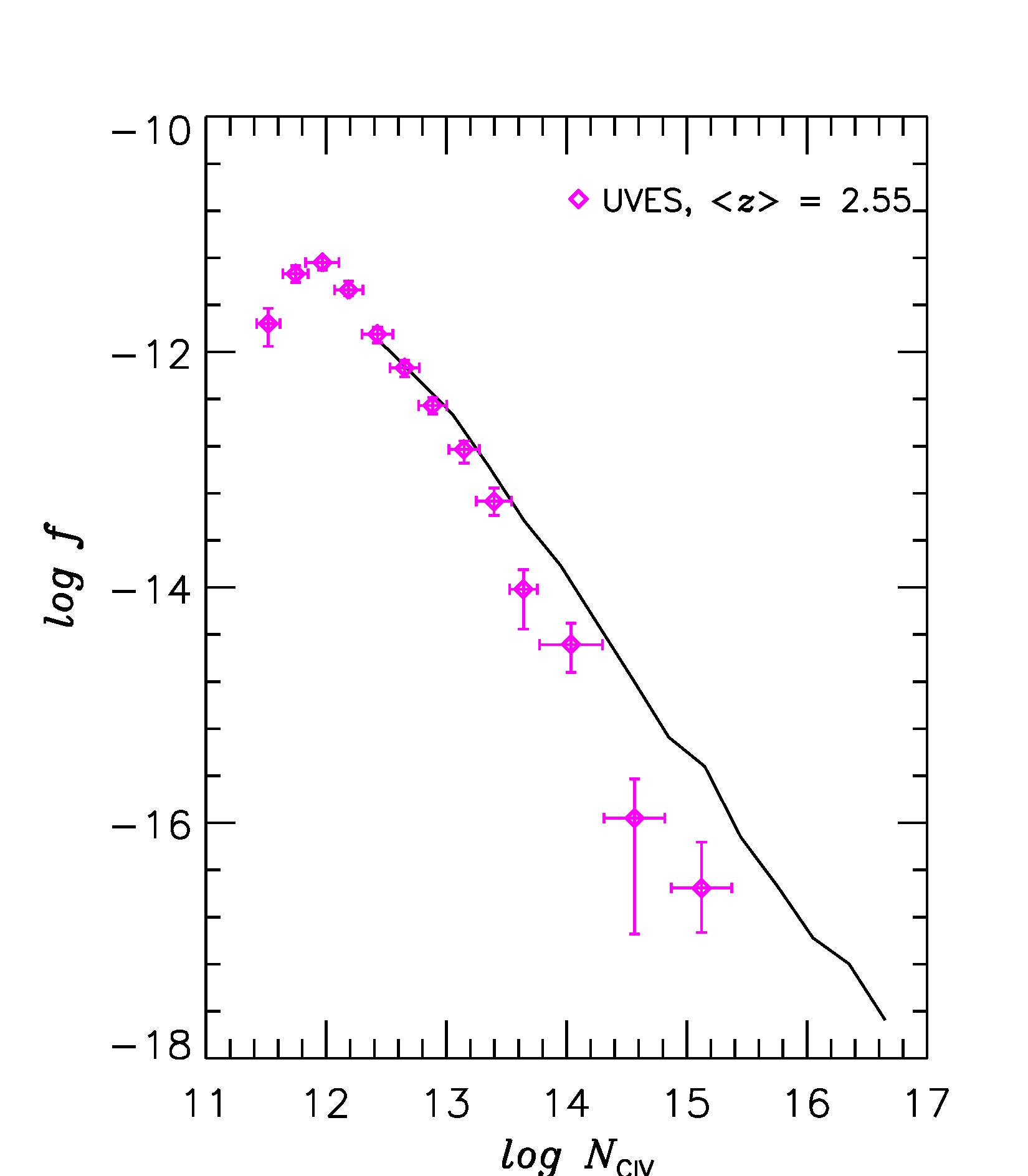}}
  
  }
   \caption{{\it Left-hand panel}: \mbox{H \, \sc{i}} CDDF of the absorption systems in the lines of sight of our MUPPI simulation.
   {\it Right-hand panel}: \mbox{C\,{\sc iv}} CDDF of absorption systems in the lines of sight of our MUPPI simulation. Observed \mbox{H\,{\sc i}} and \mbox{C\,{\sc iv}}
   CDDFs (magenta points with errorbars) are both taken from \citet{Kim2013}.}.
  \label{fig:MUPPI_cddf}
\end{figure*}
\begin{figure*}
 \centering
  \makebox[\textwidth][c]{
  \subfloat{\includegraphics[width=0.45\textwidth]{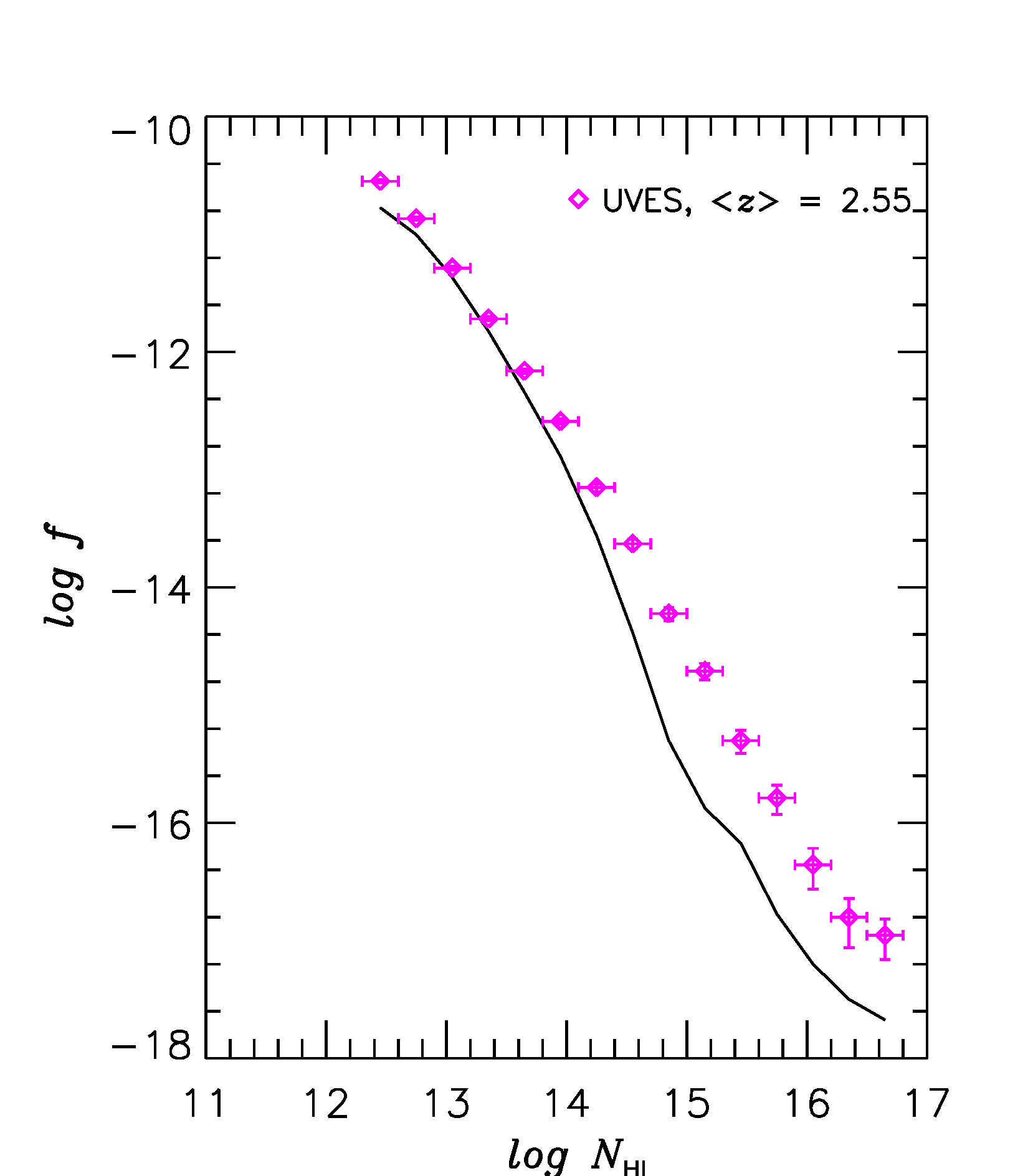}} 
  \hspace{0.5cm}
  \subfloat{\includegraphics[width=0.45\textwidth]{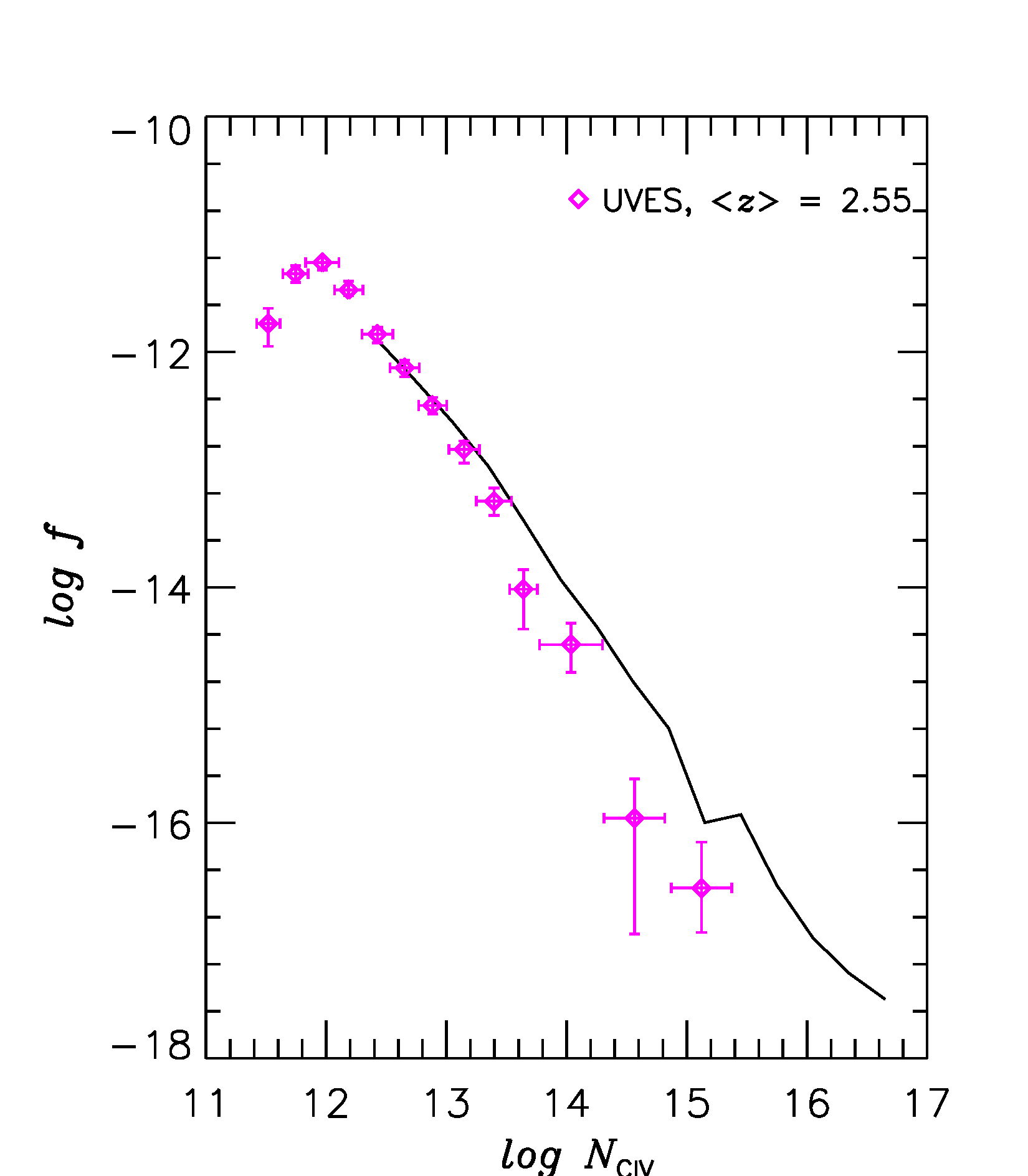}}
  
  }
   \caption{Same as Fig.~\ref{fig:MUPPI_cddf}, but for the Effective model.}
  \label{fig:EFF_cddf}
\end{figure*}
\section{Looking at O VI in the CGM}
\label{s:figures}

 We report here the comparison between simulated and observed \mbox{O \,{\sc vi}} absorption systems. The observational sample is taken from \citet{Muzahid2012} and
they use a slightly smaller window of integration (v$_{link}$=100 km s$^{-1}$) than this work for the construction of absorption systems. 

\begin{figure*}
 \centering
  \makebox[\textwidth][c]{
  \subfloat{\includegraphics[width=0.31\textwidth]{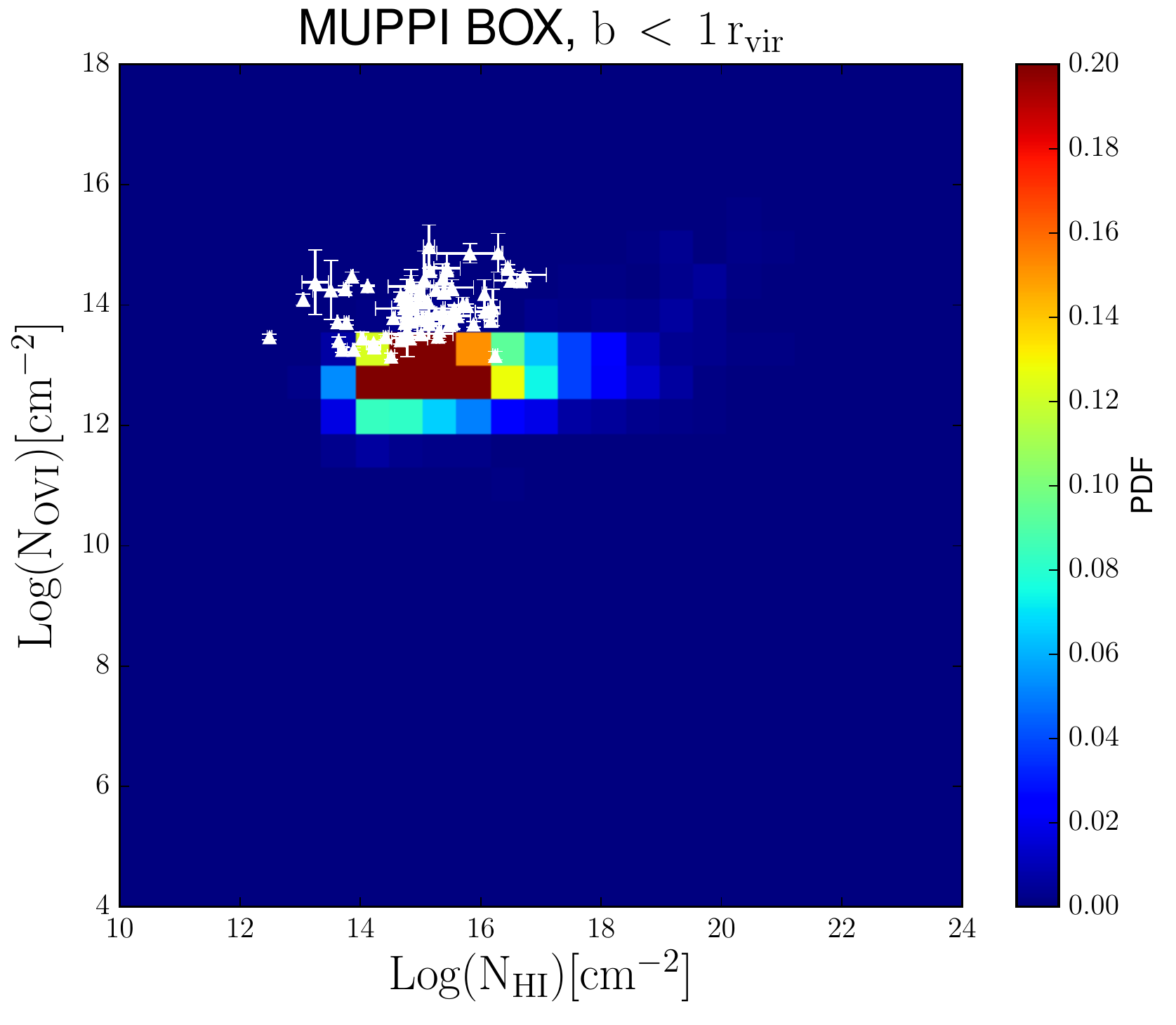}} 
  \hspace{0.5cm}
  \subfloat{\includegraphics[width=0.31\textwidth]{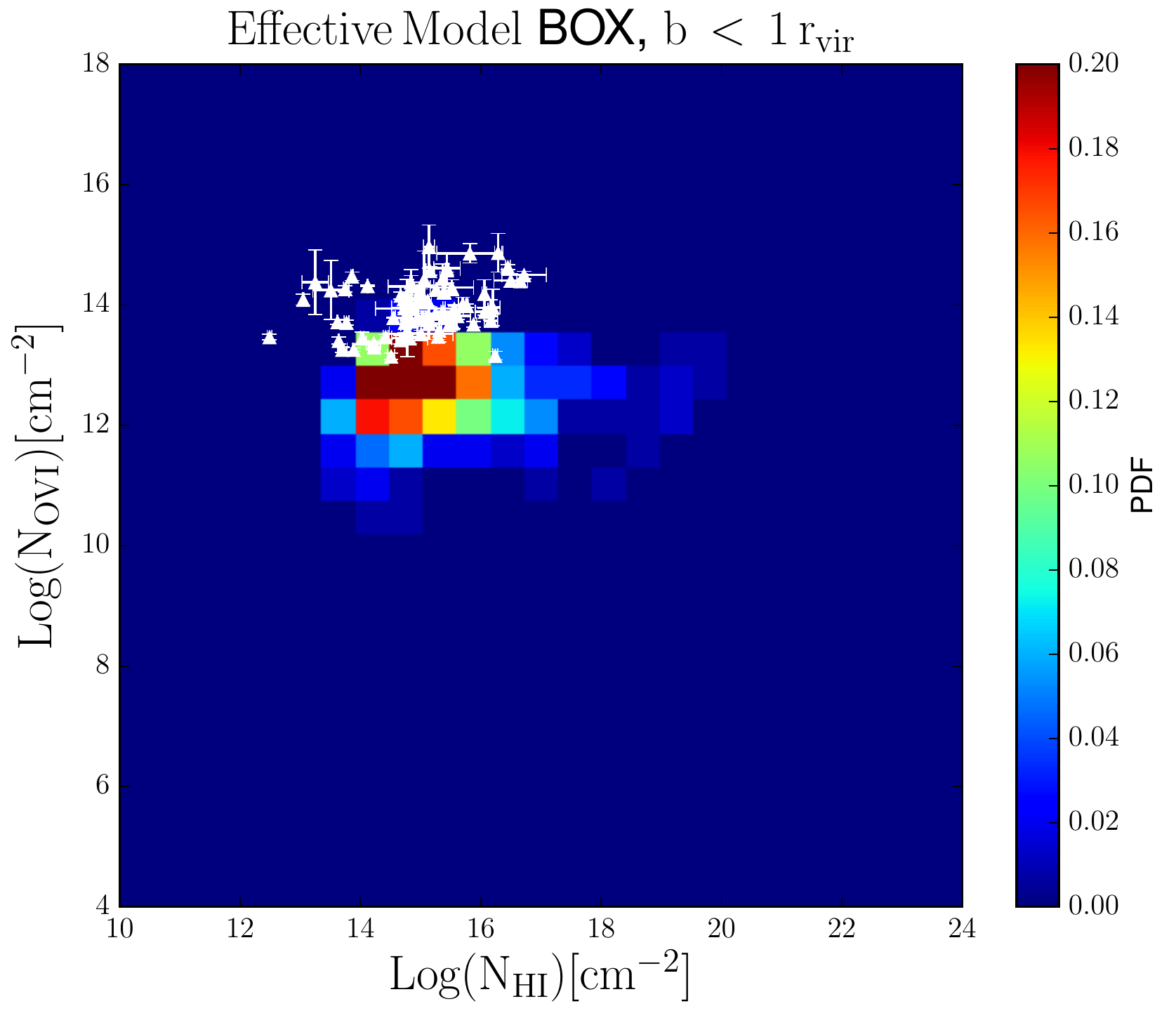}}
  
  }
   \vspace{0.1cm}
   \makebox[\textwidth][c]{
  \subfloat{\includegraphics[width=0.31\textwidth]{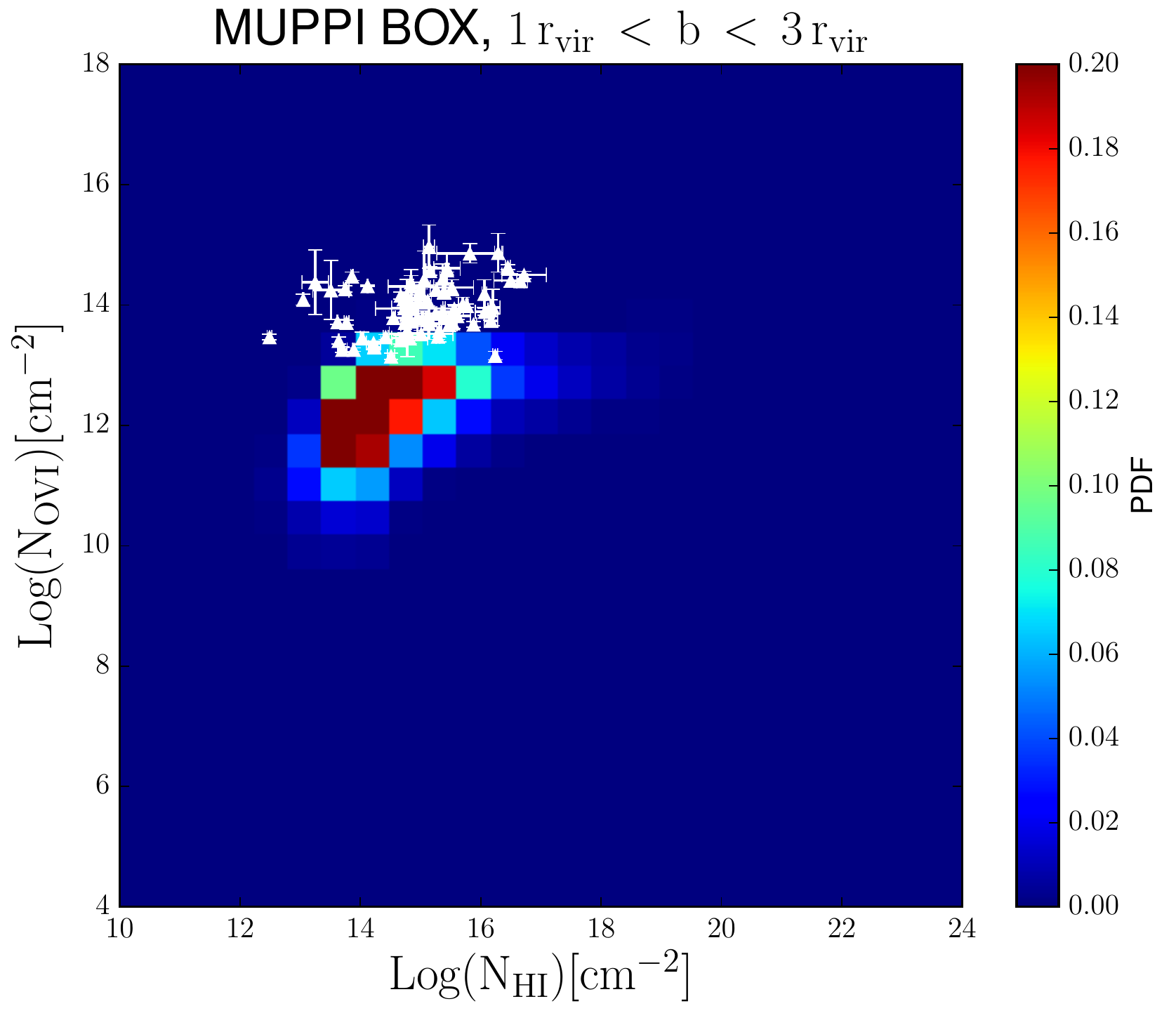}} 
  \hspace{0.5cm}
  \subfloat{\includegraphics[width=0.31\textwidth]{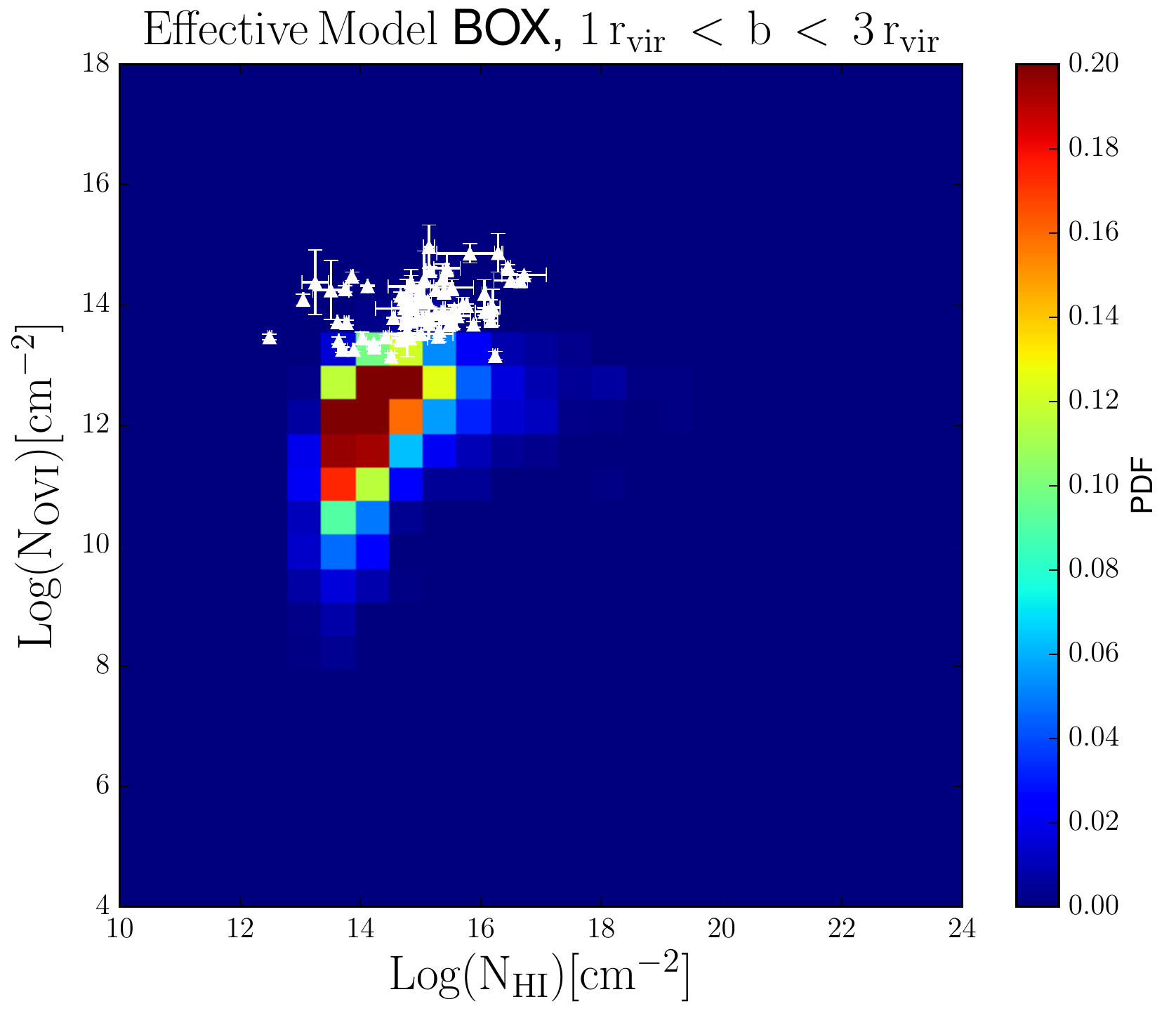}}
 
  }
      \vspace{0.1cm}
    \makebox[\textwidth][c]{
   \subfloat{\includegraphics[width=0.31\textwidth]{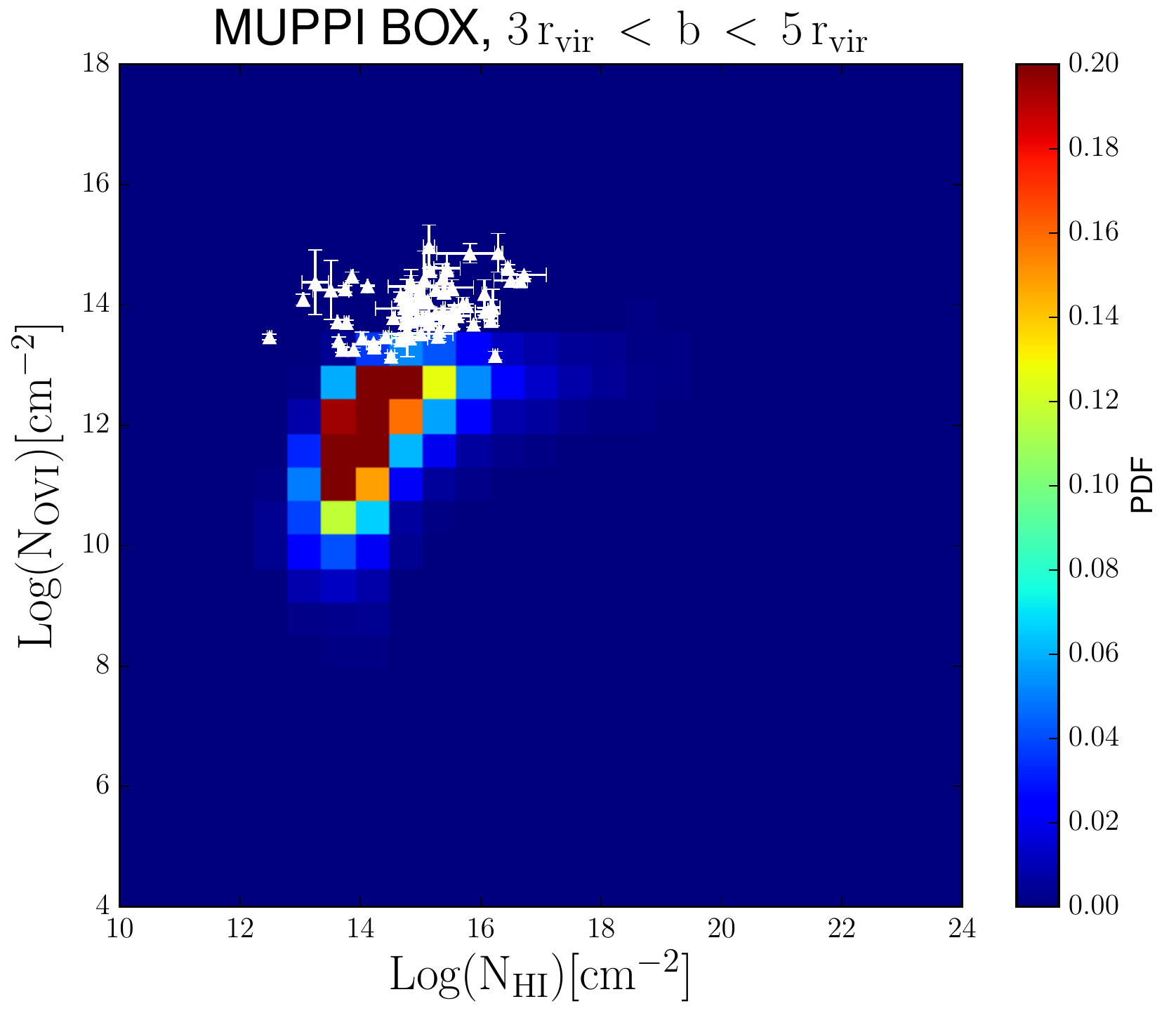}} 
  \hspace{0.5cm}
   \subfloat{\includegraphics[width=0.31\textwidth]{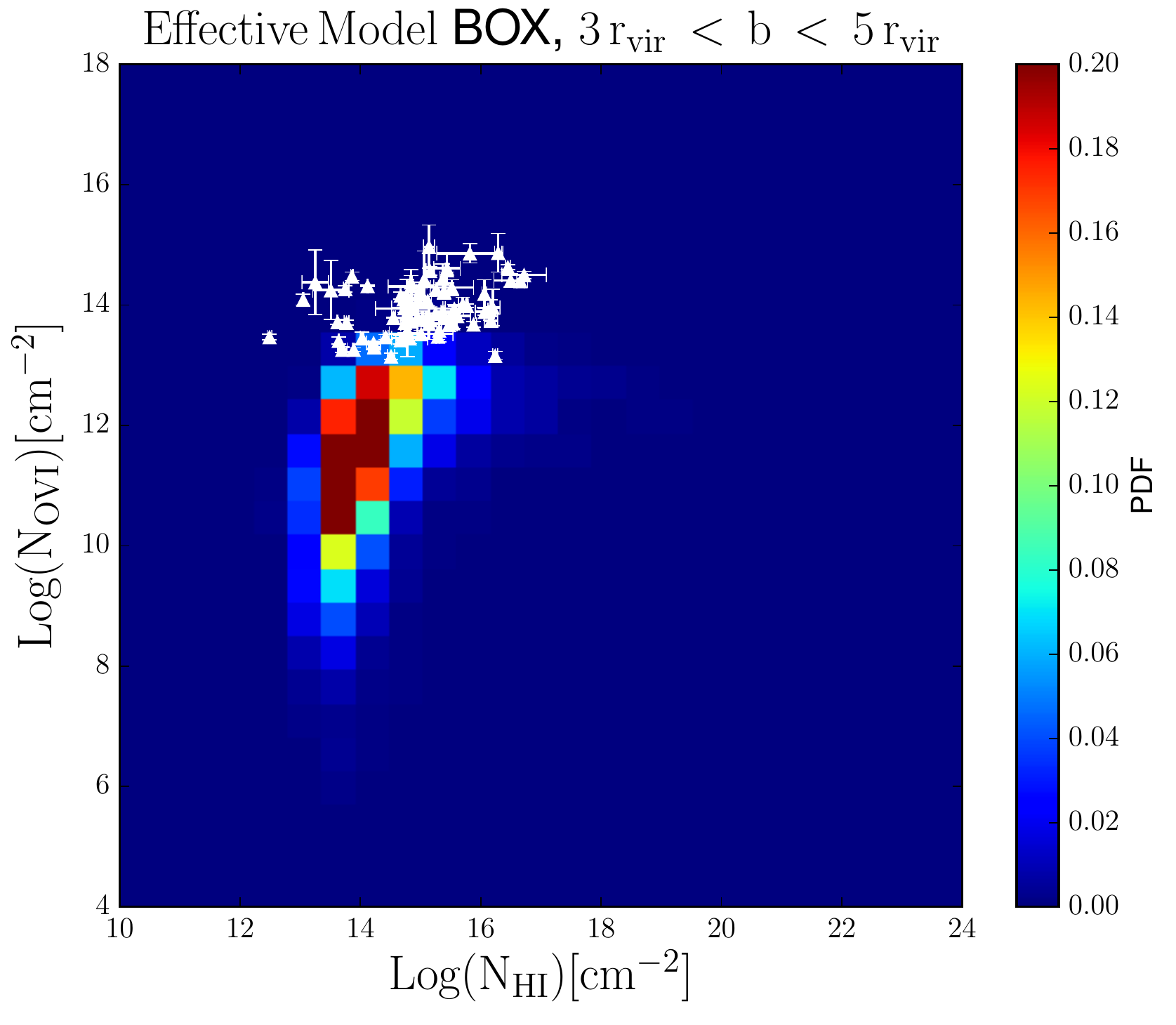}}
   }
     \vspace{0.1cm}
   \makebox[\textwidth][c]{
  \subfloat{\includegraphics[width=0.31\textwidth]{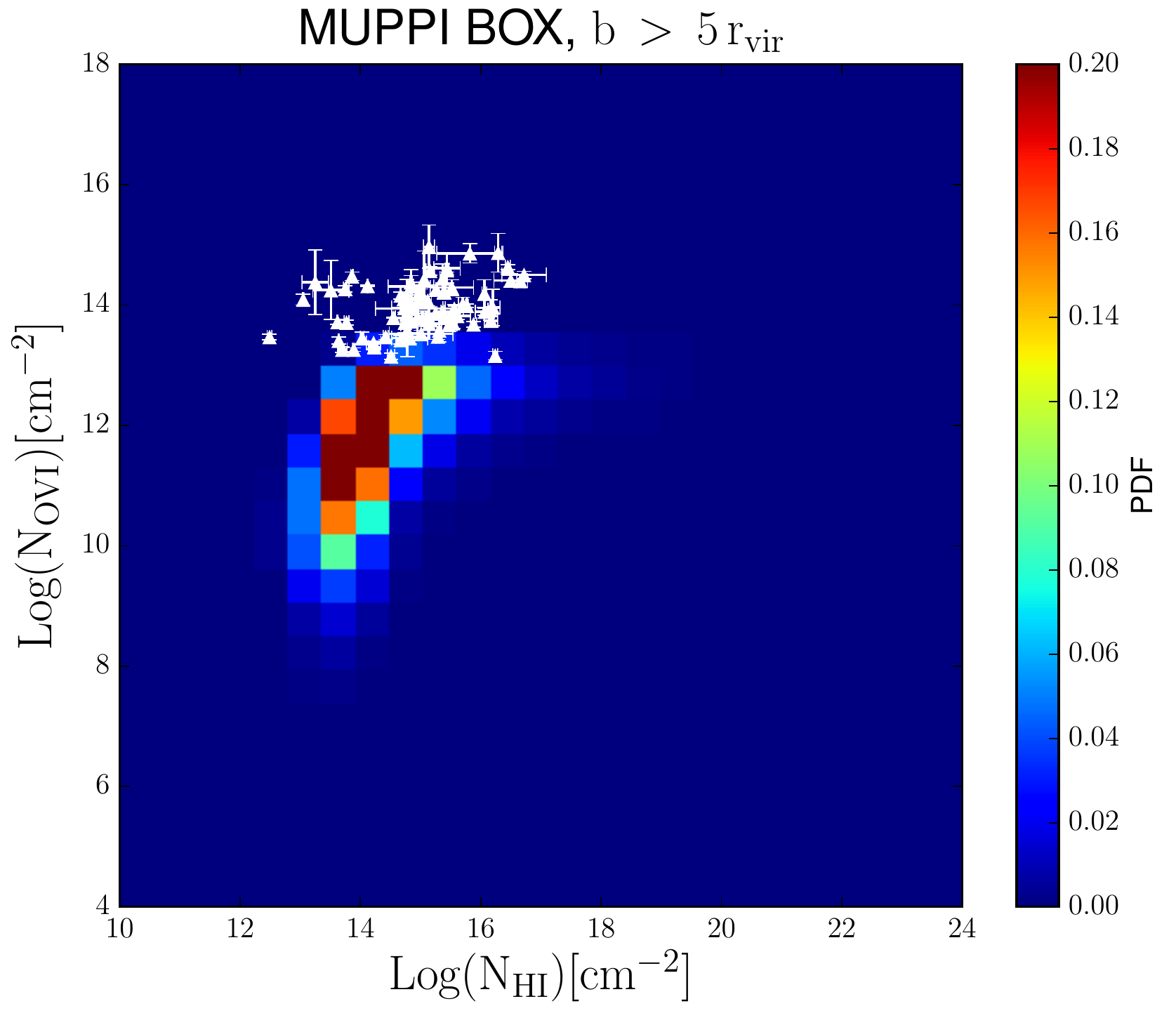}} 
  \hspace{0.5cm}
  \subfloat{\includegraphics[width=0.31\textwidth]{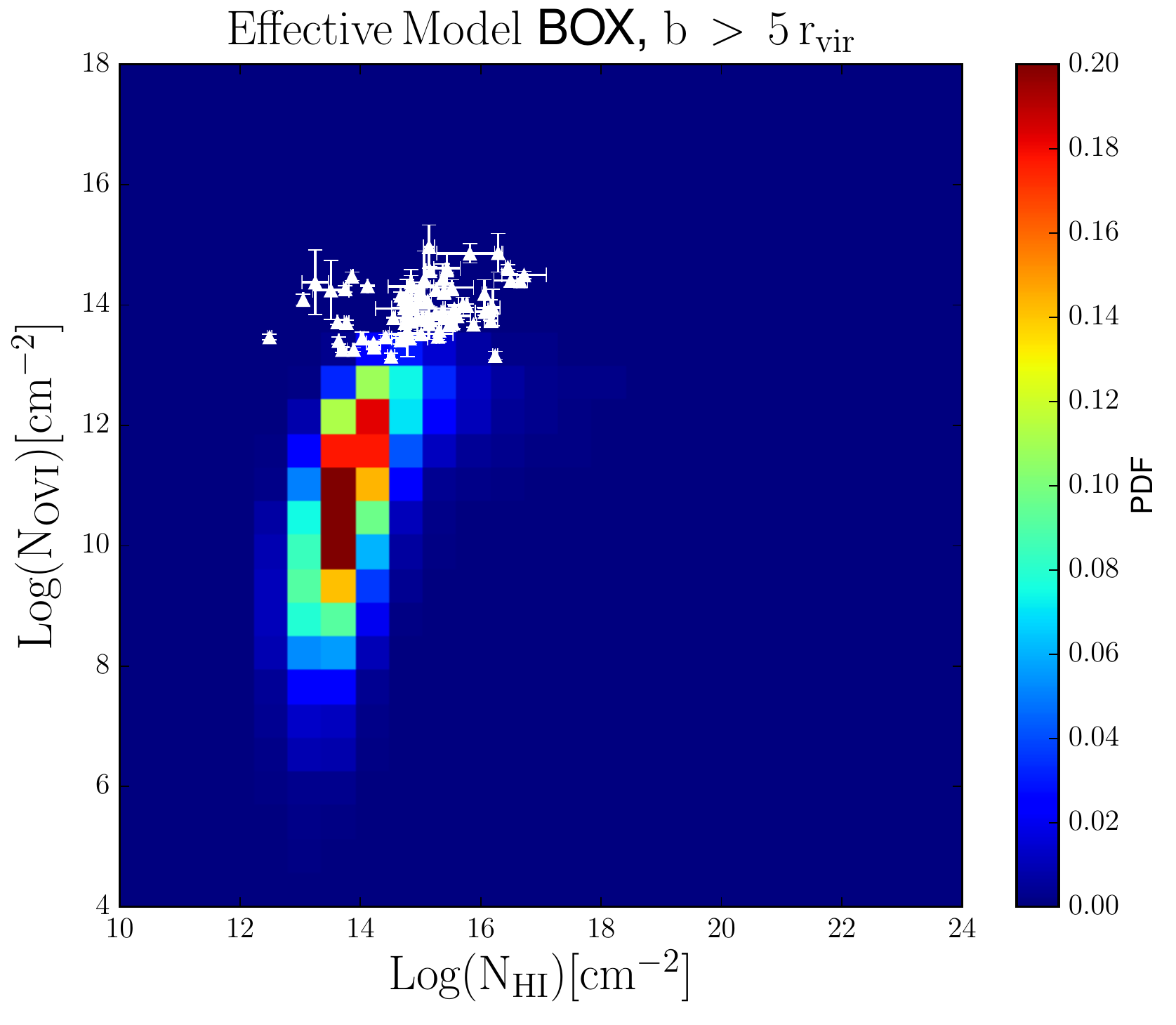}}
 
  } 
   \caption{PDF of N$_{\rm OVI}$ versus N$_{\rm{ H\, \textsc{i}}}$ relation around galaxies. 
   {\it White triangles}: observational data from \citet{Muzahid2012}.}
  \label{fig:OVICompPDF}
\end{figure*}

In Fig.~\ref{fig:OVICompPDF}, we show the PDFs of the N$_{\rm OVI}$ versus N$_{\rm{ H\, \textsc{i}}}$ relation 
for the MUPPI model and the Effective Model respectively, compared to the observational sample by \citet[white triangles with error bars]{Muzahid2012}.
Simulated absorption systems are divided according to the same distance division as previous ions.

The comparison between the two models is already discussed in Section~\ref{ss:CompMUPPIEFF}, so here we examine only the comparison with the observational sample.

Going to larger distances, we can see again that the probability distribution gradually shifts to lower values of \mbox{H\,{\sc i}} and 
\mbox{O\,{\sc vi}} column densities, as the \mbox{C\,{\sc vi}}, but in this case the simulated sample occupies a region shifted towards lower values of
\mbox{O\,{\sc vi}} column densities  with respect to observational data, unlike \mbox{C\,{\sc vi}}.

This behavior seems not to be in agreement with the evidence that the simulation produces too many metals, but these results can be ascribed to the fact that
\mbox{O\,{\sc vi}} could be due both by photoionization and collisional ionization, which is not treated in our simulation. In this case, the problem
of ionization has a stronger importance than the metal production.
Also other works find the same effect (see e.g. \citet{Hummels2013}).

\end{document}